\documentclass[11pt,a4paper]{article}
\pdfoutput=1

\usepackage{jheppub}
\usepackage{graphicx}
\usepackage{epstopdf}
\usepackage[T1]{fontenc}
\usepackage{extarrows}

\usepackage{dcolumn}
\usepackage{bm}
\usepackage{hyperref}
\usepackage{color}
\usepackage[utf8]{inputenc}
\usepackage{morefloats}
\usepackage{subcaption}

\title{All order effective action for  charge diffusion from Schwinger-Keldysh holography}
\author[a]{Yanyan Bu,}
\author[b]{Tuna Demircik,}
\author[c]{and Michael Lublinsky}

\affiliation[a]{School of Physics, Harbin Institute of Technology, Harbin 150001, China}
\affiliation[b]{Asia Pacific Center for Theoretical Physics, Pohang, 37673, Korea}
\affiliation[c]{Department of Physics, Ben-Gurion University of the Negev,
Beer-Sheva 84105, Israel}

\emailAdd{yybu@hit.edu.cn}
\emailAdd{tuna.demircik@apctp.org}
\emailAdd{lublinm@bgu.ac.il}

\abstract{An effective action for diffusion of a conserved $U(1)$ charge is derived to all orders in the derivative expansion within a holographic model dual to the Schwinger-Keldysh closed time path. A systematic approach to solution of the 5D Maxwell equations in a doubled Schwarzschild-AdS$_5$ black brane geometry is developed. Constitutive relation for the stochastic charge current is shown to have a term induced by thermal fluctuations (coloured noise). All transport coefficient functions parameterising the effective action and constitutive relations are computed analytically in the hydrodynamic expansion, and then numerically for finite momenta.}


\arxivnumber{2012.08362}
\begin{document}
\maketitle

\allowdisplaybreaks

\flushbottom

\section{Introduction}\label{intro}

Hydrodynamics~\cite{landau,forster,1963AnPhy..24..419K} is an effective long-time long-distance description of  many-body systems at nonzero temperature. Within the hydrodynamic approximation, the entire dynamics of a microscopic theory is reduced to that of conserved macroscopic currents, such as expectation values of energy-momentum tensor or  charge current operators computed in a locally near equilibrium thermal state. An essential element of any hydrodynamics is a constitutive relation which relates the macroscopic currents to fluid-dynamic variables (fluid velocity, conserved charge densities, etc), and to external forces. Derivative expansion in the fluid-dynamic variables  accounts for deviations from thermal equilibrium.  At each order, the derivative expansion is fixed by thermodynamics and symmetries, up to a finite number of transport coefficients (TCs) such as viscosity and diffusion coefficients. The latter are not calculable from hydrodynamics itself, but have to be determined from underlying microscopic theory or extracted from experiments. In general, relativistic hydrodynamics truncated to any fixed order has a well-known major conceptual problem---it violates causality.
To restore causality one has to introduce higher order gradient terms. Generically,  causality is restored only after all (infinite) order gradients are resummed,
in a way providing a UV completion of  the ``old" hydrodynamics. A compact way of organising the resummation is by introducing, instead of order by order transport coefficients,
{\it  momenta-dependent transport coefficient functions} (TCFs)  \cite{Lublinsky:2009kv}.

The focus of the present paper will be on $U(1)$ charge diffusion. The all order constitutive relation  for the spatial current density $J^i$ has the following general form
\begin{equation}
J^i= -{\mathcal D} \partial_i J^v + \sigma_e \mathcal F_{iv} + \sigma_m \partial_k \mathcal F_{ ik}, \label{Ji}
\end{equation}
where $J^v$ is the charge density, and $\mathcal F$ is field strength of  external $U(1)$ field $\mathcal A_\mu$.
The coefficients $\mathcal D$, $\sigma_e$, and $\sigma_m$ are generalised  diffusion constant, electric and magnetic conductivities. These coefficients are not constants
but rather TCFs, that is, they are functions of four-momentum in Fourier space or
functionals  of space-time derivatives in the real space:
\begin{equation}\label{Dss}
\mathcal D[\partial_v, \vec\partial^{\,2}]\rightarrow \mathcal D[\omega, q^2],~~~~~~~~~\sigma_e[\partial_v, \vec\partial^{\,2}]\rightarrow \sigma_e[\omega, q^2],~~~~~~~\sigma_m[\partial_v, \vec\partial^{\,2}]\rightarrow \sigma_m[\omega, q^2].
\end{equation}
Generically, \eqref{Ji} is a non-local constitutive relation expressible in terms of {\it memory functions}, the inverse Fourier transforms of the TCFs \cite{Bu:2015ika}.

AdS/CFT correspondence \cite{Maldacena:1997re,Gubser:1998bc,Witten:1998qj} is the only known framework, which provides a tractable approach to strongly coupled regime of non-Abelian gauge theories at finite temperature and opens a possibility to explore their transport properties exactly, at least for a class of  gauge theories for which gravity duals can be constructed.
The holographic duality  maps hydrodynamic fluctuations of a boundary fluid into  gravitational perturbations of
a stationary black brane in an asymptotic AdS space \cite{Kovtun:2004de,Policastro:2001yc,Policastro:2002se,Policastro:2002tn,Son:2002sd,
Bhattacharyya:2008jc}.
The original papers on the subject focused on  shear viscosity over entropy density ratio and two-point retarded correlators.
For the latter, Son and Starinets  proposed a computational prescription \cite{Son:2002sd,Son:2007vk}, to be discussed below.
Since then,  the field has developed in different directions.  Second and higher order  TCs were computed for various bulk models, while our team has focused on development of {\it all order resummation} technique \cite{Bu:2014sia,Bu:2014ena,Bu:2015ika}. Particularly, in \cite{Bu:2015ame} we used the Maxwell theory probing the Schwarzschild-AdS$_5$ background to compute  the TCFs introduced in \eqref{Ji} and \eqref{Dss}.

Classical hydrodynamics is dissipative with the TCs, or more generally TCFs, parameterising the rate of dissipation in the fluids.   Yet, dissipation in
non-equilibrium dynamics is  tightly related to  thermal fluctuations via
fluctuation-dissipation relations (FDRs). The latter originate from the energy momentum conservation in a closed system, which includes  an open subsystem and a thermal bath.
A canonical example is Brownian motion of a particle in a thermal bath. FDR renders the  diffusive motion of the particle
into a stochastic process described by Langevin equation.
Similarly,  proper account of thermal fluctuations in fluids, and particularly in relativistic fluids, should render classical  hydrodynamics  into stochastic one.
These ideas have sparked  many interesting developments in formulating an effective field theory (EFT) approach to dissipative hydrodynamics \cite{Dubovsky:2011sj,Dubovsky:2011sk,Endlich:2012vt,Grozdanov:2013dba,Nicolis:2013lma,
Kovtun:2014hpa,Harder:2015nxa,Haehl:2015foa,Haehl:2015uoc,Crossley:2015evo,Crossley:2015tka,
deBoer:2015ija,Glorioso:2016gsa,Glorioso:2017fpd,Glorioso:2017lcn,Jensen:2017kzi, Haehl:2018lcu,Jensen:2018hse}, from which constitutive relations for the
currents could be straightforwardly derived. In  presence of fluctuations the conserved current is expected to take the form\footnote{Strictly speaking, the constitutive relation \eqref{Jinoise} holds for quadratic EFTs only. Beyond that, the constitutive relation is non-linear.}
\begin{equation}
J_i= -\mathcal D \partial_i J^v + \sigma_e \mathcal F_{iv} + \sigma_m \partial_k \mathcal F_{ ik} +\  J_i^{\rm noise}.  \label{Jinoise}
\end{equation}
Here $J_i^{\rm noise}$ is a noise term representing  thermal force.

There is also a phenomenological interest in fluctuating hydrodynamics,  largely driven by studies of quark gluon plasma (QGP), for which relativistic
hydrodynamics is instrumental.   Phenomenological implications  of fluctuating hydrodynamics
for realistic systems such as QGP have been discussed in, e.g., \cite{Kovtun:2003vj,Kovtun:2011np,Young:2014pka,Akamatsu:2016llw,Chen-Lin:2018kfl,
Jain:2020fsm,Bluhm:2020mpc,Jain:2020hcu}. Particularly, in a model independent way, thermal fluctuations can
be integrated out, resulting in emergence of ``effective'' TCs and  shifts in positions of the hydrodynamic poles.
This idea was first considered in \cite{Kovtun:2003vj,Kovtun:2011np,Kovtun:2012rj} and recently  revisited within an
EFT framework in \cite{Chen-Lin:2018kfl}.
While in most phenomenological applications
the noise is assumed to be white,  it is generically non-Gaussian and momenta dependent (coloured), see e.g \cite{Stephanov:2008qz,Kapusta:2014dja}.
The discussion in \cite{Kapusta:2014dja} was based on an Israel-Stewart-type model for  causal diffusion.
The goal of the present paper is to put the above mentioned ideas into a more firm ground by learning
 about the noise structure to all orders in the gradient expansion from a holographic model, in which
such questions can be addressed via a first principle calculation.

While traditional holographic approach based on black hole in AdS (BH-AdS) captures  dissipative effects of the boundary dynamics, it does not include any fluctuation.
In finite temperature QFT, the unified  framework that includes both the fluctuation and dissipation is a closed time path (CTP) integral, also referred to as Schwinger-Keldysh (SK) formalism \cite{Chou:1984es,Kamenev2011,Calzetta-Hu2009}.
From the holographic perspective,  a dual geometry must have SK contour at its conformal boundary
\cite{Herzog:2002pc,Skenderis:2008dg,Skenderis:2008dh,Barnes:2010jp,Nickel:2010pr,
Son:2009vu,deBoer:2008gu,Sonner:2012if,deBoer:2015ija,deBoer:2018qqm,Crossley:2015tka,
Glorioso:2018mmw,CaronHuot:2011dr,Chesler:2011ds,Botta-Cantcheff:2018brv}. In contrast to the single BH-AdS geometry, this is achieved via patching two Lorentzian BH-AdS geometries with an Euclidean BH-AdS geometry. Proper matching conditions for the bulk fields should be imposed at  space-like surfaces at which the geometries are glued  \cite{Skenderis:2008dh,Skenderis:2008dg,vanRees:2009rw,deBoer:2018qqm}.

An alternative prescription has been proposed in \cite{Glorioso:2018mmw}, in which, instead of gluing geometries, the radial (holographic) coordinate has been complexified and analytically continued around the event horizon, forming a  geometry with two copies of BH-AdS space. This latter approach will be referred to as SK holography.
Over the last couple of years,  the SK holography was applied to open quantum systems. The questions about non-Gaussian noise and KMS relations for fermionic degrees of freedom were addressed in \cite{Chakrabarty:2019aeu,Jana:2020vyx,Loganayagam:2020eue,Loganayagam:2020iol,Chakrabarty:2020ohe}. However, the open systems considered in \cite{Chakrabarty:2019aeu,Jana:2020vyx,Loganayagam:2020eue,Loganayagam:2020iol,Chakrabarty:2020ohe} do not involve hydrodynamical low energy
degrees of freedom, for which an EFT formalism to be discussed below is required.

After this general introduction, we  briefly review our setup. We are going to study the $U(1)$ charge diffusion in a thermal plasma in 4d. This will be derived from a probe Maxwell theory in the doubled Schwarzschild-AdS$_5$ geometry.
For the holographic SK formalism we will closely follow \cite{Glorioso:2018mmw}, which derived an effective action for diffusion, up to second order in the derivative expansion. One of our results will be the effective action computed to all orders in the derivative expansion.  We will demonstrate that, thanks to linearity of the Maxwell equations in the bulk,
the resulting effective action is quadratic in the dynamical fields and  takes the precise form proposed in \cite{Crossley:2015evo} (see \eqref{effectiveaction}). The latter was derived from
general symmetry-based considerations.  In the next section we will flash the relevant results from  \cite{Crossley:2015evo}.
The core of our calculation is in  solving the bulk equations of motion (EOMs) in the doubled Schwarzschild-AdS$_5$.
Following the idea introduced by two of us in  \cite{Bu:2014sia,Bu:2014ena}, we will be solving the dynamical equations only, leaving the constraint  aside. This makes it possible to
construct the {\it ``off-shell'' constitutive relations} and {\it ``off-shell'' hydrodynamic effective action}. This approach is nowadays referred to as ``off-shell'' holography  \cite{Crossley:2015tka,deBoer:2015ija}.
At a technical level, our treatment of the bulk EOMs will be somewhat different and more systematic compared to  that of \cite{Glorioso:2018mmw}: we will
first search for a complete set of independent solutions in a single copy of the doubled Schwarzschild-AdS$_5$, and then will carefully match the two segments of the doubled Schwarzschild-AdS$_5$ near the event horizon. In this respect our formalism is more in spirit of
\cite{Skenderis:2008dh,Skenderis:2008dg}.  The latter, however, glued geometries along the space-like surfaces.
When expanded to second order in the derivatives, our results could be compared with those of \cite{Glorioso:2018mmw,deBoer:2018qqm}.  While
most of the coefficients are found to match, there are also some disagreements between all three results. The comparison and discussion  are presented  in subsection \ref{analy_results}.

The main results of this paper are

\noindent $\bullet$ Derivation from the SK holography of the effective action  \cite{Crossley:2015evo} for the charge diffusion,
from which the constitutive relation with the noise term in the form \eqref{Jinoise} follows straightforwardly.

\noindent $\bullet$ Computation of all the TCFs parameterising the effective action. These are computed analytically up to the second order in the derivative expansion and then numerically for finite (large) momenta. All the TCFs are analytically shown to satisfy the symmetry-imposed relations introduced in \cite{Crossley:2015evo}.

\noindent $\bullet $ The noise-noise correlator is computed, showing non-locality in the space-time.

\noindent $\bullet$ Derivation  of the prescription \cite{Son:2002sd} for  retarded two-point correlators, starting from the SK holography, as opposed to
the original work  based on a single  BH-AdS geometry \footnote{The prescription \cite{Son:2002sd} was derived in \cite{Herzog:2002pc,vanRees:2009rw,Iqbal:2009fd} for a probe scalar field. Yet, to the best of our knowledge,  it has not been derived for a bulk gauge field.}(see Appendix \ref{Son_Starinets_prescription}).

The paper is structured as follows. In Section \ref{diffusion_action_review}, the effective action \cite{Crossley:2015evo}  for the diffusion at quadratic order and the TCFs parameterising it are reviewed. This Section also introduces the symmetry-induced relations among the TCFs and a discussion of the constitutive relations for the current with noise. The SK holography is introduced in Section \ref{holo_setup}. Solutions to the Maxwell's equation in the bulk are presented in Section \ref{solve_bulk_dynamics}. The results for the TCFs as well as noise-noise correlator are presented in Section \ref{results}. A brief summary and outlook is presented in Section \ref{summary_outlook}. The effective action for the charge diffusion proposed in \cite{Crossley:2015evo} is derived in Appendix \ref{general_structure}. A subtle point regarding the near-horizon matching condition for the time component of the bulk gauge field is further clarified in Appendix \ref{crosscheck_dCv_Cv}. In Appendix \ref{Son_Starinets_prescription}, the prescription \cite{Son:2002sd} for  retarded current-current correlators is derived starting from the SK holography. In Appendix \ref{numTCFs}, the numerical results for independent TCFs (say, $w_{5,7,8,9}$) parameterising the effective action are presented.

{\bf Note added:} While preparing this paper for release, we got aware of the recent work \cite{Ghosh:2020lel}.
Just like us, \cite{Ghosh:2020lel} considers the Maxwell's theory within the SK holography
and constructs an EFT for stochastic diffusion. Both papers employ the time-reversal symmetry to relate the ingoing modes (dissipation) with the outgoing modes
(fluctuation/Hawking radiation).  Our first impression is that \cite{Ghosh:2020lel}  constructed EFT on-shell only, whereas we have obtained results for both on-shell and off-shell EFTs. Particularly,
if our understanding is correct,  Chapter 8 of \cite{Ghosh:2020lel} is quite similar to our  Appendix \ref{Son_Starinets_prescription}.
Admittedly, a much more careful study of \cite{Ghosh:2020lel} would be needed to fully appreciate the degree of overlap and agreement between the two papers.

\section{Effective field theory for charge diffusion} \label{diffusion_action_review}

In this section, we review the hydrodynamic effective action derived in \cite{Crossley:2015evo} and the symmetry properties of the TCFs parameterising it. We will also address the constitutive relation for the  fluctuating $U(1)$ current. Finally, we present the general structure of the momenta-dependent (coloured) noise and the noise-noise correlator.

\subsection{Effective action}

At  quadratic level in the dynamical fields, the most general form of the effective action for the $U(1)$ charge diffusion was derived in \cite{Crossley:2015evo}:
\begin{align}\label{effectiveaction}
S_{\rm eff}= \int d^4x \mathcal{L}_{\rm eff}(x),
\end{align}
where the effective Lagrangian is
\begin{align} \label{Leff3}
\mathcal{L}_{\rm eff}= & \frac{i}{2}B_{av}(x) w_1B_{av}(x) + \frac{i}{2} B_{ak}(x) w_2B_{ak}(x) + \frac{i}{2} \partial_k B_{ak}(x) w_3 \partial_l B_{al}(x) \nonumber \\
+& i B_{av}(x) w_4 \partial_k B_{ak}(x) + B_{av}(x) w_5 B_{rv}(x) + B_{av}(x) w_6 \partial_v \partial_k B_{rk}(x) \nonumber \\
+& \partial_k B_{ak}(x) w_7 B_{rv}(x) + B_{ak}(x) w_8 \partial_v B_{rk}(x) + \frac{1}{2} \mathcal{F}_{akl}(x) w_9 \mathcal{F}_{rkl}(x),
\end{align}
where
\begin{align}
B_{r\mu}= \frac{1}{2}(B_{1\mu}+ B_{2\mu}), \qquad  B_{a\mu}= B_{1\mu} - B_{2\mu}.
\end{align}
In \eqref{Leff3}, $\mathcal{F}_{r\mu\nu}$ and $\mathcal{F}_{a\mu\nu}$ are the field strengths of $B_{r\mu}$ and $B_{a\mu}$, respectively.
Here, $B_{1\mu}$ and $B_{2\mu}$ live on the upper and lower branches of the SK contour, and are defined as independent $U(1)$ gauge transforms of the background gauge fields $\mathcal{A}_{1\mu}$ and $\mathcal{A}_{2\mu}$ \cite{Crossley:2015evo}
\begin{align}
B_{1\mu}\equiv \mathcal{A}_{1\mu}+ \partial_\mu \varphi_1, \qquad  B_{2\mu}\equiv \mathcal{A}_{2\mu}+ \partial_\mu \varphi_2,
\end{align}
where the gauge transformation parameters $\varphi_1$ and $\varphi_2$ are treated as low energy hydrodynamical modes.

The parameters $w_{1\cdots9}$ are the TCFs: they are $SO(3)$ scalar functionals of the space-time derivatives. As explained in section \ref{intro}, in momentum space these TCFs  become  functions of frequency $\omega$ and spatial momentum $\vec q$.

The generating functional $W[\mathcal A_{a\mu}, \mathcal A_{r\mu}]$ is obtained by integrating over the dynamical fields $\varphi_1$ and $\varphi_2$ or, alternatively, over $\varphi_r$ and $\varphi_a$:
\begin{align}\label{WAA}
e^{W[\mathcal A_{a\mu}, \mathcal A_{r\mu}]}\equiv \int D \varphi_r D \varphi_a \ e^{iS_{\rm eff}[B_{a\mu}, B_{r\mu}]}.
\end{align}
Normalisation  of $W$ is such that $W[\mathcal A_{a\mu}=0, \mathcal A_{r\mu}] =0$.

The  hydrodynamic effective action \eqref{effectiveaction} could be thought of as being obtained by integrating out the gapped modes of an underlying microscopic theory defined on the CTP (SK contour). While the microscopic theory is formulated on the SK contour,  the low energy EFT \eqref{effectiveaction} (also \eqref{WAA}) is defined with  time running forward only, along the real axes.

Two currents defined as
\begin{align}\label{currents}
J_r^\mu(x) = \frac{\delta S_{\rm eff}}{\delta {\mathcal A}_{a\mu}(x)}, \qquad \qquad
J_a^\mu(x) = \frac{\delta S_{\rm eff}}{\delta {\mathcal A}_{r\mu}(x)}
\end{align}
are conserved by  the EOMs for the dynamical fields $\varphi_r$ and $\varphi_a$, which are derived from variation of the effective action \eqref{effectiveaction}.

\subsection{Discrete symmetries}\label{ds}

The effective action $S_{\rm eff}$  possesses several discrete symmetries, including parity $\mathcal{P}$ and time reversal $\mathcal{T}$, inherited from the underlying microscopic theory.
These symmetries impose relations among the TCFs $w_i$'s, which we review here,  see \cite{Crossley:2015evo} for details.

\noindent $\bullet$  $Z_2$-reflection symmetry:
\begin{align}
S_{\rm eff}[B_{1\mu}; B_{2\mu}]= - S_{\rm eff}[B_{2\mu}; B_{1\mu}],
\end{align}
which implies that the coefficients of the leading terms in the derivative expansion of $w_i$'s must be non-negative.

\noindent $\bullet$ $\mathcal{T}$-symmetry of \eqref{Leff3} translates into  local KMS conditions,
\begin{align}
&w_1= -\frac{i}{2} \coth \frac{\beta \omega}{2} (w_5-w_5^*), \label{w1_w5} \\
&w_2+w_3q^2= -\frac{\omega}{2} \coth \frac{\beta \omega}{2} (w_8+w_8^*), \quad w_3= \frac{i}{2} \coth \frac{\beta \omega}{2} (w_9-w_9^*), \label{w2_w8+w3_w9} \\
&w_4= -\frac{1}{2} \coth \frac{\beta \omega}{2} (\omega w_6- iw_7^*). \label{w4_w6w7}
\end{align}

\noindent $\bullet$ $\mathcal{PT}$-symmetry leads to Onsager relations,
\begin{align} \label{w6_w7}
w_4=-w_4^*, \qquad \omega w_6= -i w_7,
\end{align}
which makes it possible to rewrite the relation \eqref{w4_w6w7} as
\begin{align} \label{w4_w7+w4_w6}
w_4= \frac{i}{2} \coth \frac{\beta \omega}{2} (w_7+w_7^*)= - \frac{\omega}{2} \coth \frac{\beta \omega}{2} (w_6 - w_6^*).
\end{align}
The TCFs $w_1,w_2,w_3$ are real functions of  $\omega$ and $q$, while $w_4$ is purely imaginary.
Overall, there are  {\it four} independent parameters in \eqref{Leff3}, which could be taken as $w_5$, $w_7$ (or equivalently $w_6$), $w_8$ and $w_9$.
The non-fluctuating $U(1)$ current has  three TCFs only (see (\ref{transports})). Yet,
stochastic $U(1)$ current  possesses an additional TCF.
While the effective action/constitutive relations are parameterised by four (independent) coefficients, the number of  independent
two-point correlators is only two, with the others  being related by the FDR.

The main goal of the present  paper   is to derive (\ref{Leff3}) from a holographic model and compute
 $w_i$'s to all orders in the derivative expansion. It will be demonstrated analytically that all the
symmetry induced relations introduced above are automatically satisfied by the holographic construction.

\subsection{  $U(1)$ current with thermal noise}

From \eqref{currents} (see also equations (4.21)-(4.23)  in \cite{Crossley:2015evo}),
\begin{align}
& J_r^v= w_5 B_{rv} + w_6 \partial_v \partial_k B_{rk} + i  w_1 B_{av}  + i w_4 \partial_k B_{ak}\nonumber \\
& J_r^i= -w_7 \partial_i B_{rv} + w_8 \partial_v B_{ri} +w_9 \partial_k \mathcal{F}_{r ik}  +   i w_2B_{ai}  - iw_3 \partial_i \partial_l B_{al}  - i w_4^* \partial_i  B_{av} \nonumber \\
& J_a^v= w_5^* B_{av} + w_7^* \partial_k B_{ak}\nonumber \\
& J_a^i=w_6^* \partial_v \partial_i B_{av} - w_8^* \partial_v B_{ai}+w_9^* \partial_k \mathcal{F}_{a ik}
\label{JrJa}
\end{align}
When $B_a=0$, $J_a^\mu$ vanishes while $J_r^\mu$ becomes  the hydrodynamic current $J_{hydro}^\mu$ \cite{Crossley:2015evo}:
\begin{align}
& J_{hydro}^v= w_5 B_{rv} + w_6 \partial_v \partial_k B_{rk}= (w_5 + w_6 \vec\partial^{\,2}) \mu - w_6 \partial_k \mathcal F_{rkv}, \nonumber \\
& J_{hydro}^i= -w_7 \partial_i B_{rv} + w_8 \partial_v B_{ri} +w_9 \partial_k \mathcal{F}_{r ik} = (w_8-w_7) \partial_i \mu - w_8 \mathcal F_{riv} +w_9 \partial_k \mathcal F_{rik}\label{Jrmu}
\end{align}
$J_{hydro}^v$ is the charge density and   $\mu = B_{rv}$ is identified with the chemical potential.
With $\mu$ replaced by the charge density, the current density $\vec J_{hydro}$ is cast into the same form (\ref{Ji}) as that of \cite{Bu:2015ame}
\begin{align}
J_{hydro}^i= -\mathcal D \partial_i J_{hydro}^v + \sigma_e \mathcal F_{riv} + \sigma_m \partial_k \mathcal F_{rik}, \label{Jhydro}
\end{align}
where
\begin{align}
\mathcal D= \frac{w_7- w_8}{ w_5 + w_6 \vec\partial^{\,2}}, \quad \sigma_e = \frac{w_8- w_7}{ w_5 + w_6 \vec\partial^{\,2}} w_6 \vec\partial^{\,2} -w_8, \quad \sigma_m= w_9- \frac{w_8- w_7}{ w_5 + w_6 \vec\partial^{\,2}} w_6 \partial_v. \label{transports}
\end{align}

Thermal fluctuations are turned on by relaxing $B_a=0$ approximation. We can still set
$\mathcal{A}_{1\mu}=\mathcal{A}_{2\mu}=\mathcal{A}_{\mu}$, since $\mathcal{A}_{\mu}$ is an external field, which is not necessarily assumed to be fluctuating:
\begin{equation}
B_{r\mu}\,=\,\mathcal{A}_{\mu}\,+ \,\partial_\mu {\varphi_r}\, ~~~~~~~~~~B_{a\mu}\,=\partial_\mu {\varphi_a}
\end{equation}
The $\varphi_a$  field acts as a source of noise both
 for the charge density and  hydrodynamic current $\vec J_{hydro}$:
\begin{align}
& J_r^v= J_{hydro}^v + i (w_1 \partial_v + w_4\vec\partial^{\,2})  \varphi_a, \qquad \quad
J_r^i= J_{hydro}^i +i (w_2 -w_3 \vec\partial^{\,2}-w_4^*\partial_v)  \partial_i \varphi_a, \nonumber \\
& J_a^v=  (w_5^*\partial_v +w_7^* \vec\partial^{\,2}) \varphi_a, \qquad \qquad \qquad \quad
 J_a^i= (w_6^* \partial_v^2\partial_i -w_8^* \partial_v\partial_i)\varphi_a
\label{JrJa1}
\end{align}
The first line of (\ref{JrJa1}) is an all order stochastic constitutive relation for the conserved current $J_r^\mu$, which can be recast into
\begin{equation}\label{Jri}
J_r^i= -\mathcal D \partial_i J_r^v + \sigma_e \mathcal F_{iv} + \sigma_m \partial_k \mathcal F_{ ik} + \Xi\partial_i \varphi_a,
\end{equation}
where the $\Xi$-term acts as a thermal force:
\begin{equation}
\Xi= i\mathcal D  (w_1 \partial_v + w_4 \vec\partial^{\,2}) +i (w_2 -w_3 \vec\partial^{\,2} - w_4^*\partial_v)\label{Xi}
\end{equation}
While $J_r^\mu$ is conserved, in presence of thermal fluctuations, the hydrodynamical current $J_{hydro}^\mu$ is not:
\begin{equation}
\partial_\mu J_{hydro}^\mu= \xi, ~~~~~~~~~~~ \xi\equiv  G_0\varphi_a
\end{equation}
with
\begin{align}
G_0= &-i \left[ w_1 \partial_v^2 + w_4 \partial_v \vec\partial^{\,2} + ( w_2-w_3\vec \partial^{\,2})\vec \partial^{\,2} -w_4^* \partial_v \vec \partial^{\,2} \right] \nonumber \\
&= i \coth \frac{\beta\omega}{2}\left\{ \omega^2 Im(w_5) -\omega q^2 Re(w_8) + 2\omega q^2 Re(w_7) \right\} \nonumber \\
&= i \coth \frac{\beta\omega}{2}\left\{ \omega^2 Im(w_5) -\omega q^2 Re(w_8) - 2\omega^2 q^2 Im(w_6) \right\} \label{G0}
\end{align}
$G_0$ is clearly purely imaginary.
We have recast the continuity equation into the usual stochastic form. $G_0$ is related to the retarded current-current correlator $G_R$ \eqref{G_R}.
Up to $\coth\frac{\beta \omega}{2}$ pre-factor, $G_0$ is the real part of the denominator in the expression for $G_R$  obtained from \eqref{transports} (see \eqref{G_R})
\begin{align}
G_R^{vv} = \frac{q^2\,\sigma_e}{-i\omega + q^2 \mathcal D} = \frac{i\omega q^4 w_6^2 - q^2w_5 w_8}{-i\omega w_5 + 2i\omega q^2 w_6 -q^2 w_8}
\end{align}
Hence $G_0$ vanishes at the poles of the $G_R$ correlator, which in holography are determined by the quasi-normal modes.

It is worth noticing that the noise is a scalar, that is, only the longitudinal sector is fluctuating. This  reflects the fact that
physically the quantity that actually fluctuates is the charge density. There are no fluctuations in the transverse sector, in which the current is induced  by the external fields, assuming that the latter are not fluctuating.

The noise is Gaussian but coloured as it depends on four-momentum. Changing variable from $\varphi_a$ to $\xi$ in the action $S_{\rm eff}$ results in the following noise-noise correlator
\begin{equation}
\langle \xi(x)\,\xi(0)  \rangle\, =\tilde G_0(x)
\end{equation}
where $ \tilde G_0(x)$ is the inverse Fourier transformation of $-i\,G_0$.
Since $-i\,G_0$ is a real function in the momentum space, the noise-noise correlator is symmetric as expected.
Numerical results for $ \tilde G_0$ will be presented in subsection \ref{noise_plot}. Contrary to the white noise behaviour
($\delta$-functional form for $\tilde G_0$)  we will observe non-local space-time effects in the noise sector.

\section{Holographic setup} \label{holo_setup}

\subsection{The geometry}

The metric of  Schwarzschild-AdS$_5$ in the ingoing Eddington-Finkelstein (EF) coordinate system $x^M=(r,v,x^i)$ is given by the line element
\begin{align} \label{metric_EF}
ds^2= g_{MN} dx^M dx^N= -f(r)dv^2 +2 dvdr + r^2 \delta_{ij} dx^i dx^j, \qquad i,j=1,2,3,
\end{align}
where $f(r)=r^2-r_h^4/r^2$. We will also use the Schwarzschild coordinate system $\tilde x^M=(r,t,x^i)$, for which the metric \eqref{metric_EF} is
\begin{align} \label{metric_Schw}
ds^2= \tilde g_{MN} d \tilde x^M d \tilde x^N =\frac{dr^2}{f(r)}-f(r)dt^2 +r^2 \delta_{ij} dx^i dx^j, \qquad i,j=1,2,3.
\end{align}
In both \eqref{metric_EF} and \eqref{metric_Schw}, the curvature radius of the AdS space is set to unity.

\begin{figure}[htbp]
\centering
\includegraphics[width=0.8\textwidth]{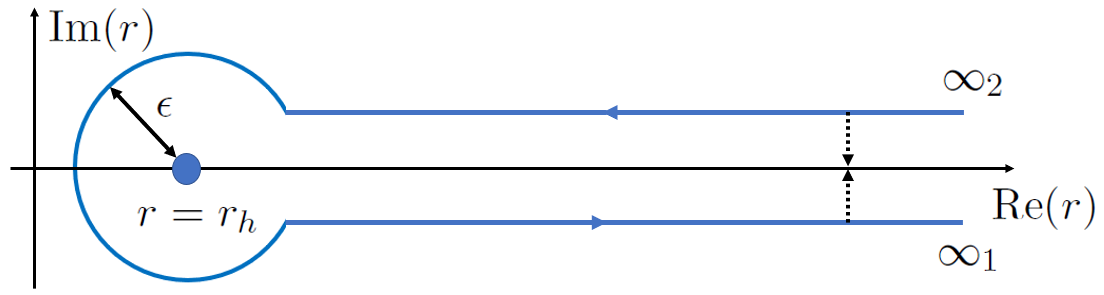}
\caption{The holographic SK contour of \cite{Glorioso:2018mmw}: the complexified radial coordinate analytically continued around the event horizon $r=r_h$. The dashed arrows indicate that the horizontal segments are infinitesimally close to the real axis.}
\label{holographic_SK_contour}
\end{figure}

The holographic geometry dual to thermal  state with the SK contour at the boundary is a doubled Schwarzschild-AdS$_5$.
We will closely follow   the holographic prescription of \cite{Glorioso:2018mmw}, which doubled the geometry (\ref{metric_EF}) by complexifying  the radial coordinate $r$ along the contour illustrated in Figure \ref{holographic_SK_contour}.

The two ends of the $r$ contour are identified with the SK contour of the boundary theory \cite{Kamenev2011,Calzetta-Hu2009}, the infinitesimally small horizon circle is mapped into
 initial thermal state,  while the  horizontal segments reflect the CTP.
The holographic contour of Figure \ref{holographic_SK_contour} is obtained by taking  two exteriors of an eternal AdS$_5$
black hole \cite{Maldacena:2001kr} and identifying their future horizons.

The EF time $v$ is related to the Schwarzschild time $t$ by
\begin{align}
& t=v- \zeta_2(r), \qquad \zeta_2(r) \equiv \int_{\infty_2}^r \frac{dy}{f(y)}, \qquad r \in [r_h-\epsilon,\infty_2), \nonumber \\
& t=v- \zeta_1(r), \qquad \zeta_1(r) \equiv \int_{\infty_1}^r \frac{dy}{f(y)}, \qquad r \in [r_h-\epsilon,\infty_1),  \label{t_v_zeta12}
\end{align}
where the integration constants are fixed by  requirement that $t$ and $v$  coincide on the AdS boundaries. An interesting observation is that viewed in the ingoing EF coordinate
(that is the EF time $v$ is identical everywhere along the radial contour), the Schwarzschild time $t$ is  discontinuous at $r=r_h-\epsilon$,
\begin{align}
t^{\rm up}(r_h-\epsilon)- t^{\rm dw}(r_h-\epsilon)= \int_{\infty_1}^{\infty_2} \frac{dy}{f(y)}= -\frac{i\pi }{2r_h}= -\frac{i\beta}{2},
\end{align}
where $\beta$ is  inverse of the black brane temperature $T$.
This  becomes important when gluing bulk fields of the upper  and  lower segments of the contour in Figure \ref{holographic_SK_contour}.

\subsection{Maxwell field in the bulk}

The holographic model for the $U(1)$ diffusion is  a probe Maxwell field in the above described geometry.
The bulk action  is
\begin{align}
S_0&=- \frac{1}{4} \int d^4x \int_{\infty_2}^{\infty_1} dr \sqrt{-g}F_{MN} F^{MN} \label{bulk_action_EF}\\
&= - \frac{1}{4} \int d^4 \tilde x \int_{\infty_2}^{\infty_1} dr \sqrt{-\tilde g} \tilde F_{MN} \tilde F^{MN}, \label{bulk_action_Schw}
\end{align}
where $C_N$ and $\tilde C_N$ are the $U(1)$ bulk gauge fields in  EF and Schwarzschild coordinate systems respectively;
$F_{MN}=\nabla_M C_N- \nabla_N C_M$ and $\tilde F_{MN}=\tilde \nabla_M \tilde C_N- \tilde\nabla_N \tilde C_M $. To remove the UV divergences near the AdS boundaries $r=\infty_1$ and $r=\infty_2$, the bulk action \eqref{bulk_action_EF} (or \eqref{bulk_action_Schw}) should be supplemented  with a counter-term action:
\begin{align} \label{counter}
S_{\rm c.t.}=\frac{1}{4}\log r \int d^4x \sqrt{-\gamma} F_{\mu\nu} F^{\mu\nu} \bigg|_{r=\infty_1} - \frac{1}{4}\log r \int d^4x \sqrt{-\gamma} F_{\mu\nu} F^{\mu\nu} \bigg|_{r=\infty_2},
\end{align}
where the indices are contracted with the induced metric $\gamma_{\mu\nu}$
\begin{align}
ds^2|_{\Sigma}= \gamma_{\mu\nu} dx^\mu dx^\nu = -f(r)dv^2+r^2\delta_{ij}dx^idx^j.
\end{align}
Here $\Sigma$ denotes either the hypersurface  $r=\infty_1$ or $r=\infty_2$. The counter-term action in the Schwarzschild coordinate system is the same as \eqref{counter},
since $v=t$ at the AdS boundaries. The minimal subtraction scheme \eqref{counter}
differs from that used in \cite{Glorioso:2018mmw}.

Transformation rule for the fields from the EF coordinate system to the Schwarzschild  can be easily derived from the coordinate-invariants
\begin{align}
\tilde C_M d\tilde x^M = C_M dx^M,   \label{CM_dxM}
\end{align}
leading to
\begin{align}
\tilde C_t(r,t,\vec x)= C_v(r,v,\vec x), \qquad  \tilde C_i(r,t,\vec x)= C_i(r,v,\vec x). \label{Cmu_Schw_EF_coordinate}
\end{align}
The radial components of the bulk gauge field differ
\begin{align}
&\tilde C_r(r,t,\vec x)- \frac{\tilde C_t(r,t,\vec x)}{f(r)}= C_r(r,v,\vec x)
\Longrightarrow  \tilde C_r(r,t,\vec x)= C_r(r,v,\vec x) + \frac{C_v(r,v,\vec x)}{f(r)}. \label{Cr_Schw_EF_coordinate}
\end{align}
Thus it is important to distinguish between two radial gauge choices:
\begin{align}
{\rm Schwarzschild ~ radial ~ gauge}: \qquad &\tilde C_r=0 \Longleftrightarrow C_r= - \frac{C_v}{f(r)}, \nonumber \\
{\rm EF ~ radial ~ gauge}: \qquad & C_r=0 \Longleftrightarrow \tilde C_r = \frac{\tilde C_t}{f(r)}. \label{radial_gauge_Schw_EF}
\end{align}
 The EF radial gauge is most commonly used, including in \cite{Glorioso:2018mmw}. Yet, for reasons related to time-reversal symmetry  which will be explained in the next section,
 we chose to perform  calculations in  the Schwarzschild  radial gauge.

Bulk EOMs are derived by variation of  \eqref{bulk_action_EF} and  \eqref{bulk_action_Schw}:
\begin{align}
&\nabla_M F^{MN} =0 \Longrightarrow \frac{1}{\sqrt{-g}}\partial_M(\sqrt{-g} F^{MN})=0, \label{bulk_EOM_EF} \\
& \tilde \nabla_M \tilde F^{MN} =0 \Longrightarrow \frac{1}{\sqrt{-\tilde g}} \tilde \partial_M(\sqrt{-\tilde g} \tilde F^{MN})=0. \label{bulk_EOM_Schw}
\end{align}
With the help of \eqref{Cmu_Schw_EF_coordinate} and \eqref{Cr_Schw_EF_coordinate}, the two sets of the Maxwell equations can be related
\begin{align}
 \tilde \nabla_M \tilde F^{Mr} = \nabla_M F^{Mr}, \quad  \tilde \nabla_M \tilde F^{Mt}= \nabla_M F^{Mv} - \frac{1}{f(r)} \nabla_M F^{Mr}, ~~~
 \tilde \nabla_M \tilde F^{Mi}= \nabla_M F^{Mi}. \label{eom_Schw_EF}
\end{align}
The first equation (the $r$-component) is a gauge invariant constraint, which will play a special role in our construction.  When all the bulk equations are solved (i.e., on-shell holography),
\eqref{bulk_EOM_EF} and \eqref{bulk_EOM_Schw} are absolutely equivalent as is obvious from \eqref{eom_Schw_EF}. Yet,
in \cite{Bu:2015ame} we argued that in order to compute the TCFs parameterising the (off-shell) constitutive relations for the current,  it is sufficient to solve the bulk dynamical equations only, while leaving the constraint aside.   Within the holographic prescription the  constraint is mapped into  continuity equation for the current at the boundary, which
is the dynamical equation for the  low energy  modes $\varphi_1$ and $\varphi_2$. Derivation of the effective action follows the very same strategy as introduced in \cite{Bu:2014sia,Bu:2014ena,Bu:2015ika,Bu:2015ame},  now frequently  referred to   as  off-shell  holography \cite{Crossley:2015tka,deBoer:2015ija}.

We are going to solve the EOMs in the Schwarzschild coordinates and then re-express the results
 in the EF coordinates using
\eqref{Cmu_Schw_EF_coordinate} and \eqref{Cr_Schw_EF_coordinate}.
 In the  spirit of the off-shell formalism, we will not impose the constraint equation.
 Hence, the dynamical equations which will be solved are
\begin{align}
\nabla_M F^{Mv} - \frac{1}{f(r)} \nabla_M F^{Mr}=0  \Longleftrightarrow \tilde \nabla_M \tilde F^{Mt}=0, \qquad
\nabla_M F^{Mi}=0 \Longleftrightarrow \tilde \nabla_M \tilde F^{Mi}=0.  \label{dynamical_EOM}
\end{align}
Notice that the first equation differs from  $\nabla_M F^{Mv} = 0$ derived in the EF coordinates.  The seemingly freedom to modify the dynamical equation is eliminated when
the Schwarzschild radial gauge is implemented in the effective action.

We will find the Bianchi identity \cite{Glorioso:2018mmw} is quite useful:
\begin{align}
\partial_r(\sqrt{-g} \nabla_M F^{Mr}) + \partial_v (\sqrt{-g} \nabla_M F^{Mv}) + \partial_k (\sqrt{-g} \nabla_M F^{Mk})=0. \label{Bianchi}
\end{align}
The dynamical equations \eqref{dynamical_EOM} are instrumental in deriving
 a holographic RG flow-like equation for $\nabla_M F^{Mr}$:
\begin{align}
\partial_r\left(\sqrt{-g} \nabla_M F^{Mr}\right)= \frac{i\omega}{f(r)}\sqrt{-g} \nabla_M F^{Mr},
\end{align}
which is solved by
\begin{align}
& \sqrt{-g}\nabla_M F^{Mr}= \mathcal{C}^{\rm up}(k) e^{i\omega\zeta_2(r)}, \qquad r\in [r_h-\epsilon, \infty_2), \nonumber \\
& \sqrt{-g}\nabla_M F^{Mr}= \mathcal{C}^{\rm dw}(k) e^{i\omega\zeta_1(r)}, \qquad r\in [r_h-\epsilon, \infty_1). \label{constraint_identity}
\end{align}
Here, $\mathcal{C}^{\rm up, \, dw}$ are  $r$-independent integration constants in either upper or lower segments of the $r$-contour.
Both  vanish on-shell.

\subsection{Boundary effective action}

The basic procedure of how to derive a hydrodynamic effective action from gravity has been realized in \cite{Crossley:2015tka} (see also \cite{Glorioso:2018mmw}), based on early attempts of formulating a holographic Wilsonian RG flow \cite{Heemskerk:2010hk,Faulkner:2010jy}. It amounts to identifying hydrodynamical variables (gapless modes) of the boundary theory, as proposed in \cite{Nickel:2010pr}. The remaining degrees of freedom are then integrated out from the bulk action, in Wilsonian sense.
The procedure is outlined below for a free $U(1)$ gauge field in the bulk.

The starting point is the bulk partition function:
\begin{align}
Z = \int D C_r D C_\mu e^{i S_0[C_\mu, C_r] + i S_{\rm c.t.}}.  \label{path_integral}
\end{align}
No gauge-fixing has been applied at this stage.
Integrating by parts, the bulk action $S_0$ \eqref{bulk_action_EF} can be expressed as
\begin{align}
S_0=& -\frac{1}{2} \int d^4x \int_{\infty_2}^{\infty_1} dr \sqrt{-g}\left[\nabla_M \left(C_N F^{MN} \right) - C_N \nabla_M F^{MN} \right] \nonumber \\
=& \frac{1}{2} \int d^4x \int_{\infty_2}^{\infty_1} dr \sqrt{-g} \left[ C_r \nabla_M F^{Mr} + C_v \nabla_M F^{Mv} + C_k \nabla_M F^{Mk}\right] \nonumber \\
& -\frac{1}{2} \int d^4x \sqrt{-\gamma} n_M C_N F^{MN}\bigg|_{r=\infty_2}^{r=\infty_1} \nonumber \\
=& \frac{1}{2}\int d^4x \int_{\infty_2}^{\infty_1} dr \sqrt{-g} \left\{C_v \left[\nabla_M F^{Mv} - \frac{1}{f(r)} \nabla_M F^{Mr}  \right] + C_k \nabla_M F^{Mk} \right. \nonumber \\
& \left. + \left(C_r + \frac{C_v}{f(r)}\right) \nabla_M F^{Mr}\right\} -\frac{1}{2} \int d^4x \sqrt{-\gamma} n_M C_N F^{MN}\bigg|_{r=\infty_2}^{r=\infty_1},
\label{surface_term}
\end{align}
where $n_M$ is a  out-pointing unit vector normal to the hypersurface $\Sigma$.


Our goal now is to derive the effective action.
We will  demonstrate how the bulk gauge invariance is realised. Our derivation is an explicit holographic construction  that closely follows the original ideas of \cite{Crossley:2015evo}.

The  discussion below applies separately to each segment of the contour.
The asymptotic values of the gauge field
are identified with the boundary external background  field $\mathcal A_\mu$,
\begin{align}
C_\mu(r=\infty, x^\alpha) \equiv \mathcal A_\mu.
\end{align}

Consider a generic gauge transformation parameterised by $\Lambda(r,x^\alpha)$  transforming
the bulk gauge field $C_M$ as
\begin{align}
C_\mu \to C_\mu^\prime = C_\mu + \partial_\mu \Lambda(r,x^\alpha), \qquad C_r \to C_r^\prime =C_r + \partial_r \Lambda(r,x^\alpha). \label{gauge-trans}
\end{align}
 In the path integral  \eqref{path_integral},  the gauge transformation \eqref{gauge-trans} can be viewed as a change of integration variables, say, from $C_r$ to $\Lambda$. In other words, the gauge transformation
 defines a map from any given configuration of $C_r$  into a configuration with some fixed radial gauge.
For example, for $C_M^\prime$ to be in the EF radial gauge,
\begin{align}
C_r^\prime =0 \Rightarrow \Lambda(r,x^\alpha)= \int_r^{r_c} d\xi C_r(\xi, x^\alpha).
\end{align}
The boundary value of $C_\mu^\prime$:
\begin{align}
 C_\mu^\prime (r=\infty, x^\alpha)  = \mathcal A_\mu + \partial_\mu \varphi,
\end{align}
where $\varphi\equiv \Lambda(r= \infty, x^\alpha)$ is identified with the hydrodynamic field associated with the $U(1)$ charge.
$\Lambda$ can be similarly constructed  for the Schwarzschild radial gauge $C_r^\prime =- C_v^\prime /f(r)$.


Integration over $C_\mu$ in \eqref{path_integral} can be  performed in the saddle point approximation, which
is exact for the Maxwell theory. That is,
solutions of the dynamical equations for $C_\mu$ are plugged into $S_0[C_\mu, C_r]$. As long as no radial
gauge is imposed, $C_\mu$ are functionals of $C_r$.
Generically, the partition function \eqref{path_integral} turns into
\begin{align}
Z = \int D C_r e^{i S_0[C_\mu[C_r], C_r] + i S_{\rm c.t.}},  \label{path_integral_pos}
\end{align}
where $S_0[C_\mu[C_r], C_r]$ is a ``partially on-shell'' bulk action \cite{deBoer:2015ija} which depends on $C_r$ both explicitly and through  solutions for $C_\mu$. The action is gauge invariant. Since in practice $C_\mu[C_r]$ are computed
in a fixed gauge, the next step is to fix the gauge using  \eqref{gauge-trans}, while changing the integration variable from $C_r$ to $\Lambda$.   The resulting partially on-shell action
does not depend on the  whole $\Lambda$ but on $\varphi$ only.
\eqref{path_integral_pos} becomes
\begin{align}
Z \sim \int D \varphi e^{i S_0|_{\rm p.o.s.} + i S_{\rm c.t.}}  \label{path_integral_varphi}
\end{align}
where $S_0|_{\rm p.o.s.} + S_{\rm c.t.}$ is the boundary effective action $S_{eff}$ of the $\varphi$ field. Here,  an overall coefficient due to Jacobian of the change of variables has been absorbed into $D\varphi$ measure.
Below we will explicitly derive  $S_0|_{\rm p.o.s.} $
for the EF radial gauge $C_r^\prime =0$ and Schwarzschild radial gauge $C_r^\prime =- C_v^\prime/f(r)$,
and demonstrate the equivalence of the results. In fact, it is straightforward to show  that
 the result for $S_0|_{\rm p.o.s.} $ is independent of the gauge-fixing.

In the EF radial gauge $C_r^\prime =0$, the partially on-shell bulk action is evaluated as
\begin{align}
S_0|_{\rm p.o.s} &= \frac{1}{2} \int d^4x \int_{\infty_2}^{\infty_1} dr \sqrt{-g} C_r \nabla_M F^{Mr} -\frac{1}{2} \int d^4x \sqrt{-\gamma} n_M C_N F^{MN}\bigg|_{r=\infty_2}^{r=\infty_1} \nonumber \\
&= \frac{1}{2} \int d^4x \int_{\infty_2}^{\infty_1} dr \sqrt{-g} (-\partial_r \Lambda) \nabla_M F^{Mr} -\frac{1}{2} \int d^4x \sqrt{-\gamma} n_M C_N F^{MN}\bigg|_{r=\infty_2}^{r=\infty_1} \nonumber \\
&=\frac{1}{2} \int d^4x \left[ -\sqrt{-g} \Lambda \nabla_M F^{Mr} - \sqrt{-\gamma} n_M C_N F^{MN} \right]\bigg|_{r=\infty_2}^{r=\infty_1} \nonumber \\
&= -\frac{1}{2} \int d^4x \sqrt{-g} C_\mu^\prime F^{r\mu}[C_\mu^\prime, C_r^\prime=0] \bigg|_{r=\infty_2}^{r=\infty_1}. \label{S0_pos_EF_gauge}
\end{align}
Similarly, in the Schwarzschild radial gauge $C_r^\prime =- C_v^\prime/f(r)$, the partially on-shell bulk action is calculated as
\begin{align}
S_0|_{\rm p.o.s}&= \frac{1}{2}\int d^4x \int_{\infty_2}^{\infty_1} dr \sqrt{-g} \left(C_r + \frac{C_v}{f(r)}\right) \nabla_M F^{Mr} -\frac{1}{2} \int d^4x \sqrt{-\gamma} n_M C_N F^{MN}\bigg|_{r=\infty_2}^{r=\infty_1} \nonumber \\
&= -\frac{1}{2}\int d^4x \int_{\infty_2}^{\infty_1} dr \sqrt{-g} \left(\partial_r \Lambda + \frac{\partial_v \Lambda}{f(r)}\right) \nabla_M F^{Mr} -\frac{1}{2} \int d^4x \sqrt{-\gamma} n_M C_N F^{MN}\bigg|_{r=\infty_2}^{r=\infty_1} \nonumber \\
&= -\frac{1}{2}\int d^4x \int_{\infty_2}^{\infty_1} dr \left(\partial_r + \frac{\partial_v }{f(r)}\right)\left[ \Lambda \sqrt{-g}\nabla_M F^{Mr}\right] -\frac{1}{2} \int d^4x \sqrt{-\gamma} n_M C_N F^{MN}\bigg|_{r=\infty_2}^{r=\infty_1} \nonumber \\
& = \frac{1}{2}\int d^4x \int_{\infty_2}^{\infty_1} dr \left[ -\Lambda \sqrt{-g} \nabla_M F^{Mr}- \sqrt{-\gamma} n_M C_N F^{MN} \right]\bigg|_{r=\infty_2}^{r=\infty_1} \nonumber \\
& = -\frac{1}{2} \int d^4x \sqrt{-g} C_\mu^\prime F^{r\mu}[C_\mu^\prime, C_r^\prime = -C_v^\prime/f(r)]\bigg|_{r=\infty_2}^{r=\infty_1}. \label{S0_pos_Schw_gauge}
\end{align}

Both in \eqref{S0_pos_EF_gauge} and \eqref{S0_pos_Schw_gauge}, the terms representing the total derivatives at the AdS boundaries were dropped. The Bianchi identity \eqref{Bianchi} was instrumental to convert the bulk integral into the surface term. Gauge invariance of the field strength $F^{MN}[C_\mu, C_r]= F^{MN}[C_\mu^\prime, C_r^\prime]$ was used as well.  \eqref{S0_pos_EF_gauge} and \eqref{S0_pos_Schw_gauge} are identical and
become a superposition of two surface terms
\begin{align}
S_0|_{\rm p.o.s.}=-\frac{1}{2} \int d^4x \sqrt{-\gamma} n_M C_N^\prime F^{MN}[C_\mu^\prime, C_r^\prime] \bigg|_{r=\infty_2}^{r=\infty_1}.
\label{surface_term1}
\end{align}
Notice that \eqref{surface_term1} could be obtained from (a primed version of) \eqref{surface_term} by dropping all the bulk terms, either because of  the dynamical equations \eqref{dynamical_EOM} or as a result of the
Schwarzschild radial gauge\footnote{In \cite{Glorioso:2018mmw}, the EF radial gauge $C_r=0$ was taken, combined with  the dynamical equations  $\nabla_M F^{Mv}=0$ and $\nabla_M F^{Mi}=0$, which also eliminated  the bulk terms in \eqref{surface_term} leading to the very same boundary action.}.

There is a subtle point in deriving \eqref{surface_term1}. As will become clear in the next section, in the off-shell formalism, the gauge potential and/or field strength develop discontinuity
near the event horizon. Hence, in principle, there might emerge an additional boundary term at the horizon surface. This term will be eliminated by a specific choice of the boundary conditions to be discussed below.

Below we will drop the prime in $C_M^\prime$ and permanently stick to the Schwarzschild radial gauge for $C_M$. The bulk EOMs \eqref{dynamical_EOM} will be solved subject to boundary conditions at two AdS boundaries $r=\infty_1$ and $r=\infty_2$:
\begin{align} \label{Cmu_bdy_cond}
C_\mu \xrightarrow[]{r\to \infty_1} B_{1\mu}(x^\alpha) \equiv \mathcal{A}_{1\mu}+ \partial_\mu \varphi_1, \qquad \qquad
C_\mu \xrightarrow[]{r\to \infty_2} B_{2\mu}(x^\alpha) \equiv \mathcal{A}_{2\mu}+ \partial_\mu \varphi_2.
\end{align}

In order to compute the boundary action \eqref{surface_term1},  near boundary asymptotic expansion of the bulk fields is required. It has the form
\begin{align}
&C_\mu \xrightarrow[]{r\to \infty_1} B_{1\mu}(x^\alpha) + \frac{\partial_v B_{1\mu} (x^\alpha)}{r} - \frac{1}{2} \partial^\nu \mathcal{F}_{1\mu\nu}(x^\alpha)\frac{\log r} {r^2} + \frac{C_{1\mu}^{(2)}(x^\alpha)} {r^2} + \cdots, \label{Cmu_inf1}\\
&C_\mu \xrightarrow[]{r\to \infty_2} B_{2\mu}(x^\alpha) + \frac{\partial_v B_{2\mu} (x^\alpha)}{r} - \frac{1}{2} \partial^\nu \mathcal{F}_{2\mu\nu}(x^\alpha)\frac{\log r} {r^2} + \frac{C_{2\mu}^{(2)}(x^\alpha)}{r^2}+\cdots, \label{Cmu_inf2}
\end{align}
where the coefficient functions $C_{1\mu}^{(2)}$ and $C_{2\mu}^{(2)}$ are functionals of both $B_{1\mu}$ and $B_{2\mu}$. They will be determined through the solution
of the dynamical equations \eqref{dynamical_EOM} over the entire
contour in Figure. \ref{holographic_SK_contour}.  This is the subject of the next section.

Once \eqref{Cmu_inf1} and \eqref{Cmu_inf2} are substituted into the total action $S_{\rm eff}=S_0|_{\rm p.o.s.}+S_{\rm c.t.}$, the latter takes the form (\ref{effectiveaction})
with the effective Lagrangian $\mathcal{L}_{\rm eff}(x)$ being a quadratic functional of the boundary
fields $B_{1\mu}$ and $B_{2\mu}$ (in the $(r,a)$-basis)
\begin{align}
\mathcal{L}_{\rm eff}=& -B_{rv} (C_{1v}^{(2)}- C_{2v}^{(2)}) - \frac{1}{2} B_{av} (C_{1v}^{(2)} + C_{2v}^{(2)}) + B_{rk} (C_{1k}^{(2)}- C_{2k}^{(2)}) \nonumber \\
&+ \frac{1}{2} B_{ak} (C_{1k}^{(2)} + C_{2k}^{(2)}) {+ \frac{1}{2} \partial_k B_{ak} \partial_v B_{rv} - \frac{1}{2} B_{av} \partial_v \partial_k B_{rk}} \nonumber \\
& -\frac{1}{2} B_{ak} \partial_v^2 B_{rk} + \frac{1}{4} \mathcal{F}_{akj} \mathcal{F}_{rkj}  + \frac{1}{2} B_{av}\vec\partial^{\,2} B_{rv} {+ B_{av} \partial_v^2 B_{rv}}. \label{Leff1}
\end{align}
With the help of a basis decomposition procedure introduced in \cite{Bu:2015ame}, the effective Lagrangian \eqref{Leff1} can be recast into the form
(\ref{Leff3}), thus providing a holographic derivation of the latter, which is fully consistent with
the general analysis of  \cite{Crossley:2015evo}. This derivation is presented in Appendix \ref{general_structure}.

\section{ Bulk dynamics: solutions and analysis} \label{solve_bulk_dynamics}

This section is devoted to solutions of  the bulk EOMs \eqref{dynamical_EOM} over the contour displayed in Figure \ref{holographic_SK_contour}.
Our strategy will be different from that of \cite{Glorioso:2018mmw}. Instead of integrating the bulk EOMs \eqref{dynamical_EOM} along the entire $r$-contour,  we split the radial contour at $r=r_h-\epsilon$ into two segments,
the upper and  lower one. Each segment ``lives'' in a single copy of the doubled Schwarzschild-AdS$_5$ geometry. In each segment, there is a set of independent solutions to the dynamical EOMs \eqref{dynamical_EOM} forming a basis.
Full solutions  obeying the respective boundary conditions \eqref{Cmu_bdy_cond}
will be constructed as  linear superpositions of the basis solutions.
Thus-constructed piecewise solutions will be carefully glued
 at the cutting slice $r=r_h-\epsilon$, under proper matching conditions to be derived in subsection \ref{match_cond_Cmu}.  One of the advantages of our approach is that it
avoids the subtleties related to non-commutativity between  two limits: the hydrodynamic derivative expansion and $\epsilon\rightarrow  0$, the latter has to be taken first.

\subsection{Discrete symmetries} \label{T-symmetry}

Symmetries in classical theories are used in order to generate solutions, if one is already known. Maxwell's theory in the bulk (single copy Schwarzschild-AdS)
has a number of discrete symmetries, such as parity and time reversal, which will
be employed in our quest after a full set of independent solutions. However, in different coordinate systems,
the symmetries are represented  differently. Particularly,
while  in the Schwarzschild coordinates the time reversal symmetry is realised trivially,  its representation in the EF coordinates is much less transparent.  This is essentially the main reason we
have chosen to first find all the solutions in the Schwarzschild coordinates and then translate those into the EF system.

In the Schwarzschild coordinates (without any gauge fixing yet), the Fourier mode $\tilde C_M(r,k^\mu)$  defined by
\begin{align}
\tilde C_M(r,t,x^i)= \int \frac{d^4k}{(2\pi)^4} e^{ik\cdot x} \tilde C_M(r, k^\mu), \qquad k^\mu=(\omega,\vec q),
\end{align}
satisfies the following system of ODEs:
\begin{align}
\tilde \nabla_M \tilde F^{Mr}=0\Rightarrow 0= &(i\omega \partial_r \tilde C_t - \omega^2 \tilde C_r) + r^{-2} f(r) (q^2 \tilde C_r+ iq_k \partial_r \tilde C_k),   \label{constraint_Schw} \\
\tilde \nabla_M \tilde F^{Mt}=0\Rightarrow 0= &\partial_r(r^3\partial_r \tilde C_t) + i\omega \partial_r(r^3 \tilde C_r) - \frac{r}{f(r)}(q^2 \tilde C_t + \omega q_k \tilde C_k),  \label{eom_Ct_Schw} \\
\tilde \nabla_M \tilde F^{Mi}=0\Rightarrow 0= &\partial_r[rf(r)\partial_r \tilde C_i] - iq_i \partial_r[rf(r) \tilde C_r] + \frac{r}{f(r)} (\omega^2 \tilde C_i+ \omega q_i \tilde C_t) \nonumber \\
&+r^{-1} (-q^2 \tilde C_i + q_i q_k \tilde C_k). \label{eom_Ci_Schw}
\end{align}
The full set of bulk EOMs \eqref{constraint_Schw}-\eqref{eom_Ci_Schw} is obviously
invariant under the following time-reversal transformation:
\begin{align} \label{T-reversal_linear_omega}
& \omega \to - \omega, \qquad \tilde C_t (r, k^\mu) \to  \tilde C_t(r,\bar k^\mu), \qquad \tilde C_i (r, k^\mu) \to -\tilde C_i(r,\bar k^\mu), \nonumber \\
& \tilde C_r (r, k^\mu) \to -\tilde C_r(r,\bar k^\mu), \qquad \qquad {\rm with}~~\bar k^\mu=(-\omega,\vec q),
\end{align}
which, in the coordinate space $(r,t,\vec x)$, turns into
\begin{align} \label{T-reversal_linear_t}
& t\to -t, \qquad \tilde C_t (r, t,\vec x) \to  \tilde C_t(r, -t,\vec x), \qquad \tilde C_i (r, t,\vec x) \to -\tilde C_i(r, -t,\vec x), \nonumber \\
& \tilde C_r (r, t,\vec x) \to -\tilde C_r(r, -t,\vec x).
\end{align}
Furthermore, the dynamical EOMs \eqref{eom_Ct_Schw} and \eqref{eom_Ci_Schw} are invariant under the time reversal independently, regardless of the constraint equation \eqref{constraint_Schw} being imposed or not.
Transformations \eqref{T-reversal_linear_omega} and \eqref{T-reversal_linear_t} can be recognised as a {\it linear} realisation of the time-reversal symmetry \cite{Chakrabarty:2019aeu}.

In the ingoing EF coordinates \eqref{metric_EF}, the Fourier mode $C_M(r,k^\mu)$ defined by
\begin{align}
C_M(r, x^\mu) = \int \frac{d^4k}{(2\pi)^4} e^{ik\cdot x} C_M(r,k^\mu), \qquad k^\mu=(\omega,\vec q),
\end{align}
obeys another system of ODEs:
\begin{align}
&\nabla_M F^{Mr}=0 ~ \Rightarrow~  0=r^3(\omega^2 C_r- i\omega \partial_r C_v) - rf(r) (q^2 C_r + iq_k \partial_r C_k) - r(q^2 C_v + \omega q_k C_k), \label{constraint_EF}\\
&\nabla_M F^{Mv}- \frac{1}{f(r)} \nabla_M F^{Mr}=0~ \Rightarrow ~ 0=\partial_r(r^3 \partial_r C_v + i\omega r^3 C_r) +r(q^2 C_r +iq_k \partial_r C_k) \nonumber \\
&~~~~~~~~~~~~~~~~~~~+ \frac{r^3}{f(r)} (\omega^2 C_r- i\omega \partial_r C_v) -\frac{r}{f(r)} (q^2 C_v + \omega q_k C_k) - r (q^2 C_r + iq_k \partial_r C_k),  \label{eom_Cv_EF} \\
&\nabla_M F^{Mi}=0 ~\Rightarrow ~0=\partial_r[rf(r) (\partial_r C_i- iq_i C_r)] -\partial_r[r(i\omega C_i + iq_i C_v)] \nonumber \\
&~~~~~~~~~~~~~~~~~~~~~~~~~~~~~~~~~~~~~~~~~~~~~~- r(i\omega \partial_r C_i +\omega q_i C_r) + r^{-1} (-q^2 C_i +q_i q_k C_k). \label{eom_Ci_EF}
\end{align}
Apparently, the transformation like \eqref{T-reversal_linear_omega} is not a symmetry of EOMs \eqref{constraint_EF}-\eqref{eom_Ci_EF}. This is related to the fact that $v\to -v$ is not a symmetry, because the
transformation does not leave  the metric  \eqref{metric_EF} invariant. This point was made clear in \cite{Chakrabarty:2019aeu}, which carried out a somewhat similar analysis  of a probe string in Schwarzschild-AdS background.

What is a {\it nonlinear} realisation of the underlying time-reversal symmetry in the ingoing EF coordinates?
This could be worked out with the help of \eqref{Cmu_Schw_EF_coordinate} and \eqref{Cr_Schw_EF_coordinate}. The Fourier modes in the Schwarzschild  and ingoing EF coordinates are related,
\begin{align}
&C_\mu(r,k)= \tilde C_\mu (r,k) e^{i\omega \zeta_s(r)}, \qquad C_r(r,k)= \left[\tilde C_r(r,k) - \frac{\tilde C_t(r,k)}{f(r)} \right] e^{i\omega \zeta_s(r)}, \nonumber \\
&r\in [r_h-\epsilon, \infty_s), \qquad s=1~ {\rm or} ~2, \label{CM_Schw_EF_Fourier}
\end{align}
where $\zeta_1(r)$ and $\zeta_2(r)$ are introduced in \eqref{t_v_zeta12}. Thus, in the ingoing EF coordinates, the time-reversal symmetry is realized as
\begin{align}
&\omega \to -\omega, \quad C_v(r,k) \to C_v(r,\bar k) e^{2i\omega\zeta_s(r)}, \quad C_i(r,k) \to - C_i(r,\bar k)e^{2i\omega\zeta_s(r)}, \nonumber \\
&C_r(r,k) \to -\left[ C_r(r,\bar k) + \frac{2C_v(r,\bar k)}{f(r)}\right] e^{2i\omega \zeta_s(r)}, \qquad s=1,2.
\end{align}

So far, no gauge choice has been specified. In \cite{Glorioso:2018mmw}, the EF radial gauge $C_r=0$ was chosen, which is  in tension with the linear
realisation of the time-reversal symmetry \eqref{T-reversal_linear_omega}. In fact,  a time reversed solution is gauge transformed with respect to $C_r=0$.
Hence, in order to benefit from the simplicity of \eqref{T-reversal_linear_omega},  we search for solutions in the Schwarzschild radial gauge $\tilde C_r=0$.

The bulk EOMs \eqref{constraint_Schw}-\eqref{eom_Ci_Schw} are also invariant under $\mathcal{P}$-symmetry (space inversion):
\begin{align}
& \vec q \to - \vec q, \qquad  \tilde C_t(r, k^\mu) \to \tilde C_t(r, -\bar k^\mu), \qquad \tilde C_i(r, k^\mu) \to - \tilde C_i(r, -\bar k^\mu), \nonumber \\
&~~~~~~~~~~~~ \tilde C_r(r, k^\mu) \to \tilde C_r(r, -\bar k^\mu), \label{P_symmetry_q}
\end{align}
or, alternatively,  in the coordinate space $(r,t,\vec x)$
\begin{align}
& \vec x \to - \vec x, \qquad  \tilde C_t(r,t,\vec x) \to \tilde C_t(r,t,-\vec x), \qquad \tilde C_i(r,t,\vec x) \to - \tilde C_i(r,t,-\vec x), \nonumber \\
& ~~~~~~~~~~~~~\tilde C_r(r,t,\vec x) \to \tilde C_r(r,t,-\vec x).  \label{P_symmetry_coordinate}
\end{align}
Since the coordinate transformation  from \eqref{metric_EF} to \eqref{metric_Schw} does not involve  spatial directions, the $\mathcal P$-symmetry in the EF coordinates
takes exactly the same form as  \eqref{P_symmetry_q} and \eqref{P_symmetry_coordinate},
which can be straightforwardly checked from the bulk EOMs \eqref{constraint_EF}-\eqref{eom_Ci_EF}.

The time-reversal $\mathcal T$ symmetry and $\mathcal P$-symmetry help to examine the symmetry relations for TCFs in the effective action, as reviewed in subsection \ref{ds}.

\subsection{Horizon matching conditions} \label{match_cond_Cmu}

As has been already mentioned, we first derive independent solutions in the upper and lower segments, and then
 glue them at the surface $r=r_h-\epsilon$. In
this way a complete solution valid along the whole radial contour in Figure \ref{holographic_SK_contour} is constructed.  A necessary element of this construction is a set of matching conditions
for the bulk fields to be discussed in this subsection.

The contour in Figure \ref{holographic_SK_contour} is cut along the surface $r=r_\pm \equiv r_h-\epsilon$, where the subscripts $_+$ and $_-$ are used to distinguish the upper and lower segments.  In the spirit of \cite{Skenderis:2008dg}, the matching conditions at $r=r_h-\epsilon$ can be obtained by demanding the total bulk action to be extremal with respect to variation of the horizon data $C_M(r_h-\epsilon,x^\mu)$. Consider  variation of the bulk action\footnote{The counter-term action $S_{\rm c.t.}$ does not contribute to the variational problem near horizon. } \eqref{bulk_action_EF}
\begin{align}
\delta S_0= & - \int d^4x\sqrt{-\gamma} n_M \delta C_N F^{MN} \bigg|_{r_-}^{\infty_1} + \int d^4x \sqrt{-\gamma} n_M \delta C_N F^{MN}\bigg|_{r_+}^{\infty_2} \nonumber \\
&+ \int d^4x \int_{\infty_2}^{\infty_1} dr \sqrt{-g} \left[ \delta C_r \nabla_M F^{Mr} + \delta C_v \frac{\nabla_M F^{Mr}}{f(r)} \right] \label{deltaS}
\end{align}
where we imposed the dynamical equations \eqref{dynamical_EOM}. The extremum condition (at the surface $r=r_h-\epsilon$) gives
\begin{align}
&\frac{\delta S}{\delta C_v(r_h-\epsilon)} =0 \Rightarrow F^{rv}(r_+)- F^{rv}(r_-) = \lim_{\Delta \to 0}\int_{r_+ + \Delta}^{r_- -\Delta} dr \frac{\nabla_M F^{Mr}}{f(r)}, \label{glue_dCv}\\
&\frac{\delta S}{\delta C_i(r_h -\epsilon)} =0 \Rightarrow F^{ri}(r_{+})- F^{ri}(r_{-})=0, \label{glue_dCi}
\end{align}
where $\Delta$ is an infinitesimal interval along the circle in Figure \ref{holographic_SK_contour}.
The field strength components $F^{ri}$ are  continuous through the cutting surface. Yet,   the $F^{rv}$ component,  being continuous for on-shell theory,
 may develop  discontinuity if  the constraint equation is relaxed, that is,  $\nabla_M F^{Mr}\ne 0$.
When implementing  the Schwarzschild radial gauge $\tilde C_r=0$, the matching condition \eqref{glue_dCi} translates into
\begin{align}
f(r) \partial_r C_i\big|_{r=r_+} = f(r) \partial_r C_i \big|_{r=r_-}. \label{glue_dCi_simple}
\end{align}
Since $f(r)$ vanishes at the horizon, $\partial_r C_i$ may not  be continuous. Similarly, $C_r$ component is discontinuous.

The condition \eqref{glue_dCv} is very non-trivial. It suggests that $\partial_r C_v$ is discontinuous too. Yet, this conclusion is based on  assumption that
$C_\mu$ is continuous across the cutting slice:
\begin{align}
C_\mu(r_-)= C_\mu(r_+). \label{continuity_Cmu}
\end{align}
 \eqref{continuity_Cmu} is a natural choice that could be realised via a residual gauge transformation implementable on each segment independently. Furthermore, thanks to the residual gauge freedom, we could set
 \begin{align}
C_v(r_h-\epsilon,x^\mu)=0. \label{glue_Cv=0_horizon}
\end{align}
This choice has been also implemented in  \cite{Glorioso:2018mmw},  though through a somewhat different chain of arguments. Once the condition
\eqref{glue_Cv=0_horizon} is imposed,  \eqref{glue_dCv} fixes the discontinuity of $\partial_r C_v$ uniquely.  Below we will construct the full solution of EOMs
with the matching condition \eqref{glue_Cv=0_horizon} and then check that
 \eqref{glue_dCv}  is indeed satisfied.  This consistency check will be presented in Appendix \ref{crosscheck_dCv_Cv}.

Finally, an added value of the choice \eqref{glue_Cv=0_horizon} is that it makes the horizon contribution to the effective action $S_0$ vanish.
Hence, \eqref{surface_term1} is correct \footnote{Working with another residual gauge would lead us to  different matching conditions
and, as a consequence, different solutions of the EOMs, and also to a modified expression for the effective action. The final action is however gauge invariant and hence should not depend on a particular choice of the residual gauge. }.

\subsection{Linearly independent solutions} \label{sol_single_AdS}

In this subsection, we derive and analyse all  linearly independent solutions of  the dynamical EOMs \eqref{dynamical_EOM} in a single Schwarzschild-AdS$_5$.
 The solutions are equally valid both in the upper and lower segments in Figure \ref{holographic_SK_contour}.
 As argued previously, in order to benefit from the linear realisation \eqref{T-reversal_linear_omega} of the time-reversal symmetry,  we temporarily work in the Schwarzschild coordinate system combined with the Schwarzschild radial gauge.
Eventually, linearly independent solutions in the ingoing EF coordinates will be deduced via the transformation rule \eqref{CM_Schw_EF_Fourier}. Without loss of generality, the spatial momentum $\vec q$ is taken to be along the $x$-direction. Then, the bulk fields $\tilde C_\mu$  decouple between two sectors:  the transverse sector $\tilde C_\perp=\{\tilde C_y, \tilde C_z\}$ and the longitudinal sector $\tilde C_\parallel=\{\tilde C_t, \tilde C_x\}$.

~

\noindent{\bf Transverse sector.}

The transverse mode $\tilde C_\perp$ obeys a single ODE:
\begin{align}
0= \partial_r[rf(r)\partial_r \tilde C_\perp] + \frac{\omega^2r}{f(r)}\tilde C_\perp -q^2 r^{-1}\tilde C_\perp, \qquad \perp=y,z. \label{eom_Cperp_Schw}
\end{align}
Eq. \eqref{eom_Cperp_Schw} has two independent solutions distinguishable by their near horizon behaviour.

Near the horizon $r=r_h$, the ingoing solution $\tilde C_\perp^{\rm ig}(r,k^\mu)$ behaves as
\begin{align}
\tilde C_\perp^{\rm ig}(r,k^\mu) \xrightarrow[]{r\to r_h} (r-r_h)^{-i\omega/(4r_h)} \left[\tilde C_\perp^h + \tilde C_\perp^1(r-r_h)+ \tilde C_\perp^2 (r-r_h)^2 + \cdots\right],
\end{align}
where $\tilde C_\perp^h$ is an integration constant (initial condition), which we refer to as horizon data. The remaining coefficients $\tilde C_\perp^1, \tilde C_\perp^2,\cdots$ are uniquely fixed in terms of  $\tilde C_\perp^h$.

The outgoing solution\footnote{The outgoing solution can be identified with the Hawking radiation.}
 $\tilde C_\perp^{\rm og} (r,k^\mu)$ is obtained from the ingoing one by the time-reversal symmetry \eqref{T-reversal_linear_omega}
\begin{align}
\tilde C_\perp^{\rm og}(r,k^\mu)= - \tilde C_\perp^{\rm ig}(r,\bar k^\mu). \label{Cperp_ig_og_relation_Schw}
\end{align}
Both solutions are functions of $q^2$ as is obvious from \eqref{eom_Cperp_Schw}, and hence they are $\mathcal P$-invariant.
From \eqref{CM_Schw_EF_Fourier}, the solutions  in the ingoing  EF coordinates read:
\begin{align}\label{Cperp_EF_Schw}
C_\perp^{\rm ig}(r,k^\mu)= \tilde C_\perp^{\rm ig}(r,k^\mu) e^{i\omega \zeta_s(r)}, \quad C_\perp^{\rm og}(r,k^\mu)= \tilde C_\perp^{\rm og}(r,k^\mu) e^{i\omega \zeta_s(r)}, \quad r\in [r_h-\epsilon, \infty_s).
\end{align}
Thus,  the ingoing and outgoing solutions are related to each other via
\begin{align}
C_\perp^{\rm og}(r,k^\mu)= - C_\perp^{\rm ig}(r,\bar k^\mu) e^{2i\omega\zeta_s(r)}, \qquad r\in[r_h-\epsilon, \infty_s), \qquad s=1 ~ {\rm or} ~2, \label{Cperp_ig_og}
\end{align}
where \eqref{Cperp_ig_og_relation_Schw} is used.

Near the AdS boundary the ingoing solution $C_\perp^{\rm ig}$ can be expanded
\begin{align}\label{Cperpexpand}
C_\perp^{\rm ig}(r,k^\mu) \xrightarrow[]{r\to \infty}& C_\perp^{\rm ig(0)} (k^\mu) - \frac{i\omega C_\perp^{\rm ig(0)}(k^\mu)}{r} + \frac{1}{2}(\omega^2-q^2) C_\perp^{\rm ig(0)}(k^\mu) \frac{\log r}{r^2} + \frac{C_\perp^{\rm ig(2)}(k^\mu)}{r^2} + \cdots.
\end{align}
In principle, one could tune $\tilde C_\perp^h$
so that  $C_\perp^{\rm ig(0)}(k^\mu)=1$, though for a while we prefer to keep it unspecified.

~

\noindent{\bf Longitudinal sector.}

The EOMs for the longitudinal sector $\tilde C_\parallel=\{\tilde C_t, \tilde C_x\}$ are
\begin{align}
&0= \partial_r(r^3\partial_r \tilde C_t) - \frac{r}{f(r)}(q^2 \tilde C_t + \omega q \tilde C_x), \nonumber \\
&0= \partial_r[rf(r)\partial_r \tilde C_x] + \frac{r}{f(r)} (\omega^2 \tilde C_x+ \omega q \tilde C_t). \label{eom_CtCx_Schw}
\end{align}
Compared to the transverse case,
the coupling between $\tilde C_t$ and $\tilde C_x$ makes the longitudinal sector much more involved.
The system of coupled equations \eqref{eom_CtCx_Schw} has four linearly independent solutions, which could be differentiated by their near horizon  behaviour.

$\bullet$ The first solution is the ingoing solution $\{\tilde C_t^{\rm ig}, \tilde C_x^{\rm ig}\}$,
\begin{align}
& \tilde C_t^{\rm ig}(r,k^\mu) \xrightarrow[]{r\to r_h} (r-r_h)^{1-i\omega/(4r_h)}  \left[\frac{4iq \tilde C_x^{{\rm ig}\, h}}{4r_h^2 -i\omega r_h} + \cdots\right], \nonumber \\
& \tilde C_x^{\rm ig}(r,k^\mu) \xrightarrow[]{r\to r_h} (r-r_h)^{-i\omega/(4r_h)}  \left[\tilde C_x^{{\rm ig}\,h}+ \frac{i\omega \tilde C_x^{{\rm ig}\,h}(8r_h^2+2i\omega r_h + \omega^2 -4q^2)} {8r_h^2(8r_h^2- 6i\omega r_h -\omega^2)}(r-r_h)+ \cdots\right], \label{CtCx_ingoing}
\end{align}
where $\cdots$ are higher powers of $(r-r_h)$.  Both functions $\{\tilde C_t^{\rm ig}, \tilde C_x^{\rm ig}\}$ are
 uniquely determined in terms of  single horizon data $\tilde C_x^{{\rm ig} \,h}$.
It is important to observe that  $\tilde C_t^{\rm ig}$ is an odd function of $q$ while $\tilde C_x^{\rm ig}$ is even.

$\bullet$ The second solution of \eqref{eom_CtCx_Schw} is the outgoing solution $\{\tilde C_t^{\rm og}, \tilde C_x^{\rm og}\}$, which is
 obtained from the ingoing solution by the time-reversal transformation \eqref{T-reversal_linear_omega}
\begin{align}
\tilde C_t^{\rm og}(r,k^\mu)= \tilde C_t^{\rm ig}(r,\bar k^\mu), \qquad \tilde C_x^{\rm og}(r,k^\mu)=- \tilde C_x^{\rm ig}(r,\bar k^\mu).  \label{CtCx_outgoing}
\end{align}

$\bullet$ The third solution is the pure gauge (pg) solution \cite{Policastro:2002tn}:
\begin{align}
\tilde C_t^{\rm pg}(r,k^\mu)= -i\omega \tilde\Lambda (k^\mu), \qquad \tilde C_x^{\rm pg}(r,k^\mu)= iq \tilde\Lambda (k^\mu), \label{CtCx_pure_gauge}
\end{align}
where $\tilde \Lambda(k^\mu)$ is an $r$-independent  gauge parameter of the residual gauge symmetry.  As we have argued above, $\tilde C_t=0$ at the horizon can be imposed as a residual gauge fixing.
This choice of the gauge is equivalent to setting $\tilde \Lambda=0$.

$\bullet$ The fourth solution is the polynomial (pn) solution $\{\tilde C_t^{\rm pn}, \tilde C_x^{\rm pn}\}$.
Near  horizon it has a  Taylor expansion in powers of $(r-r_h)$:
\begin{align}
&\tilde C_t^{\rm pn}(r,k^\mu)\xrightarrow[]{r\to r_h} (r-r_h)\left[\tilde C_t^{{\rm pn}\,h}+ \frac{\tilde C_t^{{\rm pn}\,h} (4q^2-48r_h^2- 3\omega^2)} {2r_h(16r_h^2+ \omega^2)} (r-r_h) + \cdots\right], \label{CtCx_polynomial} \\
&\tilde C_x^{\rm pn}(r,k^\mu)\xrightarrow[]{r\to r_h}(r-r_h)\left[- \frac{\tilde C_t^{{\rm pn}\,h} \omega q}{16 r_h^2 +\omega^2}- \frac{\tilde C_t^{{\rm pn}\, h} \omega q (4q^2-32 r_h^2- 3\omega^2)}{2r_h(16r_h^2 +\omega^2) (64 r_h^2 +\omega^2)}(r-r_h)+ \cdots\right], \nonumber
\end{align}
where  $\cdots$ refer to terms that are uniquely fixed in terms of the horizon data $\tilde C_t^{{\rm pn}\, h}$.
Thus the boundary values of $\tilde C_t^{\rm pn}$ and $ \tilde C_x^{\rm pn}$ are not independent.
$\tilde C_t^{\rm pn}$ is a function of $\omega^2$  and $q^2$, that is, it is both $\mathcal{T}$ and $\mathcal{P}$ even.
Similarly, because $\tilde C_x^{\rm pn}$ has an overall extra factor $\omega q$, it is $\mathcal{T}$ and $\mathcal{P}$ invariant too.

Thus we have found all four linearly independent solutions.  So far, they have been identified by their near horizon behaviours. In subsection \ref{num_results}, we will also construct them numerically, for finite momenta.
Under the rule \eqref{CM_Schw_EF_Fourier},  the linearly independent solutions in the ingoing EF coordinate are
\begin{align}
& C_\parallel^{\rm ig}(r,k^\mu)= \tilde C_\parallel^{\rm ig}(r,k^\mu) e^{i\omega \zeta_s(r)}, \qquad
C_\parallel^{\rm og}(r,k^\mu)= \tilde C_\parallel^{\rm og}(r,k^\mu) e^{i\omega \zeta_s(r)}, \nonumber \\
& C_\parallel^{\rm pg}(r,k^\mu)= \tilde C_\parallel^{\rm pg}(r,k^\mu) e^{i\omega \zeta_s(r)}, \qquad
C_\parallel^{\rm pn}(r,k^\mu)= \tilde C_\parallel^{\rm pn}(r,k^\mu) e^{i\omega \zeta_s(r)}, \label{CvCx_EF_Schw}
\end{align}
where $s=2$ when $r\in[r_h-\epsilon, \infty_2)$ and $s=1$ when $r\in[r_h-\epsilon, \infty_1)$. Here, the subscript $_\parallel$  collectively denotes the time component and $x$ component of the bulk gauge field. Similarly as
in the transverse case \eqref{Cperp_ig_og}, the ingoing solution $\{C_v^{\rm ig}, C_x^{\rm ig}\}$ and the outgoing
solution $\{C_v^{\rm og}, C_x^{\rm og}\}$ are related:
\begin{align}
& C_v^{\rm og}(r,k^\mu)= C_v^{\rm ig}(r,\bar k^\mu) e^{2i\omega \zeta_s(r)}, \qquad C_x^{\rm og}(r,k^\mu)= - C_x^{\rm ig}(r,\bar k^\mu) e^{2i\omega \zeta_s(r)}, \nonumber \\
& r\in[r_h-\epsilon, \infty_s), \qquad s=1 ~ {\rm or} ~2.
\end{align}
Near the AdS boundary the linearly independent solutions can be expanded
\begin{align}\label{Cparexpand}
C_v^{\rm S}(r,k^\mu) \xrightarrow[]{r\to \infty}& C_v^{\rm S(0)} (k^\mu) - \frac{i\omega C_v^{\rm S(0)}(k^\mu)}{r} + \frac{1}{2}\partial^\mu{\cal F}_{v\mu}^{\rm S(0)}(k^\mu) \frac{\log r}{r^2}  + \frac{C_v^{\rm S(2)}(k^\mu)}{r^2} + \cdots, \nonumber \\
C_x^{\rm S}(r,k^\mu) \xrightarrow[]{r\to \infty}& C_x^{\rm S(0)} (k^\mu) - \frac{i\omega C_x^{\rm S(0)}(k^\mu)}{r} + \frac{1}{2}\partial^\mu{\cal F}_{x\mu}^{\rm S(0)}(k^\mu) \frac{\log r}{r^2}  + \frac{C_x^{\rm S(2)}(k^\mu)}{r^2} +  \cdots,
\end{align}
where $\rm S$ stands for any of the three solutions, $\rm S=(ig,\, og,\,pn)$ and ${\cal F}^{\rm S(0)}$ is the corresponding field strength.
The expansion \eqref{Cparexpand} is not valid for the pure gauge solution.
In principle, one could tune the horizon data for each solution independently,
so that  $C_v^{\rm S(0)}(k^\mu)=1$. Then, there is no freedom left to also set $C_x^{\rm S(0)}(k^\mu)$ to one: the value of $C_x^{\rm S(0)}(k^\mu)$ would have to be determined from the dynamical equations.

It is important to notice that the solutions $\{\tilde C_t^{\rm ig}, \tilde C_x^{\rm ig}\}$, $\{\tilde C_t^{\rm og}, \tilde C_x^{\rm og}\}$, and $\{\tilde C_t^{\rm pg}, \tilde C_x^{\rm pg}\}$
 satisfy the constraint equation \eqref{constraint_Schw} automatically, which makes it possible to relate the near boundary expansions of these functions.
Particularly,
\begin{align}\label{relation}
& \omega \tilde C_t^{\rm S(2)}(k^\mu) + q \tilde C_x^{\rm S(2)}(k^\mu)=0, \nonumber \\
& C_\mu^{\rm S(0)}(k^\mu)=\tilde C_\mu^{\rm S(0)}(k^\mu),\qquad C_\mu^{\rm S(2)}(k^\mu)= \tilde C_\mu^{\rm S(2)}(k^\mu)- \frac{1}{2}\omega^2 \tilde C_\mu^{\rm S(0)}(k^\mu),
\end{align}
where the last two relations follows from \eqref{CM_Schw_EF_Fourier}.
The polynomial
solution, on the other hand, does not satisfy the constraint \eqref{constraint_Schw}.
Consequently, having the theory put on-shell is equivalent to setting the coefficient of
the polynomial solution to zero.

\subsection{Solutions over the entire radial contour: gluing at the horizon} \label{sol_contour}

In the previous subsection, we have found the independent solutions to dynamical  EOMs \eqref{dynamical_EOM}  for a single copy of the doubled Schwarzschild-AdS$_5$.
Our next task is to construct a full solution  over the entire contour in Figure \ref{holographic_SK_contour}. To this end, the independent  solutions on the upper and lower segments will be glued at the horizon, employing the matching conditions \eqref{glue_dCi_simple}, \eqref{continuity_Cmu} and \eqref{glue_Cv=0_horizon} derived in subsection \ref{match_cond_Cmu}.

~

\noindent $\bullet$ {\bf Transverse sector.}

The most general solution for $C_\perp$  expressed in a piecewise form is:
\begin{align}
& C_\perp^{\rm up}(r,k^\mu)= c_\perp^{\rm up} C_\perp^{\rm ig}(r,k^\mu) - h_\perp^{\rm up} C_\perp^{\rm ig}(r,\bar k^\mu) e^{2i\omega\zeta_2(r)}, \qquad r\in[r_h-\epsilon, \infty_2), \nonumber\\
& C_\perp^{\rm dw}(r,k^\mu)= c_\perp^{\rm dw} C_\perp^{\rm ig}(r,k^\mu) - h_\perp^{\rm dw} C_\perp^{\rm ig}(r,\bar k^\mu) e^{2i\omega\zeta_1(r)}, \qquad r\in[r_h-\epsilon, \infty_1). \label{Cperp_up_dw}
\end{align}
Here $c_\perp^{\rm up,dw}=c_\perp^{\rm up,dw}(k_\mu)$ and $h_\perp^{\rm up,dw}=h_\perp^{\rm up,dw}(k_\mu)$ are   linear superposition coefficients.
Near the horizon, $C_\perp^{\rm ig}$ is regular while $C_\perp^{\rm og}$ oscillates as $e^{2i\omega \zeta_1(r)}$  in the upper segment and as $e^{2i\omega \zeta_2(r)}$ in the lower one. The matching condition \eqref{glue_dCi_simple} and the continuity condition \eqref{continuity_Cmu} imply
\begin{align}
h_\perp^{\rm up}= h_\perp^{\rm dw} e^{\beta \omega},
~~~~~~~~~~~~~~
c_\perp^{\rm up}= c_\perp^{\rm dw}.
\end{align}
Eventually, the solution for the transverse mode is
\begin{align}
& C_\perp^{\rm up}(r,k^\mu)= c_\perp C_\perp^{\rm ig}(r,k^\mu) - h_\perp C_\perp^{\rm ig}(r,\bar k^\mu) e^{2i\omega\zeta_2(r)}, \qquad \qquad r\in[r_h-\epsilon, \infty_2), \nonumber \\
& C_\perp^{\rm dw}(r,k^\mu)= c_\perp C_\perp^{\rm ig}(r,k^\mu) - h_\perp e^{-\beta \omega} C_\perp^{\rm ig}(r,\bar k^\mu) e^{2i\omega\zeta_1(r)}, \qquad r\in[r_h-\epsilon, \infty_1), \label{Cperp_up_dw1}
\end{align}
where  $c_\perp^{\rm up} \to c_\perp$ and $h_\perp^{\rm up} \to h_\perp$ relabelling was made. The piecewise solution \eqref{Cperp_up_dw1}
could be put into a more compact form:
\begin{align}
C_\perp(r,k^\mu)= c_\perp C_\perp^{\rm ig}(r,k^\mu) - h_\perp C_\perp^{\rm ig}(r, \bar k^\mu) e^{2i\omega \zeta(r)}, \qquad r\in(\infty_2, \infty_1), \label{Cperp_final}
\end{align}
where $\zeta(r)$ is defined similarly to $\zeta_2(r)$ but with the $r$  interval extended over the whole contour:
\begin{align}
\zeta(r)\equiv \int_{\infty_2}^r \frac{dy}{f(y)}, \qquad r\in(\infty_2,\infty_1).
\end{align}
The decomposition coefficients $c_\perp$ and $h_\perp$ are fixed
from the  boundary conditions at $r=\infty_1$ and $r=\infty_2$ (see \eqref{Cperpexpand}).
\begin{align}
& c_\perp C_\perp^{\rm ig(0)}(k^\mu)- h_\perp C_\perp^{\rm ig(0)}(\bar k^\mu) = B_{2\perp}(k^\mu), \nonumber \\
& c_\perp C_\perp^{\rm ig(0)}(k^\mu) - h_\perp e^{-\beta \omega} C_\perp^{\rm ig(0)}(\bar k^\mu)  = B_{1\perp}(k^\mu), \nonumber \\
\Rightarrow & c_\perp = \frac{1}{2} \coth\frac{\beta \omega}{2} \frac{B_{a\perp}(k^\mu)} {C_\perp^{\rm ig(0)}(k^\mu)} + \frac{B_{r\perp}(k^\mu)} {C_\perp^{\rm ig(0)}(k^\mu)}
 \qquad h_\perp = \frac{B_{a\perp}(k^\mu)}{(1-e^{-\beta \omega})C_\perp^{\rm ig(0)}(\bar k^\mu)}. \label{chperp}
\end{align}
As mentioned earlier, $C_\perp^{\rm ig(0)}$ can be set to one without loss of generality.

~

\noindent$\bullet$ {\bf Longitudinal sector}

With the four independent solutions presented in subsection \ref{sol_single_AdS}, we
construct the most general solution for the longitudinal sector $\{C_v, C_x\}$ in a
piecewise form:
\begin{align}
C_v^{\rm up}(r,k^\mu)= & c_\parallel^{\rm up} C_v^{\rm ig}(r,k^\mu) + h_\parallel^{\rm up} C_v^{\rm ig}(r,\bar k^\mu)e^{2i\omega \zeta_2(r)} + p_\parallel^{\rm up} C_v^{\rm pg}(r,k^\mu) \nonumber \\
&+ n_\parallel^{\rm up} C_v^{\rm pn}(r,k^\mu), \qquad r\in[r_h-\epsilon, \infty_2), \nonumber \\
C_x^{\rm up}(r,k^\mu)= & c_\parallel^{\rm up} C_x^{\rm ig}(r,k^\mu) - h_\parallel^{\rm up} C_x^{\rm ig}(r,\bar k^\mu)e^{2i\omega \zeta_2(r)} + p_\parallel^{\rm up} C_x^{\rm pg}(r,k^\mu) \nonumber \\
&+ n_\parallel^{\rm up} C_x^{\rm pn}(r,k^\mu), \qquad r\in[r_h-\epsilon, \infty_2), \nonumber \\
C_v^{\rm dw}(r,k^\mu)= & c_\parallel^{\rm dw} C_v^{\rm ig}(r,k^\mu) + h_\parallel^{\rm dw} C_v^{\rm ig}(r,\bar k^\mu)e^{2i\omega \zeta_1(r)} + p_\parallel^{\rm dw} C_v^{\rm pg}(r,k^\mu) \nonumber \\
&+ n_\parallel^{\rm dw} C_v^{\rm pn}(r,k^\mu), \qquad r\in[r_h-\epsilon, \infty_1), \nonumber \\
C_x^{\rm dw}(r,k^\mu)= & c_\parallel^{\rm dw} C_x^{\rm ig}(r,k^\mu) - h_\parallel^{\rm dw} C_x^{\rm ig}(r,\bar k^\mu)e^{2i\omega \zeta_1(r)} + p_\parallel^{\rm dw} C_x^{\rm pg}(r,k^\mu) \nonumber \\
&+ n_\parallel^{\rm dw} C_x^{\rm pn}(r,k^\mu), \qquad r\in[r_h-\epsilon, \infty_1), \label{CvCx_up_dw}
\end{align}
where $\{C_v^{\rm ig}, C_x^{\rm ig}\}$, $\{C_v^{\rm pg}, C_x^{\rm pg}\}$ and $\{C_v^{\rm pn}, C_x^{\rm pn}\}$
are related to the solutions in the Schwarzschild coordinates by the rule \eqref{CvCx_EF_Schw}.
Next, the piecewise solutions \eqref{CvCx_up_dw} are glued via the matching conditions \eqref{continuity_Cmu}, \eqref{glue_dCi_simple}, and \eqref{glue_Cv=0_horizon}.
At the horizon surface, both $C_v^{\rm ig}$ and $C_v^{\rm pn}$ vanish, while $C_v^{\rm pg}$ is generically nonzero. Thus, the condition \eqref{glue_Cv=0_horizon} requires\footnote{In fact, $p_\parallel^{\rm up, dw}$ can be absorbed into redefinition of $\tilde \Lambda$ in \eqref{CtCx_pure_gauge}.}
\begin{align}
p_\parallel^{\rm up}= p_\parallel^{\rm dw}=0.
\end{align}
The condition \eqref{glue_dCi_simple} implies
\begin{align}
h_\parallel^{\rm up} = h_\parallel^{\rm dw} e^{\beta \omega},
\end{align}
where we have  used  the fact that
\begin{align}
&f(r) \partial_r C_x^{\rm ig}(r,k^\mu) \xrightarrow[]{r\to r_h} 0, \qquad \qquad
f(r) \partial_r C_x^{\rm pn}(r,k^\mu) \xrightarrow[]{r\to r_h} 0, \nonumber \\
&f(r) \partial_r C_x^{\rm og}(r,k^\mu) \xrightarrow[]{r\to r_h} -2i\omega C_x^{\rm ig} (r,\bar k^\mu) e^{2i\omega \zeta_1(r)}, \qquad r\in[r_h-\epsilon, \infty_1), \nonumber \\
&f(r) \partial_r C_x^{\rm og}(r,k^\mu) \xrightarrow[]{r\to r_h} -2i\omega C_x^{\rm ig} (r,\bar k^\mu) e^{2i\omega \zeta_2(r)}, \qquad r\in[r_h-\epsilon, \infty_2).
\end{align}
Finally, the matching condition \eqref{continuity_Cmu} implies
\begin{align}
c_\parallel^{\rm up}= c_\parallel^{\rm dw}.
\end{align}
Eventually, the entire solution for the  longitudinal sector is
\begin{align}
C_v^{\rm up}(r,k^\mu)= & c_\parallel C_v^{\rm ig}(r,k^\mu) + h_\parallel C_v^{\rm ig}(r,\bar k^\mu)e^{2i\omega \zeta_2(r)}  + n_\parallel^{\rm up} C_v^{\rm pn} (r,k^\mu), \nonumber \\
C_x^{\rm up}(r,k^\mu)= & c_\parallel C_x^{\rm ig}(r,k^\mu) - h_\parallel C_x^{\rm ig}(r,\bar k^\mu)e^{2i\omega \zeta_2(r)} + n_\parallel^{\rm up} C_x^{\rm pn}(r,k^\mu), \nonumber \\
C_v^{\rm dw}(r,k^\mu)= & c_\parallel C_v^{\rm ig}(r,k^\mu) + h_\parallel e^{-\beta \omega} C_v^{\rm ig}(r,\bar k^\mu)e^{2i\omega \zeta_1(r)} + n_\parallel^{\rm dw} C_v^{\rm pn}(r,k^\mu), \nonumber \\
C_x^{\rm dw}(r,k^\mu)= & c_\parallel C_x^{\rm ig}(r,k^\mu) - h_\parallel e^{-\beta \omega} C_x^{\rm ig}(r,\bar k^\mu)e^{2i\omega \zeta_1(r)} + n_\parallel^{\rm dw} C_x^{\rm pn}(r,k^\mu), \label{CvCx_up_dw1}
\end{align}
where  $c_\parallel^{\rm up} \to c_\parallel, h_\parallel^{\rm up} \to h_\parallel$ relabelling is made.
Due to  the presence of the polynomial solution $\{C_v^{\rm pn}, C_x^{\rm pn}\}$, it is not possible to cast the final result \eqref{CvCx_up_dw1} into a more compact form, similar to \eqref{Cperp_final}.

Recall that near the AdS boundary all the linearly independent solutions have asymptotic expansions similar to the general case \eqref{Cmu_inf1} and \eqref{Cmu_inf2}, cf. \eqref{Cparexpand}.
The  coefficients $c_\parallel, h_\parallel, n_\parallel^{\rm dw}, n_\parallel^{\rm up}$  in \eqref{CvCx_up_dw1} are fixed by the AdS boundary conditions
\begin{align}
& c_\parallel C_v^{\rm ig(0)}(k^\mu) + h_\parallel C_v^{\rm ig(0)}(\bar k^\mu) + n_\parallel^{\rm up} C_v^{\rm pn(0)}(k^\mu) = B_{2v}(k^\mu), \nonumber \\
& c_\parallel C_x^{\rm ig(0)}(k^\mu) - h_\parallel C_x^{\rm ig(0)}(\bar k^\mu) + n_\parallel^{\rm up} C_x^{\rm pn(0)}(k^\mu) = B_{2x}(k^\mu), \nonumber \\
& c_\parallel C_v^{\rm ig(0)}(k^\mu) + h_\parallel e^{-\beta \omega} C_v^{\rm ig(0)} (\bar k^\mu) + n_\parallel^{\rm dw} C_v^{\rm pn(0)}(k^\mu) = B_{1v}(k^\mu), \nonumber \\
& c_\parallel C_x^{\rm ig(0)}(k^\mu) - h_\parallel e^{-\beta \omega} C_x^{\rm ig(0)} (\bar k^\mu) + n_\parallel^{\rm dw} C_x^{\rm pn(0)}(k^\mu) = B_{1x}(k^\mu), \label{AdS_condition_CvCx}
\end{align}
which yield
\begin{align}
& c_\parallel= \frac{1}{2}G_1^{-1} \left\{ 2B_{rx}(k^\mu) C_v^{\rm pn(0)}(k^\mu) - 2 B_{rv}(k^\mu) C_x^{\rm pn(0)}(k^\mu) \right. \nonumber\\
& \qquad \qquad \left.+ \coth\frac{\beta \omega}{2} \left[B_{ax}(\omega,q) C_v^{\rm pn(0)}(k^\mu) - B_{av}(k^\mu) C_x^{\rm pn(0)} (k^\mu) \right] \right\}, \label{c_sol}\\
& h_\parallel = (1-e^{-\beta \omega})^{-1} G_2^{-1} \left[ B_{ax}(k^\mu) C_v^{\rm pn(0)} (k^\mu) - B_{av}(k^\mu) C_x^{\rm pn(0)}(k^\mu) \right], \label{h_sol}\\
& n_\parallel^{\rm dw}- n_\parallel^{\rm up}= G_2^{-1} \left[B_{ax}(k^\mu) C_v^{\rm ig(0)}(\bar k^\mu) + B_{av}(k^\mu) C_x^{\rm ig(0)} (\bar k^\mu) \right], \label{n-_sol}\\
& \frac{1}{2}(n_\parallel^{\rm dw} + n_\parallel^{\rm up}) = - G_1^{-1} \left[ B_{rx}(k^\mu) C_v^{\rm ig(0)}(k^\mu) - B_{rv}(k^\mu) C_x^{\rm ig(0)} (k^\mu) \right] - \frac{1}{2}\coth\frac{\beta \omega}{2}\nonumber \\
& \qquad \qquad \qquad \quad \times   G_1^{-1} G_2^{-1} G_3 \left[  B_{ax}(k^\mu) C_v^{\rm pn(0)}(k^\mu) - B_{av}(k^\mu) C_x^{\rm pn(0)} (k^\mu) \right], \label{n+_sol}
\end{align}
where
\begin{align}
& G_1=C_v^{\rm pn(0)}(k^\mu) C_x^{\rm ig(0)}(k^\mu)- C_v^{\rm ig(0)}(k^\mu) C_x^{\rm pn(0)}(k^\mu), \nonumber \\
& G_2=C_v^{\rm pn(0)}(k^\mu) C_x^{\rm ig(0)}(\bar k^\mu) + C_v^{\rm ig(0)}(\bar k^\mu) C_x^{\rm pn(0)}(k^\mu), \nonumber \\
& G_3= C_v^{\rm ig(0)} (\bar k^\mu) C_x^{\rm ig(0)}(k^\mu) + C_v^{\rm ig(0)}(k^\mu) C_x^{\rm ig(0)}(\bar k^\mu).
\end{align}
Notice that $G_2=G_1^*$.
Without loss of generality two of the coefficients, say, $ C_v^{\rm pn(0)}$ and $C_v^{\rm ig(0)}$ could be set to one. The remaining coefficients would have to be found from the
solutions of the EOMs. Finally, it is not difficult to verify that $G_1 \neq 0$.

\subsection{From the bulk solutions to the effective action}

With the entire solution for $C_M(r,v,\vec x)$ derived in subsection \ref{sol_contour}, we are now ready to evaluate the effective Lagrangian \eqref{Leff1},
which can be split into transverse and longitudinal parts:
\begin{align}
\mathcal{L}_{\rm eff} = \mathcal{L}_{\rm eff}^\perp + \mathcal{L}_{\rm eff}^{\parallel}
\end{align}
We will need  near-boundary expansion coefficients  $C_M^{(2)}$  and the expansion  of
 $e^{2i\omega \zeta_s(r)}$:
\begin{align}
& e^{2i\omega \zeta_s(r)} \xrightarrow[]{r\to\infty_s} 1-\frac{2i\omega}{r} - \frac{2\omega^2}{r^2} + \cdots, \qquad s=1, ~ {\rm or} ~ ~ 2.
\end{align}

\noindent $\bullet$  $\bf \mathcal{L}_{\rm eff}^\perp$

In the $(r,a)$-basis, the coefficients $C_\perp^{(2)}$  (normalizable modes) are
\begin{align}
C_{a\perp}^{(2)}(k^\mu) =& - \frac{ C_\perp^{\rm ig(2)} (\bar k^\mu)} {C_\perp^{\rm ig(0)}(\bar k^\mu)} B_{a\perp}(k^\mu), \\
C_{r\perp}^{(2)}(k^\mu) =&\frac{1}{2}\coth{\frac{\beta \omega}{2}} \left[\frac{C_\perp^{\rm ig(2)}(k^\mu)}{C_\perp^{\rm ig(0)}(k^\mu)} -\frac{C_\perp^{\rm ig(2)}(\bar k^\mu)}{C_\perp^{\rm ig(0)}(\bar k^\mu)}\right] B_{a\perp}(k^\mu) + \frac{C_\perp^{\rm ig(2)}(k^\mu)}{C_\perp^{\rm ig(0)} (k^\mu)} B_{r\perp}(k^\mu). \label{Cperp_norma_modes}
\end{align}
From \eqref{Leff1}, the transverse part of the effective Lagrangian is
\begin{align}
\mathcal{L}_{\rm eff}^\perp = B_{r\perp}(x) C_{a\perp}^{(2)}(x) + B_{a\perp}(x) C_{r\perp}^{(2)}(x) - \frac{1}{2} B_{a\perp}(x) \partial_v^2 B_{r\perp}(x) - \frac{1}{2} B_{a\perp}(x) \vec\partial^{\;2} B_{r\perp}(x),
\end{align}
which in  the momentum space reads
\begin{align}
\mathcal{L}_{\rm eff}^\perp(k)=& B_{a\perp}(-k)\left[\frac{2 C_\perp^{\rm ig(2)} (k)}{C_\perp^{\rm ig(0)}(k)} + \frac{1}{2}\omega^2 + \frac{1}{2}q^2 \right] B_{r\perp}(k) \nonumber \\
&+ B_{a\perp}(-k) \frac{1}{2} \coth\frac{\beta \omega}{2} \left[\frac{C_\perp^{\rm ig(2)} (k)}{C_\perp^{\rm ig(0)}(k)}- \frac{C_\perp^{\rm ig(2)} (\bar k)}{C_\perp^{\rm ig(0)}(\bar k)} \right] B_{a\perp}(k).  \label{Lperp1}
\end{align}
Comparing \eqref{Lperp1} with \eqref{Leff3},  the  TCFs  in \eqref{Leff3} are expressed
in terms of  the results obtained in the bulk:
\begin{align}
& -i\omega w_8(k) + q^2 w_9(k)= \frac{2 C_\perp^{\rm ig(2)} (k)}{C_\perp^{\rm ig(0)}(k)} + \frac{1}{2}\omega^2 + \frac{1}{2}q^2, \label{w8_w9}\\
& \frac{i}{2}w_2(k) = \frac{1}{2} \coth\frac{\beta \omega}{2} \left[\frac{C_\perp^{\rm ig(2)} (k)}{C_\perp^{\rm ig(0)}(k)}- \frac{C_\perp^{\rm ig(2)} (\bar k)}{C_\perp^{\rm ig(0)}(\bar k)} \right]. \label{w2}
\end{align}
While $w_2$ is determined entirely  by  the transverse sector,
 the coefficients $w_8,w_9$ will be uniquely fixed  only with addition of  the longitudinal sector.

~

\noindent$\bullet$  $\bf \mathcal{L}_{\rm eff}^\parallel$

In order to cast the results into a more compact form,  the following ratios are introduced
\begin{align}\label{ratios}
&R_v^{\rm ig}(k)= \frac{C_v^{\rm ig(2)}(k)}{C_v^{\rm ig(0)} (k)}, \qquad R_v^{\rm pn}(k)= \frac{C_v^{\rm pn(2)} (k)}{C_v^{\rm pn(0)}(k)}, \qquad
R_{xv}^{\rm ig}(k) = \frac{C_x^{\rm ig(2)}(k)}{C_v^{\rm ig(0)}(k)}, \nonumber \\  &R_{xv}^{\rm pn}(k) = \frac{C_x^{\rm pn(2)} (k)}{C_v^{\rm pn(0)}(k)}, \qquad \bar R_{xv}^{\rm ig}(k) = \frac{C_x^{\rm ig(0)} (k)}{C_v^{\rm ig(0)}(k)}, \qquad \bar R_{xv}^{\rm pn}(k) = \frac{C_x^{\rm pn(0)} (k)}{C_v^{\rm pn(0)}(k)}.
\end{align}
These ratios are determined by solving the dynamical EOMs \eqref{dynamical_EOM} in a single copy of  doubled Schwarzschild-AdS$_5$, see subsection \ref{sol_single_AdS}.
It is important to recall that $C_v^{\rm ig(2)}$ is related to $C_x^{\rm ig(2)}$ via \eqref{relation}.
Furthermore, there is still the freedom to set both
$C_v^{\rm ig(0)}$ and $C_v^{\rm pn(0)}$ to one.

Near the AdS boundaries $r=\infty_1$ and $r=\infty_2$, we extract the normalizable modes in $C_v$ and $C_x$ (in the $(r,a)$-basis):
\begin{align}
C_{rv}^{(2)}(k)=& c_\parallel C_v^{\rm ig(2)}(k) + \frac{1}{2} h_\parallel (1+ e^{-\beta\omega}) C_v^{\rm ig(2)}(\bar k) + \frac{1}{2}(n_\parallel^{\rm dw} + n_\parallel^{\rm up}) C_v^{\rm pn(2)}(k) \nonumber\\
=& \frac{R_v^{\rm ig}(k)-R_v^{\rm pn}(k)}{\bar R_{xv}^{\rm ig}(k)-\bar R_{xv}^{\rm pn} (k)} B_{rx}(k) + \frac{\bar R_{xv}^{\rm ig}(k) R_v^{\rm pn} (k) -\bar R_{xv}^{\rm pn} (k) R_v^{\rm ig}(k)}{\bar R_{xv}^{\rm ig}(k) -\bar R_{xv}^{\rm pn} (k)} B_{rv}(k) \label{Crv_norma_mode}\\
&+ \frac{1}{2} \coth\frac{\beta \omega}{2} \left\{\frac{R_v^{\rm ig}(k)-R_v^{\rm pn}(k)}{\bar R_{xv}^{\rm ig}(k)-\bar R_{xv}^{\rm pn} (k)} +\frac{R_v^{\rm ig}(\bar k)-R_v^{\rm pn} (k)}{\bar R_{xv}^{\rm ig}(\bar k) +\bar R_{xv}^{\rm pn} (k)}\right\}  \left[B_{ax}(k)- \bar R_{xv}^{\rm pn}(k) B_{av}(k)\right], \nonumber
\end{align}
\begin{align}
C_{av}^{(2)}(k)=& -h_\parallel (1-e^{-\beta \omega})C_v^{\rm ig(2)}(\bar k) +(n_\parallel^{\rm dw}- n_\parallel^{\rm up}) C_v^{\rm pn(2)}(k) \nonumber\\
=& R_v^{\rm pn}(k) \frac{B_{ax}(k) + \bar R_{xv}^{\rm ig}(\bar k) B_{av}(k)}{\bar R_{xv}^{\rm ig} (\bar k) + \bar R_{xv}^{\rm pn}(k)} + R_v^{\rm ig}(\bar k) \frac{ -B_{ax}(k) + \bar R_{xv}^{\rm pn}(k) B_{av}(k)}{\bar R_{xv}^{\rm ig}(\bar k) + \bar R_{xv}^{\rm pn} (k)}, \label{Cav_norma_mode}
\end{align}
\begin{align}
C_{rx}^{(2)}(k)= & c_\parallel C_x^{\rm ig(2)}(k) - \frac{1}{2} h_\parallel (1+ e^{-\beta\omega}) C_x^{\rm ig(2)}(\bar k) + \frac{1}{2}(n_\parallel^{\rm dw} + n_\parallel^{\rm up}) C_x^{\rm pn(2)}(k) \nonumber \\
= & R_{xv}^{\rm ig}(k) \frac{ B_{rx}(k) -\bar R_{xv}^{\rm pn}(k) B_{rv}(k)}{\bar R_{xv}^{\rm ig}(k) - \bar R_{xv}^{\rm pn}(k)} + R_{xv}^{\rm pn}(k) \frac{-B_{rx}(k) + \bar R_{xv}^{\rm ig}(k) B_{rv}(k)}{\bar R_{xv}^{\rm ig}(k) - \bar R_{xv}^{\rm pn}(k)} \label{Crx_norma_mode}\\
& + \frac{1}{2}\coth\frac{\beta \omega}{2} \left\{\frac{R_{xv}^{\rm ig}(k) - R_{xv}^{\rm pn}(k)}{\bar R_{xv}^{\rm ig}(k) - \bar R_{xv}^{\rm pn}(k)} - \frac{R_{xv}^{\rm ig}(\bar k) + R_{xv}^{\rm pn}(k)}{\bar R_{xv}^{\rm ig}(\bar k) + \bar R_{xv}^{\rm pn}(k)}  \right\}\left[ B_{ax}(k)- \bar R_{xv}^{\rm pn}(k) B_{av}(k) \right],\nonumber
\end{align}
\begin{align}
C_{ax}^{(2)}(k)=& h_\parallel (1-e^{-\beta \omega})C_x^{\rm ig(2)}(\bar k) +(n_\parallel^{\rm dw}- n_\parallel^{\rm up}) C_x^{\rm pn(2)}(k) \nonumber \\
=& R_{xv}^{\rm ig}(\bar k) \frac{B_{ax}(k) - \bar R_{xv}^{\rm pn}(k) B_{av}(k)}{\bar R_{xv}^{\rm ig}(\bar k)+ \bar R_{xv}^{\rm pn}(k)} + R_{xv}^{\rm pn}(k) \frac{B_{ax}(k) + \bar R_{xv}^{\rm ig}(\bar k) B_{av}(k)}{\bar R_{xv}^{\rm ig}(\bar k)+ \bar R_{xv}^{\rm pn}(k)}. \label{Cax_norma_mode}
\end{align}
From \eqref{Leff1}, the longitudinal part of  the effective Lagrangian is
\begin{align}
\mathcal{L}_{\rm eff}^{\parallel}=&- B_{rv}(x) C_{av}^{(2)}(x) -B_{av}(x) C_{rv}^{(2)}(x) + B_{rx}(x) C_{ax}^{(2)}(x) + B_{ax} C_{rx}^{(2)}(x) \nonumber \\
& + \frac{1}{2}\partial_x B_{ax}(x) \partial_v B_{rv}(x) - \frac{1}{2} B_{av}(x) \partial_v \partial_x B_{rx}(x) - \frac{1}{2} B_{ax}(x) \partial_v^2 B_{rx}(x) \nonumber \\
& + \frac{1}{2} B_{av}(x) \vec\partial^{\;2} B_{rv}(x) + B_{av}(x) \partial_v^2 B_{rv}(x),
\label{Lparallel_10}
\end{align}
which in the momentum space becomes
\begin{align}
\mathcal{L}_{\rm eff}^{\parallel}(k)= & B_{av}(-k) \frac{i}{2}w_1(k) B_{av}(k) + B_{ax}(-k) \left(\frac{i}{2}w_2(k) + \frac{i}{2}q^2 w_3(k)\right)B_{ax}(k) \nonumber \\
& + B_{av}(-k) [-qw_4(k)]B_{ax}(k) + B_{av}(-k) w_5(k) B_{rv}(k) \nonumber \\
& + B_{av}(-k) \omega q w_6(k) B_{rx}(k) + B_{ax}(-k)[-iq w_7(k) ]B_{rv}(k) \nonumber \\
& + B_{ax}(-k)[-i\omega w_8(k)]B_{rx}(k), \label{Lparallel_1}
\end{align}
with the TCFs given by the following expressions
\begin{align}
\frac{i}{2}w_1(k)= \frac{1}{2} \coth\frac{\beta\omega}{2} \left[ \frac{R_v^{\rm ig}(k) - R_v^{\rm pn}(k)}{\bar R_{xv}^{\rm ig} (k)- \bar R_{xv}^{\rm pn}(k)} + \frac{R_v^{\rm ig} (\bar k)- R_v^{\rm pn}(k)}{\bar R_{xv}^{\rm ig}(\bar k) + \bar R_{xv}^{\rm pn}(k)} \right] \bar R_{xv}^{\rm pn}(k), \label{w1}
\end{align}
\begin{align}
\frac{i}{2}w_2(k) + \frac{i}{2}q^2 w_3(k)= \frac{1}{2}\coth\frac{\beta\omega}{2} \left[\frac{R_{xv}^{\rm ig}(k)- R_{xv}^{\rm pn}(k)}{\bar R_{xv}(k) - \bar R_{xv}^{\rm pn}(k)} - \frac{R_{xv}^{\rm ig}(\bar k) + R_{xv}^{\rm pn}(k)}{\bar R_{xv}^{\rm ig} (\bar k) + \bar R_{xv}^{\rm pn}(k)} \right], \label{w2_w3}
\end{align}
\begin{align}
-qw_4(k)=&-\frac{1}{2} \coth\frac{\beta\omega}{2} \left[ \frac{R_v^{\rm ig}(k) -R_v^{\rm pn}(k)}{\bar R_{xv}^{\rm ig}(k) - \bar R_{xv}^{\rm pn}(k)} + \frac{R_v^{\rm ig}(\bar k) -R_v^{\rm pn}(k)}{\bar R_{xv}^{\rm ig}(\bar k) + \bar R_{xv}^{\rm pn}(k)}\right] \nonumber \\
&+\frac{1}{2} \coth\frac{\beta\omega}{2} \left[ \frac{R_{xv}^{\rm ig}(\bar k) + R_{xv}^{\rm pn}(k)}{\bar R_{xv}^{\rm ig}(\bar k) + \bar R_{xv}^{\rm pn}(k)} - \frac{R_{xv}^{\rm ig}(k) -R_{xv}^{\rm pn}(k)}{\bar R_{xv}^{\rm ig}(\bar k) - \bar R_{xv}^{\rm pn}(k)}\right] \bar R_{xv}^{\rm pn}(k), \label{w4}
\end{align}
\begin{align}
w_5(k)= -\omega^2 - \frac{1}{2}q^2 + 2\frac{R_v^{\rm ig}(k) \bar R_{xv}^{\rm pn}(k)- R_v^{\rm pn}(k) \bar R_{xv}^{\rm ig}(k)}{\bar R_{xv}^{\rm ig}(k)- \bar R_{xv}^{\rm pn}(k)}, \label{w5}
\end{align}
\begin{align}
\omega q w_6(k)=-\frac{1}{2}\omega q - \frac{R_v^{\rm ig}(k) - R_v^{\rm pn}(k)}{\bar R_{xv}^{\rm ig}(k) - \bar R_{xv}^{\rm pn}(k)} + \frac{\bar R_{xv}^{\rm ig}(k) R_{xv}^{\rm pn}(k)- \bar R_{xv}^{\rm pn}(k) R_{xv}^{\rm ig}(k)}{\bar R_{xv}^{\rm ig} (k) - \bar R_{xv}^{\rm pn}(k)}, \label{w6}
\end{align}
\begin{align}
-iq w_7(k)= \omega q w_6(k), \label{w7}
\end{align}
\begin{align}
-i\omega w_8(k)= \frac{1}{2}\omega^2 +2 \frac{R_{xv}^{\rm ig}(k) - R_{xv}^{\rm pn}(k)} {\bar R_{xv}^{\rm ig}(k) - \bar R_{xv}^{\rm pn}(k)}. \label{w8}
\end{align}
We observe  that all the TCFs are expressed in terms of the ratios \eqref{ratios}, which are extracted from the linearly dependent solutions to the dynamical EOMs in  a single copy of the doubled Schwarzschild-AdS$_5$.
In the next section  these results are presented explicitly. Finally, it is very important to realise
 and straightforward to check that all the symmetry relations imposed by the discrete symmetries (see subsection \ref{ds})
are satisfied  by
\eqref{w8_w9}, \eqref{w2}, \eqref{w1}-\eqref{w8} automatically.  Furthermore,  to demonstrate that one does not actually need to solve the bulk EOMs  at all.

\section{Results for the TCFs} \label{results}

In this section, all the  results for the parameters in the effective Lagrangian \eqref{Leff3} are presented. For completeness and consistency check, we first consider limits in which analytical calculations could be performed,
hereby recovering some of the  results available in the literature. Next we switch to  numerical analysis.

\subsection{Analytical results} \label{analy_results}

\noindent{$\bullet$   $\bf q=0$}

When $q=0$,  $SO(3)$ rotational symmetry is recovered, $\tilde C_\perp=\tilde C_x=\tilde C_i$ and the dynamical EOM \eqref{eom_CtCx_Schw}
for this component decouples from that of $\tilde C_t$.
The analytical results for the basic set of solutions are
\begin{align}
&\tilde C_t^{\rm ig}(r,\omega,q=0)=0 ~~\Rightarrow~~ \tilde C_t^{\rm ig(0)}(\omega,q)= \tilde C_t^{\rm ig(2)}(\omega,q)= 0, \nonumber \\
&\tilde C_i^{\rm pn}(r,\omega,q=0)=0~~ \Rightarrow ~~\tilde C_i^{\rm pn(0)}(\omega,q)= \tilde C_i^{\rm pn(2)}(\omega,q) =0.
\end{align}
\begin{align}
 \tilde C_t^{\rm pn}(r,\omega,q=0)=1-\frac{r_h^2}{r^2} ~ ~
\Rightarrow ~~ \tilde C_t^{\rm pn(0)}(\omega,q=0)=1, \qquad \tilde C_t^{\rm pn(2)} (\omega,q=0)=-r_h^2.
\end{align}
The only non-trivial result is related to the spatial component $\tilde C_i^{\rm ig}$, which is however well known  \cite{Horowitz:2008bn,Myers:2007we}:
\begin{align}
\tilde C_i^{\rm ig}(r,\omega,q=0)= & \left(1- \frac{r_h^2}{r^2} \right)^{-i\omega/(4r_h)} \left(1+ \frac{r_h^2}{r^2}\right)^{-\omega/(4r_h)} \left(\frac{r_h^2}{r^2}\right)^{(1+i)\omega/(4r_h)} \nonumber \\
& \times {_2}F_1\left[1 - \frac{(1+i)\omega}{4 r_h}, - \frac{(1+i)\omega}{4 r_h}, 1- \frac{i\omega}{2r_h}, \frac{1}{2}\left( 1-\frac{r^2}{r_h^2} \right) \right],
\end{align}
where ${_2}F_1$ is a hypergeometric function. Near the AdS boundary,
\begin{align}
\frac{\tilde C_i^{\rm ig(2)}(\omega,q=0)}{\tilde C_i^{\rm ig(0)}(\omega,q=0)}=& - r_h^2 \tilde \omega \left\{ i +\left[2\gamma_e - 1 + \log(2r_h^2/L^2)\right]\tilde \omega + \tilde \omega \psi\left(-\frac{(1+i)}{2}\tilde \omega\right) \right. \nonumber \\
& \qquad \qquad \left. + \tilde \omega \psi\left(\frac{(1-i)}{2}\tilde \omega\right) \right\},
\end{align}
where $\tilde \omega= \omega/(2r_h)=\omega\beta/(2\pi) $ is introduced for compactness, $\psi(z)= d\Gamma(z)/dz$, and $\gamma_e$ is the Euler constant.
Here we have  reinstalled  the AdS curvature radius $L$  in the logarithmic term.

Thus, in the limit $q=0$ we can fix $w_1$, $w_2$, $w_5$ and $w_8$ while the remaining TCFs decouple:
\begin{align}
& w_1(q=0)=0, \nonumber \\
& w_2(q=0)= 2\coth(\pi \tilde \omega) {\rm Im}\left[\frac{\tilde C_\perp^{\rm ig(2)} (\omega,q=0)} {\tilde C_\perp^{\rm ig(0)}(\omega,q=0)} \right], \nonumber \\
& w_5(q=0)= -2\frac{\tilde C_t^{\rm pn(2)}(\omega,q=0)}{\tilde C_t^{\rm pn(0)}(\omega,q=0)}= 2r_h^2, \nonumber \\
& w_8(q=0)= - \frac{1}{2}i\omega - \frac{2}{i\omega} \frac{\tilde C_x^{\rm ig(2)}(\omega,q=0)} {\tilde C_x^{\rm ig(0)}(\omega,q=0)}.
\end{align}

\noindent {$\bullet$ $\bf \omega=q$}

In this limit analytical results are available  for the  transverse mode $\tilde C_\perp$ only  \cite{CaronHuot:2006te}.
The ingoing solution of the dynamical EOM \eqref{eom_Cperp_Schw} is
\begin{align}
\tilde C_\perp^{\rm ig}(r,\omega=q)= & \left(1- \frac{r_h^2}{r^2} \right)^{-i\omega/(4r_h)} \left(1+ \frac{r_h^2}{r^2} \right)^{\omega/(4r_h)} \nonumber \\
& \times {_2}F_1\left[ 1- \frac{(1+i)\omega}{4 r_h}, - \frac{(1+i)\omega}{4 r_h}, 1- \frac{i\omega}{2r_h}, \frac{1}{2}\left( 1-\frac{r_h^2}{r^2} \right) \right].
\end{align}
Near the AdS boundary $r=\infty$,
\begin{align}
\frac{\tilde C_\perp^{\rm ig(2)}(\omega=q)}{\tilde C_\perp^{\rm ig(0)}(\omega=q)}= r_h^2\left[ 1-\tilde \omega + \left(\frac{1+i}{2}\tilde \omega-1\right) \frac{{_2}F_1(2- (1+i)\tilde\omega/2, - (1+i)\tilde\omega/2, 1-i\tilde \omega, 1/2)}{{_2}F_1(1- (1+i) \tilde\omega/2, - (1+i)\tilde\omega/2, 1-i\tilde \omega, 1/2)} \right].
\end{align}
In the absence of analytical solution in the longitudinal sector, only $w_2$ and the combination $w_8+i\omega w_9$ can be determined:
\begin{align}
& w_2(\omega=q)= 2\coth(\pi \tilde \omega) {\rm Im}\left[\frac{\tilde C_\perp^{\rm ig(2)} (\omega=q)} {\tilde C_\perp^{\rm ig(0)}(\omega=q)} \right], \nonumber \\
& w_8(\omega=q) + i\omega w_9(\omega=q)= - \frac{2}{i\omega} \frac{\tilde C_\perp^{\rm ig(2)} (\omega=q)} {\tilde C_\perp^{\rm ig(0)}(\omega=q)}.
\end{align}

\noindent {\bf $\bullet$   The hydrodynamic limit $\bf \omega\ll T\sim r_h,  q \ll T\sim r_h$.}

In the hydrodynamic limit, the results available in the literature (see \cite{deBoer:2018qqm} and \cite{Glorioso:2018mmw})
pertain to the effective Lagrangian \eqref{Leff3} up to second order in the derivatives of $B_{r\mu}$ and $B_{a\mu}$.  Since our formalism is somewhat different from the others,
it makes sense to perform a comparison. Hence we have to obtain analytical results accurate up to second order.
Naively, one would expect to achieve this accuracy  by solving the bulk EOMs also up to second order in the derivatives of $B_{r\mu}$ and $B_{a\mu}$.
However, as can be seen from \eqref{w2}, \eqref{w2_w3}, and \eqref{w4}, in fact one has to solve for the ingoing solutions keeping the third order terms in the derivative expansion.

It is  convenient to introduce  a new radial coordinate $u$:
\begin{align}
u=r_h^2/r^2 \Longrightarrow \tilde C_\mu(r,\omega,q) \to \tilde C_\mu(u,\omega,q).
\end{align}
The ingoing solution for the transverse mode $\tilde C_\perp$ is well known in the literature, see e.g. \cite{Policastro:2002se}. Up to third order in momenta the solution is
\begin{align}
\tilde C_\perp^{\rm ig}(u,\omega,q)= (1-u^2)^{-i\tilde \omega/2} \left\{ 1+ i\tilde \omega \log(1+u) + \frac{1}{24} \pi^2(3\tilde \omega^2- 2\tilde q^2) - \frac{1}{4}\tilde\omega^2 \log^22 \right. \nonumber \\
 \left. + \frac{1}{2} \tilde \omega^2 \log(1-u) \log \frac{2}{1+u} - \frac{1}{4} \log(1+u) \left[2(\tilde\omega^2- \tilde q^2) \log u + \tilde \omega^2 \log(1+u) \right] \right. \nonumber \\
\left. +\frac{1}{2} (\tilde q^2- \tilde \omega^2) \left[{\rm Li}_2(1-u) + {\rm Li}_2(-u) \right] - \frac{1}{2} \tilde \omega^2 {\rm Li}_2\left(\frac{1+u}{2}\right) + \tilde C_\perp^{\rm ig[3]}(u,\omega,q) + \cdots \right\},
\end{align}
where $\rm Li_2$ is the Polylogarithm function, and $\tilde C_\perp^{\rm ig[3]}(u,\omega,q)$ is the third order solution, which is too lengthy to be shown here.
Near the AdS boundary $u=0$,
\begin{align}
& \tilde C_\perp^{\rm ig(0)}=1, \nonumber \\
& \tilde C_\perp^{\rm ig(2)}=r_h^2\left[ i\tilde \omega -\tilde q^2 + \tilde\omega^2 -\tilde \omega^2 \log 2 + (\tilde q^2- \tilde \omega^2) \log(r_h^2/L^2) + \frac{\pi^2}{12} i\tilde \omega (2\tilde \omega^2- 3\tilde q^2) \right].
\end{align}

Most of the  results about the longitudinal sector available in the literature are based on the on-shell holography, which for this reason cannot be recycled for our study. In the off-shell formalism similar to ours, recently
the authors of \cite{deBoer:2018qqm} worked out the hydrodynamic limit of the independent solutions  $\{\tilde C_t, \tilde C_x\}$ (see appendix B there), though up to second order only.
We have computed the expansion up to third order\footnote{Our results are somewhat different from \cite{deBoer:2018qqm}. The origin of the difference is in the freedom to arbitrarily select the horizon data.}:
\begin{align}
\tilde C_t^{\rm ig}(u, \omega,q)&= (1-u^2)^{1-i\tilde \omega/2} \left[\frac{i\tilde q } {1+u} +\frac{\tilde \omega \tilde q }{1-u^2} \left( \log\frac{2}{1+u} + u\log u \right)  + \tilde C_t^{\rm ig[3]}(u,\omega,q) +\cdots\right], \nonumber \\
\tilde C_x^{\rm ig}(u,\omega,q)&= (1-u^2)^{-i\tilde \omega/2} \left\{1 + i\tilde \omega \log\frac{1+u}{2} + \frac{\pi^2 \tilde \omega^2}{24} - \frac{1}{2} \tilde\omega^2 \log\frac{1-u}{2} \log \frac{1+u}{2} \right. \nonumber \\
& \left. - \frac{1}{4}\tilde\omega^2 \log^2\frac{1+u}{2} - \frac{1}{2}\tilde \omega^2 \log u \log(1+u) - \frac{1}{2} \tilde \omega^2 {\rm Li}_2(1-u) - \frac{1}{2} \tilde \omega^2 {\rm Li}_2(-u)\right. \nonumber \\
& \left.- \frac{1}{2} \tilde \omega^2 {\rm Li}_2\left( \frac{1+u}{2}\right) + \tilde C_x^{\rm ig[3]}(u,\omega,q) +\cdots \right\},
\end{align}
where the third order solutions $\tilde C_t^{\rm ig[3]}(u,\omega,q)$ and $\tilde C_x^{\rm ig[3]}(u,\omega,q)$ have been worked out by us, but are too lengthy to be presented here.
The near AdS boundary data are
\begin{align}
\tilde C_t^{\rm ig(0)}(\omega,q)=&  i\tilde q + \tilde \omega \tilde q \log2 + \frac{1}{2}i\tilde q (-\tilde \omega^2 \log^22 +2\tilde q^2 \log2)+ \cdots, \nonumber \\
\tilde C_t^{\rm ig(2)}(\omega,q)= & r_h^2\left[-i\tilde q -\tilde\omega \tilde q + \tilde \omega \tilde q \log(r_h^2/L^2) \right] - \frac{r_h^2}{12}i\tilde q \left[12\tilde q^2 +\pi^2 \tilde \omega^2 -12 \tilde \omega^2 \log2 \right. \nonumber \\
&\left. + 6\tilde \omega^2 \log^22 +12 \tilde \omega^2 \log2 \log(r_h^2/L^2) -12 \tilde q^2 \log(2r_h^2/L^2)  \right] + \cdots, \nonumber \\
\tilde C_x^{\rm ig(0)}(\omega,q)= & 1- i\tilde \omega \log 2 - \frac{1}{12}\tilde \omega^2 (\pi^2+6\log^22)  + \frac{1}{12} i\omega \pi^2 (\tilde \omega^2 \log 2- \tilde q^2) \nonumber \\
& + \frac{1}{6} i\tilde \omega^3 \left( \log^32 -3\zeta(3) \right) + \cdots, \nonumber \\
\tilde C_x^{\rm ig(2)}(\omega,q)=&  r_h^2 \left[i\tilde \omega + \tilde \omega^2 - \tilde \omega^2 \log(r_h^2/L^2)\right]  + \frac{r_h^2}{12} i\tilde \omega \left[12\tilde q^2 + \pi^2 \tilde \omega^2 - 12 \tilde \omega^2 \log2 \right.\nonumber \\
&\left.  + 6 \tilde \omega^2 \log^22 + 12\tilde \omega^2 \log2 \log (r_h^2/L^2) - 12 q^2 \log(2r_h^2/L^2) \right] + \cdots.
\end{align}
Here we have kept terms up to third order in momenta.

Next, we present the hydrodynamic limit of the polynomial solution $\{\tilde C_t^{\rm pn}, \tilde C_x^{\rm pn}\}$:
\begin{align}
& \tilde C_t^{\rm pn}(u,\omega,q) = 2(1-u) + 2\tilde q^2 \left[u\log u +(1+u) \log \frac{2}{1+u} \right] + \cdots , \nonumber \\
& \tilde C_x^{\rm pn}(u,\omega,q) = -\tilde \omega \tilde q \left\{\frac{1}{4}(\pi^2 - 2\log^22) + \log u \log\frac{1+u}{1-u} + \frac{1}{2} \log(1+u) \log\frac{4}{1+u} \right. \nonumber \\
& \qquad \qquad \qquad \qquad \quad \left. - {\rm Li}_2\left(\frac{1-u}{2}\right) + {\rm Li}_2(-u) - {\rm Li}_2(u) \right\} + \cdots.
\end{align}
Near the AdS boundary,
\begin{align}
&\tilde C_t^{\rm pn(0)}(\omega,q)= 2 +2\tilde q^2 \log2 + \cdots,\nonumber \\
&\tilde C_t^{\rm pn(2)}(\omega,q)= r_h^2\left[-2 + 2\tilde q^2 (\log2-1) + 2\tilde q^2 \log(r_h^2/L^2)\right] + \cdots, \nonumber \\
&\tilde C_x^{\rm pn(0)}(\omega,q)= - \frac{1}{6}\pi^2 \tilde \omega \tilde q + \cdots, \nonumber \\
&\tilde C_x^{\rm pn(2)}(\omega,q)= r_h^2\left[- 2\tilde \omega \tilde q (\log2-1) - 2\tilde \omega \tilde q \log(r_h^2/L^2) \right] + \cdots.
\end{align}

Finally, we are ready to compute the TCFs in \eqref{Leff3}, based on the hydrodynamic expansion for the bulk fields. From the solution in the transverse sector, we obtain
\begin{align}
&w_8= -r_h - \frac{1}{2} i\omega \left[\log2 + \log(r_h^2/L^2)\right] + \cdots, \nonumber \\
&w_9 =\frac{1}{2} \log( r_h^2/L^2) + \cdots, \nonumber \\
&w_2= \frac{2r_h^2}{\pi} + \frac{\pi}{4}\omega^2 - \frac{\pi}{8}q^2 + \cdots.
\end{align}
From the solutions in the longitudinal sector,
\begin{align}
&w_1=0 + \mathcal{O}(\lambda^3),\nonumber \\
&w_3=\frac{\pi}{8} + \cdots, \nonumber \\
&w_4=- \frac{\pi}{24}i\omega + \cdots, \nonumber \\
&w_5= 2r_h^2 - \frac{1}{2} q^2 \left[2\log2 + \log(r_h^2/L^2)\right] + \cdots, \nonumber \\
&w_6=- \frac{\log(2r_h^2/L^2)}{2} + \cdots, \nonumber \\
&w_7= -\frac{1}{2}i\omega \log(2r_h^2/L^2) + \cdots.
\end{align}
The TCFs of \eqref{Jri} as well as the noise-noise correlator $G_0$ are expanded as
\begin{align}
&\mathcal D=\frac{1}{2r_h}+ \mathcal{O}(\lambda^2), \qquad  \sigma_e= r_h + \frac{1}{2}i\omega \log \frac{2r_h^2}{L^2} + \mathcal{O}(\lambda^2),
\qquad  \sigma_m= \frac{1}{2} \log\frac{r_h^2}{L^2} + \mathcal{O}(\lambda^1), \nonumber \\
&\Xi =   \frac{ 2i r_h^2}{\pi}+ \mathcal{O}(\lambda^2), \qquad  G_0 =  -\frac{ 2i r_h^2}{\pi}q^2+ \mathcal{O}(\lambda^4),
\end{align}
where $\lambda \sim \partial_\mu$ is the bookkeeping parameter for the derivative expansion. Notice that the relaxation time (order $\omega$  term) for the diffusion TCF vanishes.
This is not in agreement with the results of \cite{Bu:2015ame}. We postpone the comparison  with \cite{Bu:2015ame} to the end of this section.

We now compare our  analytical results in the hydrodynamic limit  with those of \cite{Glorioso:2018mmw,deBoer:2018qqm}. It is important to keep in mind
that the minimal subtraction counter-term $S_{\rm c.t.}$ introduced in the present work as well as in  \cite{deBoer:2018qqm} differs from the one used in \cite{Glorioso:2018mmw}.
Ref. \cite{deBoer:2018qqm} claimed agreement with  \cite{Glorioso:2018mmw} on  the values of the transport coefficients. Yet,  after careful examination of the results of both papers
and  taking into account differences originating from the different counter-terms, we fail to see a complete agreement. Below, we detail on the  comparison.

{\bf Comparison with \cite{deBoer:2018qqm}}.  Ref. \cite{deBoer:2018qqm} focused on the  longitudinal sector only. Hence, it does  not have any results on $w_9$, neither
separately on $w_2$ and $w_3$ (only the combination $w_2 + q^2 w_3$ was determined).
To ease the comparison,  the results of \cite{deBoer:2018qqm} are summarised below.
\begin{align}
&w_1^{\rm dBHPF} =0 + \mathcal{O}(\lambda^3), \nonumber \\
&w_2^{\rm dBHPF} + q^2 w_3^{\rm dBHPF} = \frac{2\pi^2 T^2 L}{g_A^2} \frac{1}{\pi} + \mathcal{O}(\lambda^3), \nonumber \\
&w_4^{\rm dBHPF} = 0 + \mathcal{O}(\lambda^2),\nonumber \\
&w_5^{\rm dBHPF} = \frac{2\pi^2 T^2 L}{g_A^2} - \frac{2\pi^2 T^2 L}{g_A^2} \frac{\log2}{2\pi^2 T^2}q^2 + \mathcal{O}(\lambda^3), \nonumber \\
&w_6^{\rm dBHPF} = - \frac{2\pi^2 T^2 L}{g_A^2} \frac{\log2}{4\pi^2 T^2} + \mathcal{O}(\lambda^1), \nonumber \\
&w_7^{\rm dBHPF}=  - \frac{2\pi^2 T^2 L}{g_A^2} \frac{\log2}{4\pi^2 T^2} i\omega + \mathcal{O}(\lambda^2), \nonumber \\
&w_8^{\rm dBHPF}= - \frac{2\pi^2 T^2 L}{g_A^2} \frac{1}{2\pi T} - \frac{2\pi^2 T^2 L}{g_A^2} \frac{\log2}{4\pi^2T^2}i\omega + \mathcal{O}(\lambda^2),
\end{align}
where $g_A$ is the gauge coupling constant which has been set to one in our work.
The  $\log (r_h/L)$ terms do not appear in \cite{deBoer:2018qqm}, because  they have been set to zero.

Our results are largely consistent with those of \cite{deBoer:2018qqm} except

\noindent $\bullet$ The $\omega^2$-term in $w_2+ q^2 w_3$;

\noindent $\bullet$ The $\omega$-term in $w_4$.

\noindent The differences  can be  attributed to the lack  of the third order accuracy in the
ingoing solutions $\{\tilde C_t, \tilde C_x\}$ in \cite{deBoer:2018qqm},
which is necessary for correct determination of $w_2+ q^2 w_3$ (up to second order) and $w_4$ (up to first order)\footnote{\cite{deBoer:2018qqm} adopted a derivative counting scheme,
in which $B_{r\mu} \sim \mathcal{O}(\partial^0)$ while $B_{a\mu} \sim \mathcal{O}(\partial^1)$.
Consequently, the terms under discussion appear as of higher order  and hence cannot be extracted from the effective action truncated at the second order.
}.

{\bf Comparison with \cite{Glorioso:2018mmw}}.
We quote the results of \cite{Glorioso:2018mmw}:
\begin{align}
&w_1^{\rm GCL}= 0 + \mathcal{O}(\lambda^3), \nonumber \\
&w_2^{\rm GCL}= \frac{2\pi}{\beta^2} + \frac{\pi}{4}\omega^2- \frac{\pi}{8}q^2 + \mathcal{O}(\lambda^3), \nonumber \\
&w_3^{\rm GCL}= \frac{\pi}{8} + \mathcal{O}(\lambda^1), \nonumber \\
&w_4^{\rm GCL}= - \frac{2\pi}{\beta} - \frac{48G+7\pi^2}{96\pi} i\omega + \mathcal{O}(\lambda^2), \nonumber \\
&w_5^{\rm GCL} = \frac{2\pi^2}{\beta^2} + \left(\frac{1}{2}- \log2\right)q^2 + \mathcal{O}(\lambda^3), \nonumber \\
&w_6^{\rm GCL}= \frac{1-\log2}{2} - \frac{\pi}{4} + \mathcal{O}(\lambda^1), \nonumber \\
&w_7^{\rm GCL}= \left(\frac{1-\log2}{2} + \frac{\pi}{4}\right)i\omega + \mathcal{O}(\lambda^2), \nonumber \\
&w_8^{\rm GCL}= -\frac{\pi}{\beta} + \frac{1-\log2}{2} i\omega + \mathcal{O}(\lambda^2), \nonumber \\
& w_9^{\rm GCL}= -\frac{1}{2} + \mathcal{O}(\lambda^1). \label{GCL_results}
\end{align}
In order to compare the results, we have to account for the difference between the subtraction terms used here and in \cite{Glorioso:2018mmw}.
In order to represent the results of \cite{Glorioso:2018mmw} within the minimal subtraction scheme, some $w_i^{\rm GCL}$'s in \eqref{GCL_results} have to be shifted:
\begin{align}
&w_5^{\rm GCL}\to w_5^{\rm GCL} +q^2 \left[\log (r_h/L) + 1/2\right], \qquad w_6^{\rm GCL} \to w_6^{\rm GCL} + \log (r_h/L) + 1/2, \nonumber \\
&w_7^{\rm GCL} \to w_7^{\rm GCL} + i\omega \left[\log (r_h/L) + 1/2\right], \qquad w_8^{\rm GCL} \to w_8^{\rm GCL} + i\omega \left[\log (r_h/L) + 1/2\right], \nonumber \\
&w_9^{\rm GCL} \to w_9^{\rm GCL} - \log (r_h/L) - 1/2, \qquad \qquad  {\rm others~are~not~changed.}
\end{align}
With the differences in the counter-terms taken into account, our results agree with those of \cite{Glorioso:2018mmw} except

\noindent $\bullet$  $w_4$ is completely different. The relevant result in  \cite{Glorioso:2018mmw} appears to be
 also in disagreement with \cite{deBoer:2018qqm}.
We also notice that $w_4^{\rm GCL}$   does not seem to satisfy the KMS condition \eqref{w4_w7+w4_w6}.

\noindent $\bullet$ $w_6$ and $w_7$. Both $w_6^{\rm GCL},\,w_7^{\rm GCL}$
do not obey the KMS condition \eqref{w6_w7}.

\noindent We have attempted to trace the origin of these differences.
First of all, our approach to solving the bulk dynamics is quite different from that of \cite{Glorioso:2018mmw}. Particularly,
the horizon limit ($\epsilon \to 0$) in our formalism is always taken before the hydrodynamic limit
($\partial_\mu \to 0$).  This is
in contrast to  \cite{Glorioso:2018mmw}, which performs the hydrodynamic expansion at the level of the dynamical EOMs.
These two limits do not always commute. Particularly, it is important to keep the oscillating factors like $e^{i\omega \zeta_1(r)}$ unexpanded.

\subsection{Numerical  results  at finite  $\omega$ and $q$} \label{num_results}

Except for the couple of  special cases of vanishing three-momentum ($q=0$) and light-like momenta ($\omega=q$), solutions of the ODEs (\ref{eom_Cv_EF}), (\ref{eom_Ci_EF})
at finite frequency and momentum are not known analytically.  Therefore, in order to provide
complete information about the TCFs,  we resort to numerical technique.
As has been extensively explained above, we have to solve the dynamical EOMs in a single Schwarzschild-AdS  for
the ingoing modes and also for the polynomial one, in  the longitudinal sector.
We solve the equations in the Schwarzschild coordinates and then transform to EF coordinates.  Once the solutions are found,  we first  numerically extract the coefficients of the near boundary expansion and then compute $w_i$ and other TCFs according to \eqref{w8_w9}, \eqref{w2}, \eqref{w1}-\eqref{w8} and \eqref{transports}.

In the bulk model there are in principle two independent length parameters: $r_h$ and the AdS radius $L$. In the metric \eqref{metric_EF}
$L=1$.  For the numerical results to be presented next, we also set $r_h=1$.  There are two consequences of this choice. First, the results  do not
include the logarithmic branch proportional to $\log(r_h/L)$, though it is not difficult to recover it analytically.  Second, $r_h$ becomes a unit of length
 for all dimension-full quantities  such as frequency and momentum. Thus all the results below will be shown for dimensionless $\omega\rightarrow \omega/r_h$
and $q\rightarrow q/r_h$, while we stick to the same notations to avoid introducing new ones.

While the effective Lagrangian \eqref{Leff3} is parameterised by nine TCFs, the discrete symmetries induce relations (\ref{w1_w5})-(\ref{w4_w7+w4_w6}) which
leave only four of them independent.
Specifically, we have chosen to take $w_5$, $w_7$, $w_8$ and $w_9$ as independent for which
the numerical results are presented in  Appendix \ref{numTCFs}.

The TCFs that have physical interpretation and hence are more interesting are the ones parameterising the constitutive relation for the physical current \eqref{Jri}. Those are the diffusion TCF $\mathcal{D}$, the electric conductivity $\sigma_e$,
the magnetic conductivity $\sigma_m$, and the thermal force $\Xi$,
whose expressions in terms of  $w_i$ are given in \eqref{transports} and \eqref{Xi}.

\begin{figure}
    \centering
    \begin{subfigure}[h]{0.49\textwidth}
        \includegraphics[width=\textwidth]{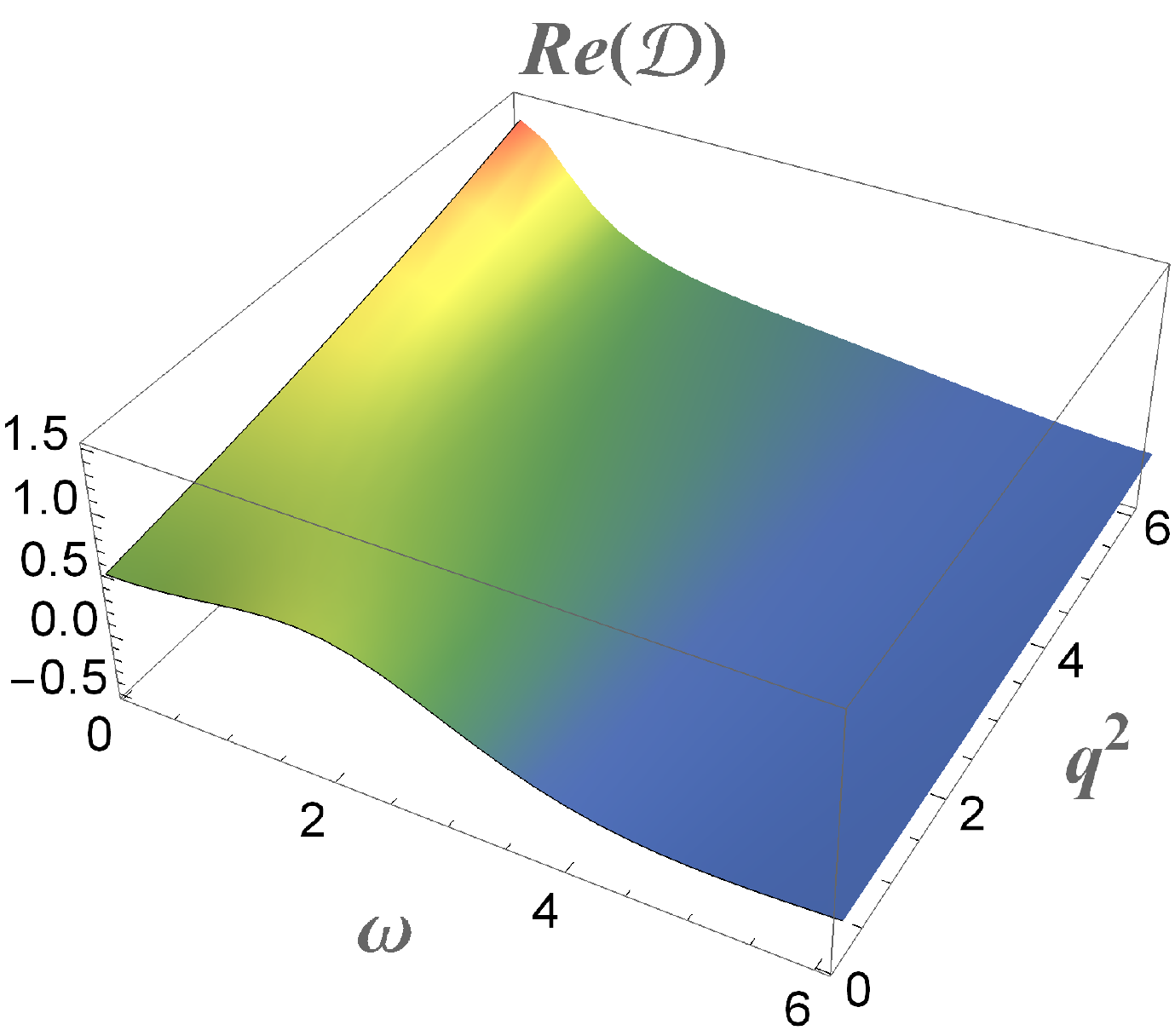}
        \caption{}
        \label{}
    \end{subfigure}
      \begin{subfigure}[h]{0.49\textwidth}
        \includegraphics[width=\textwidth]{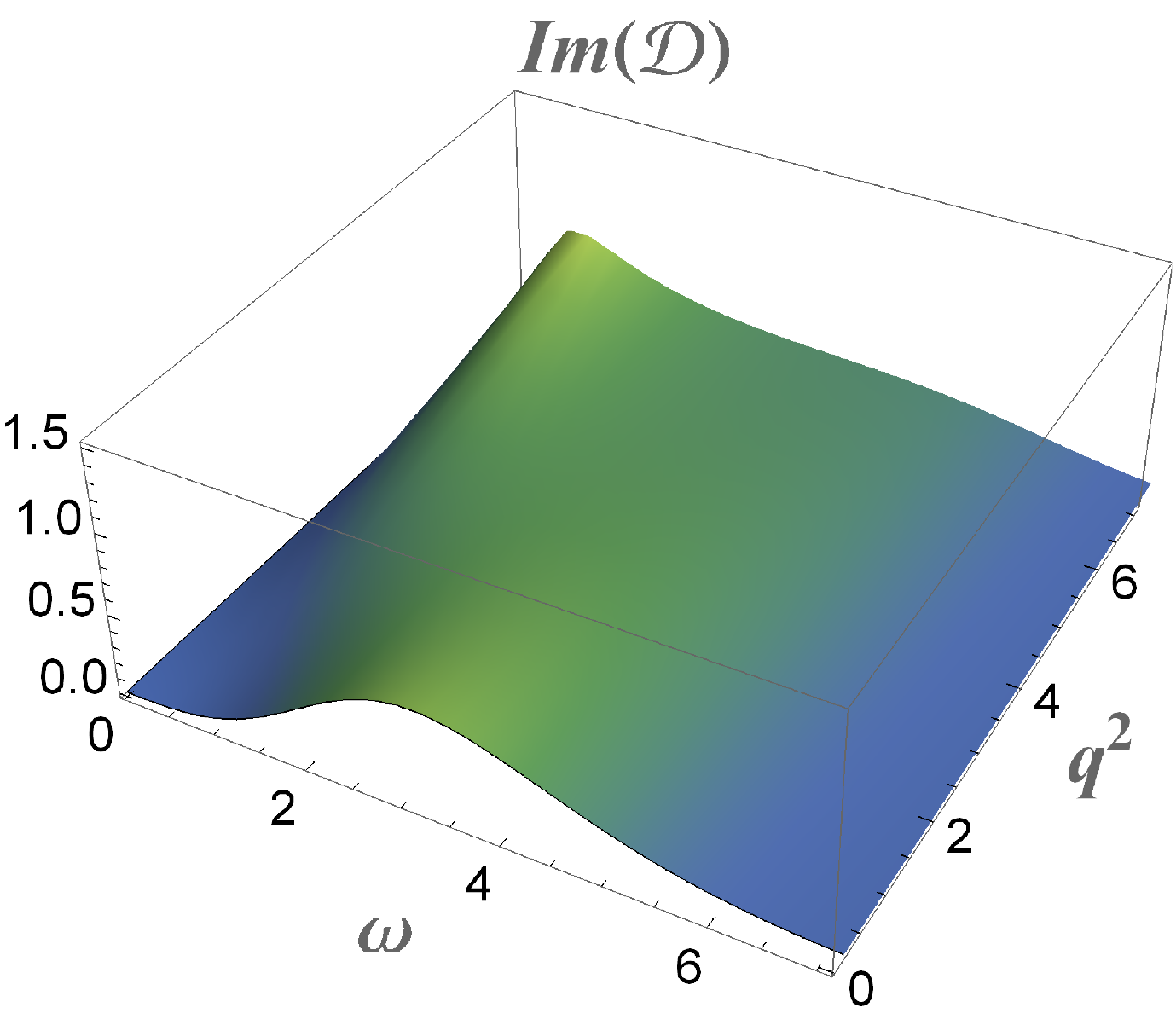}
        \caption{}
        \label{}
    \end{subfigure}

    \caption{The $\omega$-, $q$-dependence of (a) $Re(\mathcal{D}(\omega,q^2))$, (b) $Im(\mathcal{D}(\omega,q^2))$.}\label{Dp}
\end{figure}

\begin{figure}
    \centering
    \begin{subfigure}[h]{0.49\textwidth}
        \includegraphics[width=\textwidth]{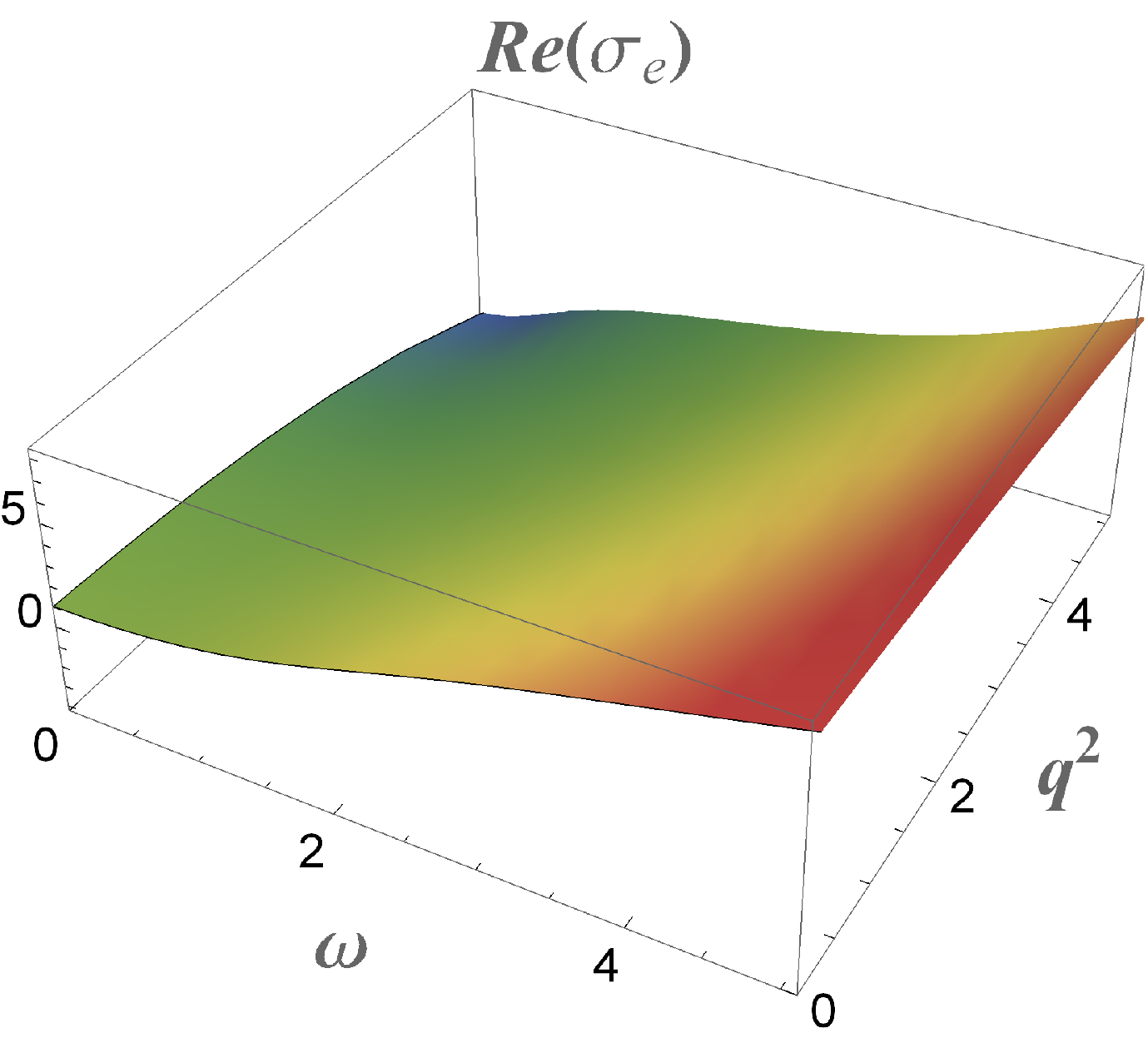}
        \caption{}
        \label{}
    \end{subfigure}
      \begin{subfigure}[h]{0.49\textwidth}
        \includegraphics[width=\textwidth]{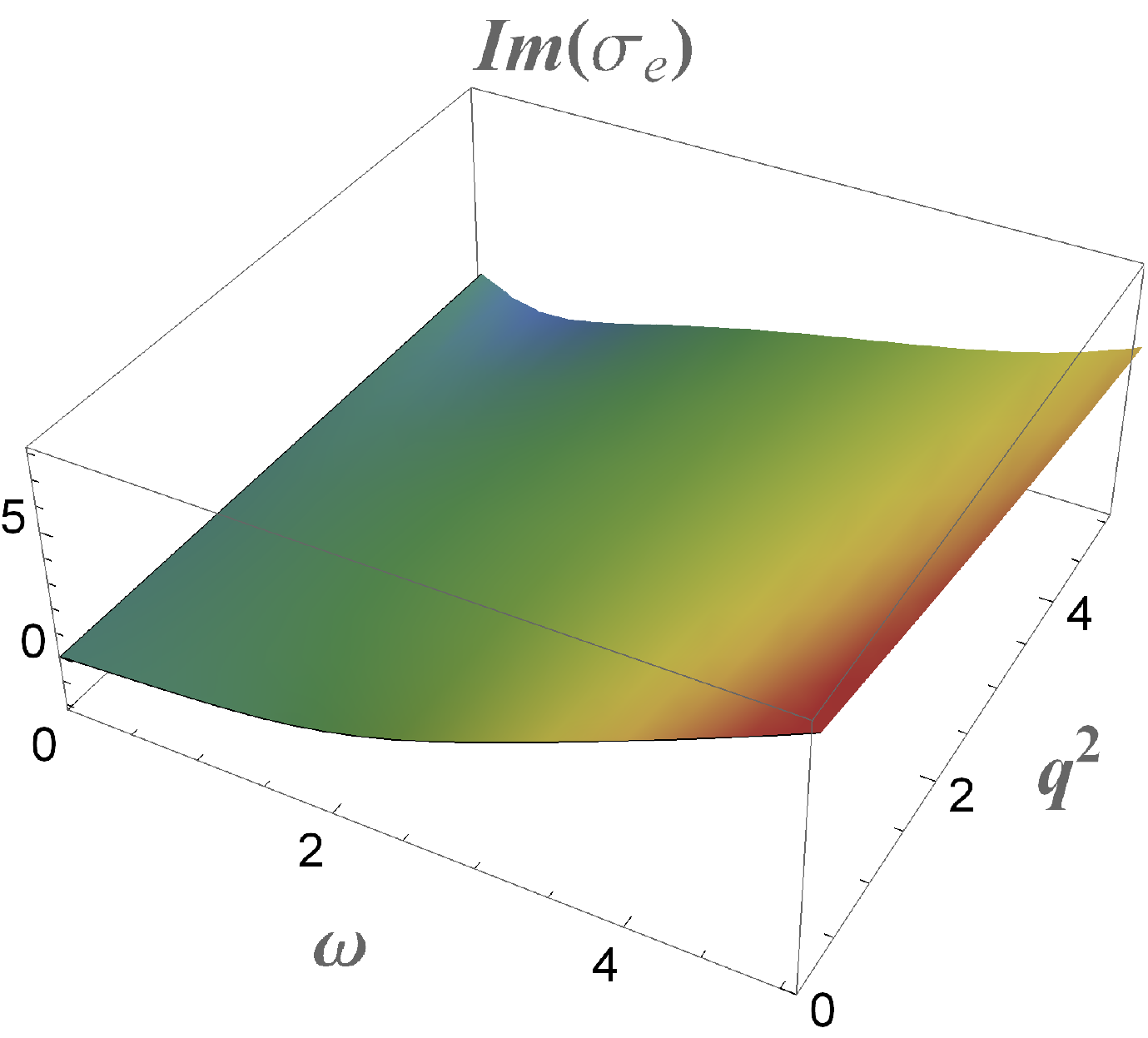}
        \caption{}
        \label{}
    \end{subfigure}

    \caption{Ther $\omega$-, $q$-dependence of (a) $Re(\sigma_{e}(\omega,q^2))$, (b) $Im(\sigma_e(\omega,q^2))$.}\label{Sep}
\end{figure}

\begin{figure}
    \centering
    \begin{subfigure}[h]{0.49\textwidth}
        \includegraphics[width=\textwidth]{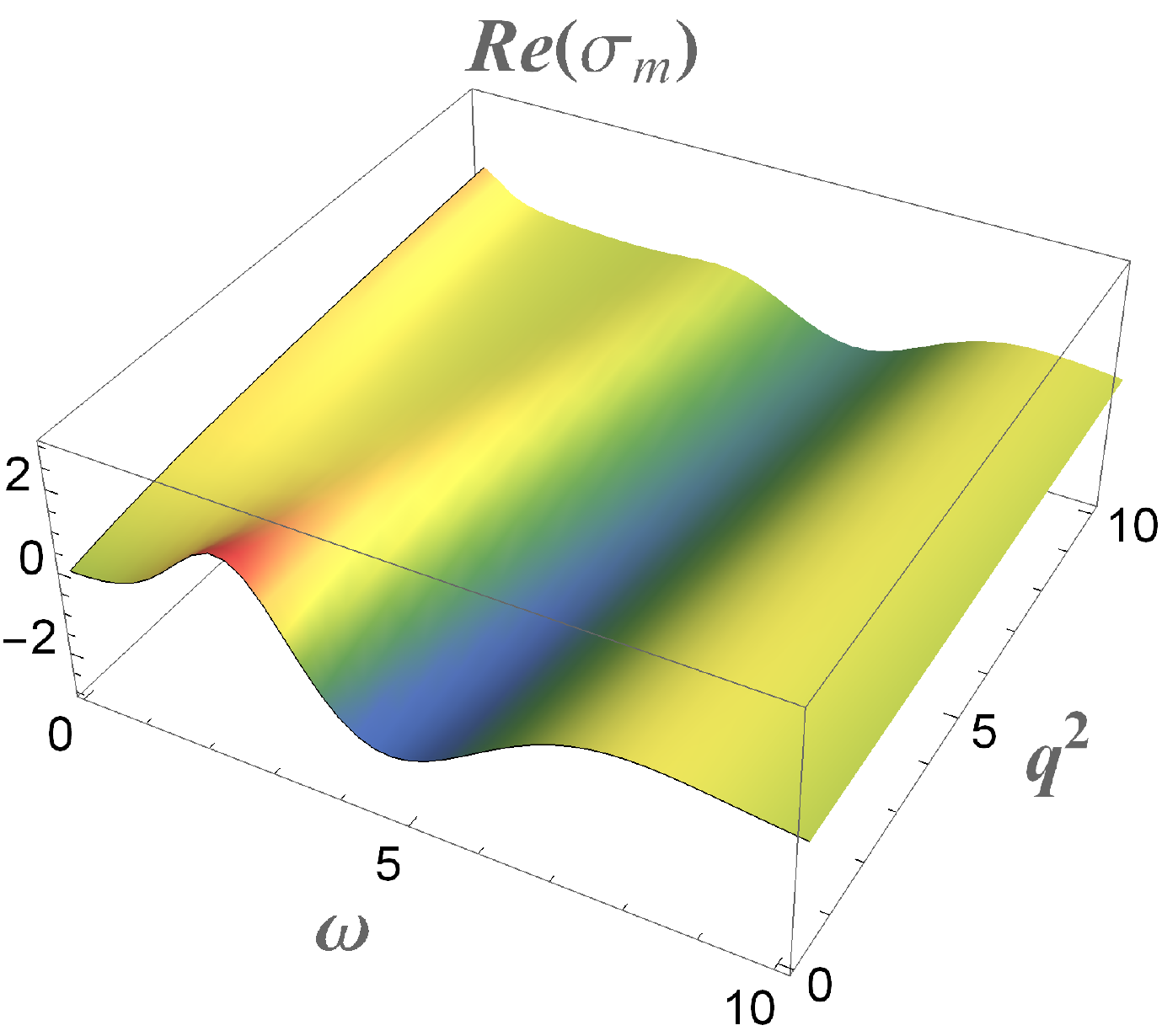}
        \caption{}
        \label{}
    \end{subfigure}
      \begin{subfigure}[h]{0.49\textwidth}
        \includegraphics[width=\textwidth]{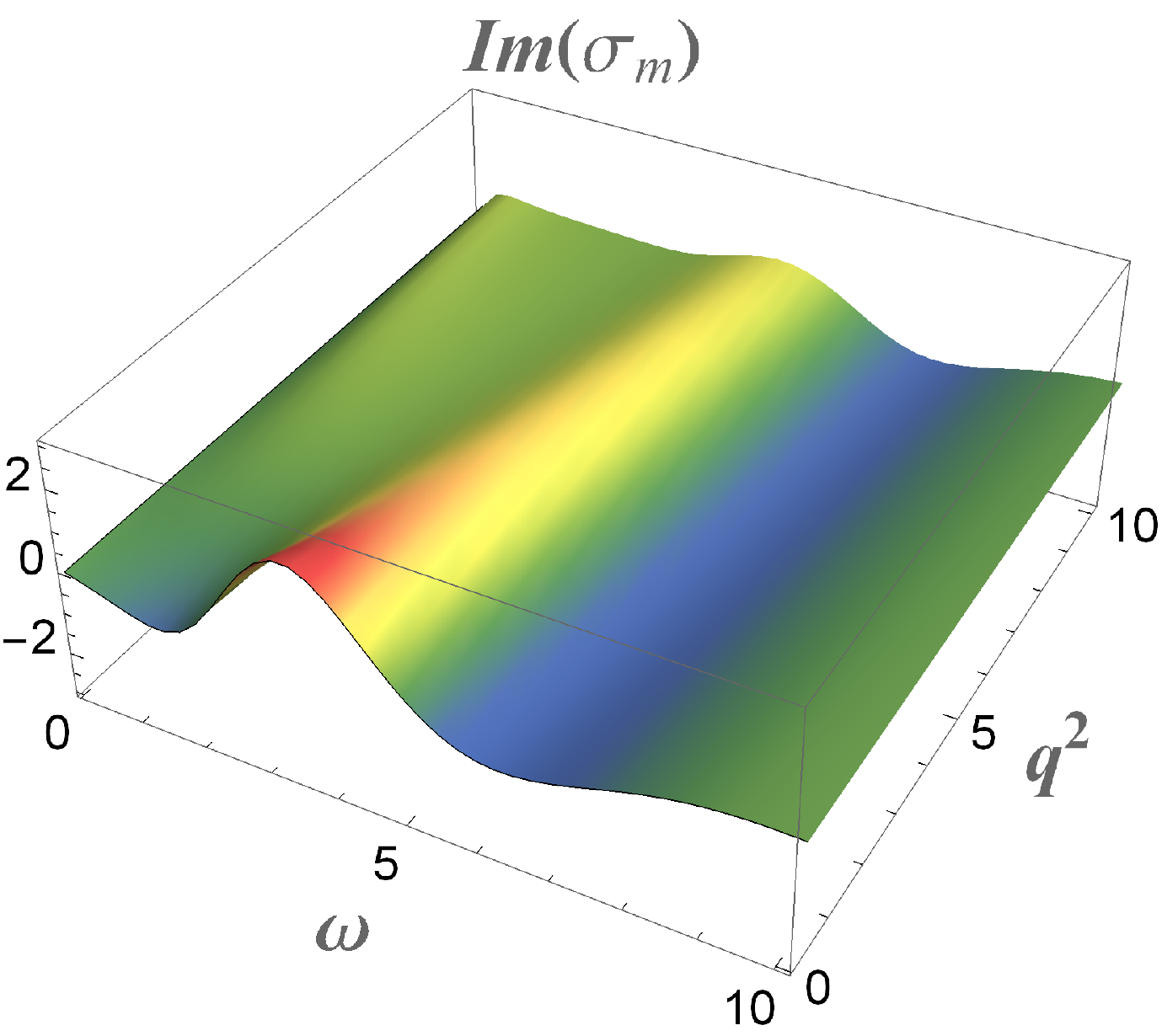}
        \caption{}
        \label{}
    \end{subfigure}

    \caption{The $\omega$-, $q$-dependence of (a) $Re(\sigma_{m}(\omega,q^2))$, (b) $Im(\sigma_m(\omega,q^2))$.}\label{Smp}
\end{figure}

\begin{figure}
    \centering
    \begin{subfigure}[h]{0.49\textwidth}
        \includegraphics[width=\textwidth]{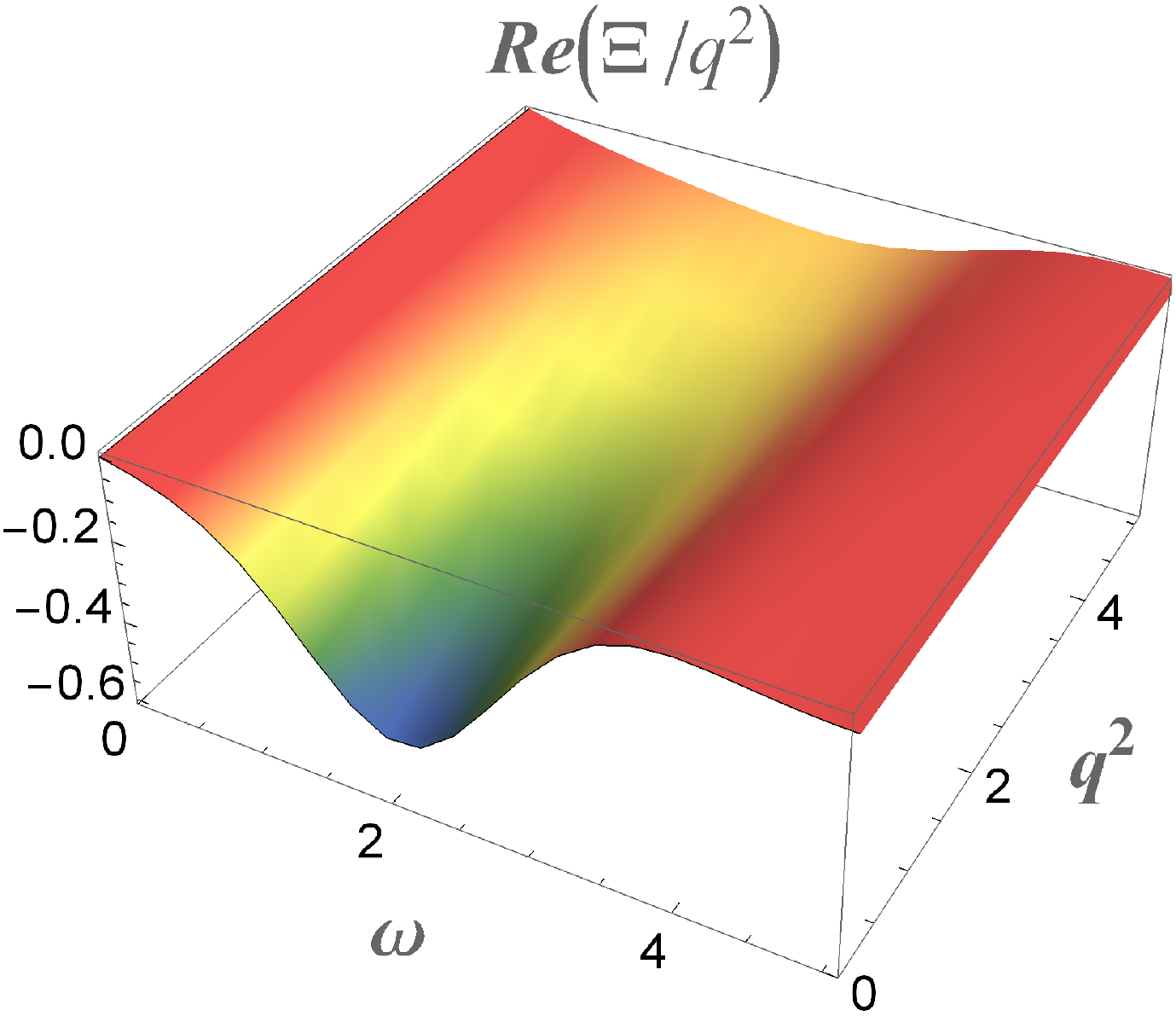}
        \caption{}

    \end{subfigure}
      \begin{subfigure}[h]{0.49\textwidth}
        \includegraphics[width=\textwidth]{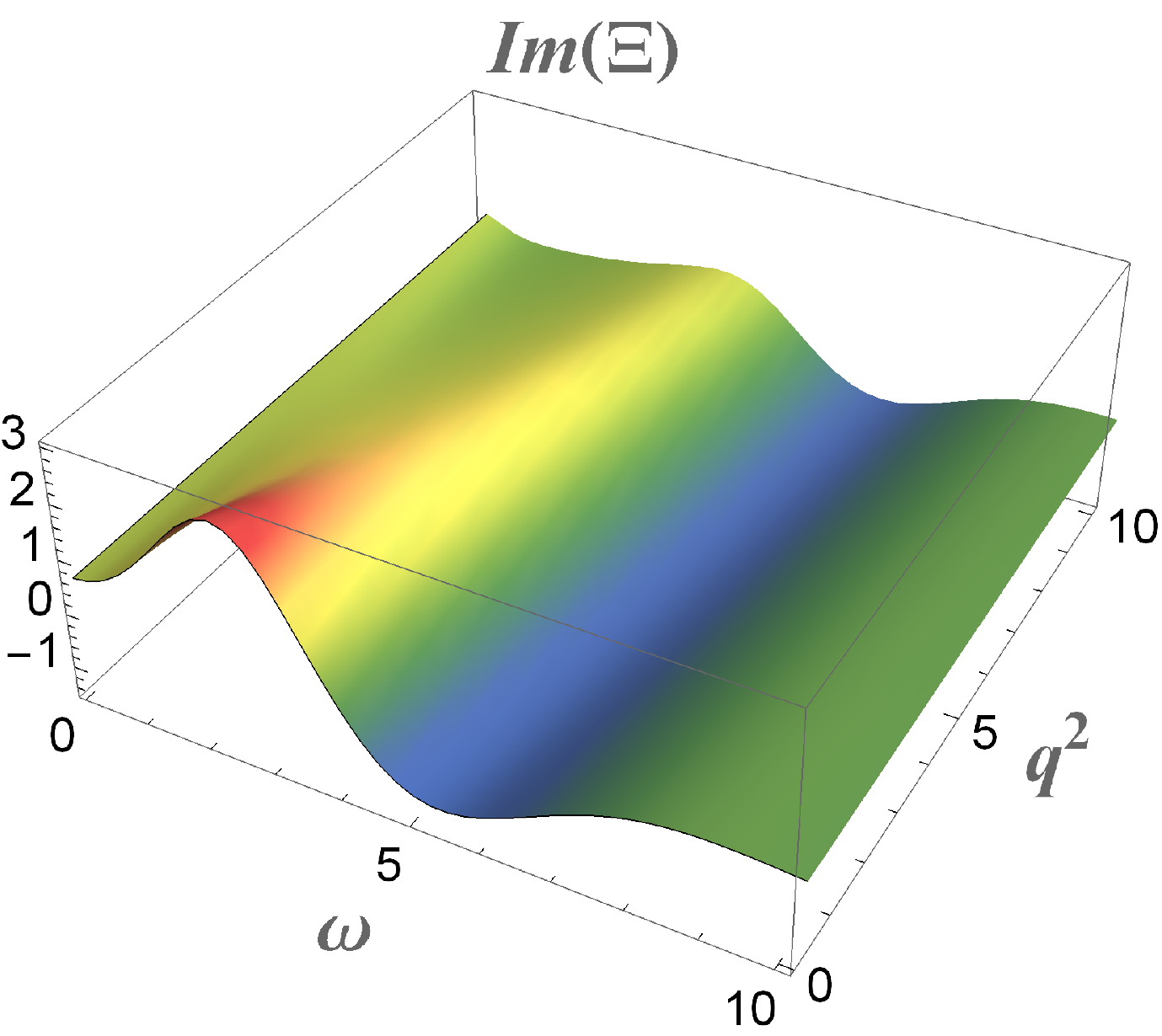}
        \caption{}

    \end{subfigure}
    \caption{The $\omega$-, $q$-dependence of the coefficients (a) $Re(\Xi(\omega,q^2))$  (b)  $Im(\Xi(\omega,q^2))$}
    \label{riXiwq}
\end{figure}

The results are summarized in Figures \ref{Dp}, \ref{Sep}, \ref{Smp} and \ref{riXiwq}.
At vanishing  frequency and momentum both the diffusion constant and  electric conductivity are well known: $\mathcal{D}[\omega=q=0]=1/2$ \cite{Policastro:2001yc} and $\sigma_e[\omega=q=0]=1$ \cite{CaronHuot:2006te}. When $q=0$, $\sigma_e$ is known analytically \cite{Horowitz:2008bn,Myers:2007we}. Our numerical results are fully consistent with all known analytical results.

Beyond the hydrodynamic limit, both the diffusion TCF $\mathcal{D}$, the magnetic conductivity $\sigma_m$, and $\Xi$ vanish at large frequencies $\omega$.
In contrast, $\sigma_e$ is monotonically increasing function of $\omega$; particularly $Re(\sigma_e)\sim \omega$ asymptotically. As functions of the three-momentum $q$,  we mostly notice a very mild dependence reflecting quasi-locality. $Re(\Xi)$ scales with $q^2$ at not too large momentum, hence we plot  $Re(\Xi/q^2)$ in Figure \ref{riXiwq}.

The TCFs $\mathcal{D}$, $\sigma_e$, $\sigma_m$ have been originally computed in \cite{Bu:2015ame}, using the off-shell holography in a single Schwarzschild-AdS$_5$ geometry. Compared with the results of \cite{Bu:2015ame}, the present results for the TCFs have different profiles. In appendix \ref{discrepancy},
we briefly review  the formalism of \cite{Bu:2015ame} compared with the present one. The discrepancy is
entirely within the longitudinal sector when the currents are taken off-shell.
Despite the disagreement in the TCFs, the current-current retarded correlators $G^{\mu\nu}_R$ \eqref{G_R} are all found to coincide. The equivalence is proven analytically in Appendix \ref{Son_Starinets_prescription}. We have also cross-checked the result numerically.


\subsection*{The noise-noise correlator} \label{noise_plot}

\begin{figure}
    \centering
        \includegraphics[width=0.49\textwidth]{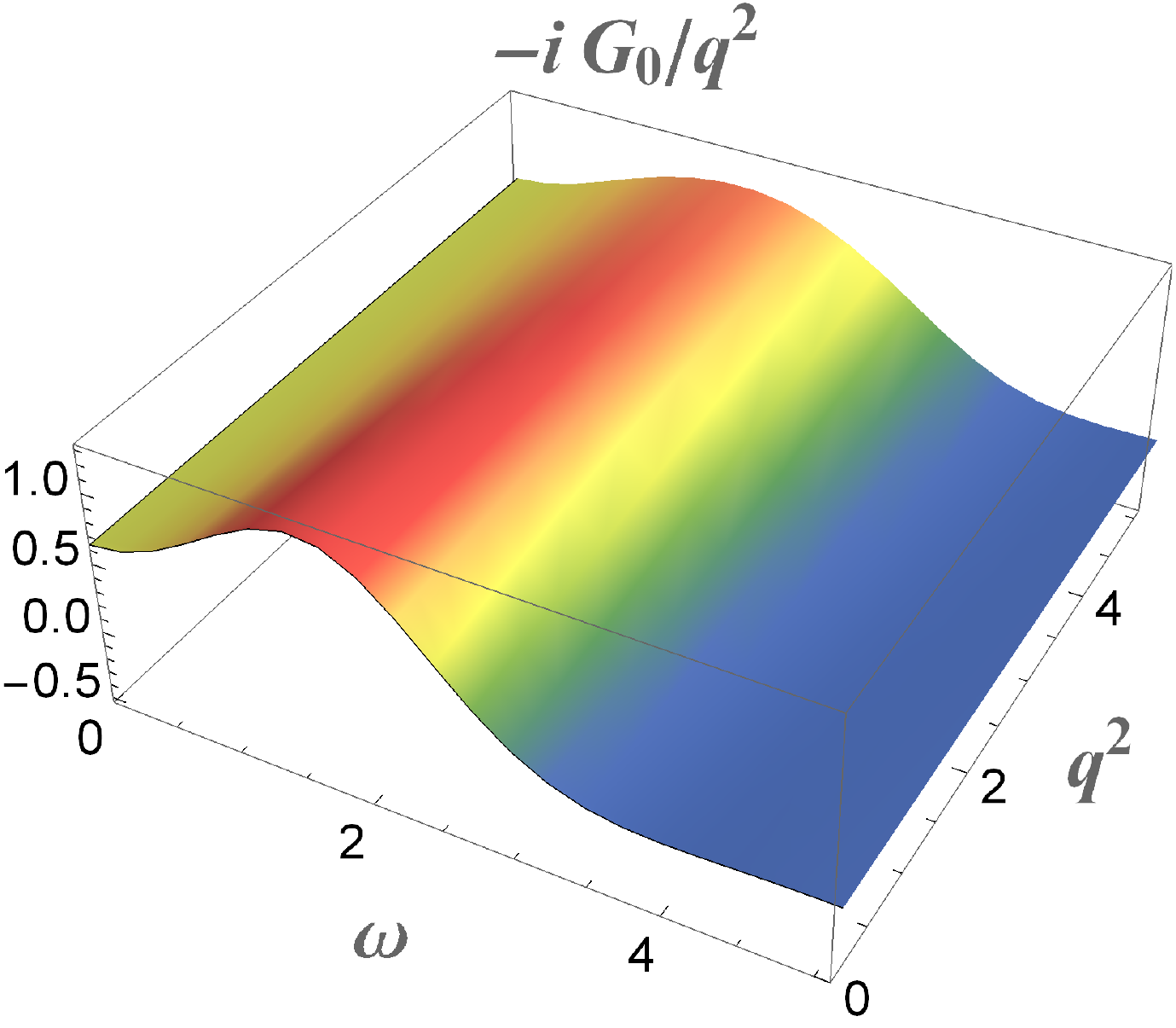}
    \caption{The $\omega$-, $q^2$-dependence of
         $-iG_0(\omega,q^2)/q^2$}
    \label{iG0wq}
\end{figure}

\begin{figure}
    \centering
    \begin{subfigure}[h]{0.49\textwidth}
        \includegraphics[width=\textwidth]{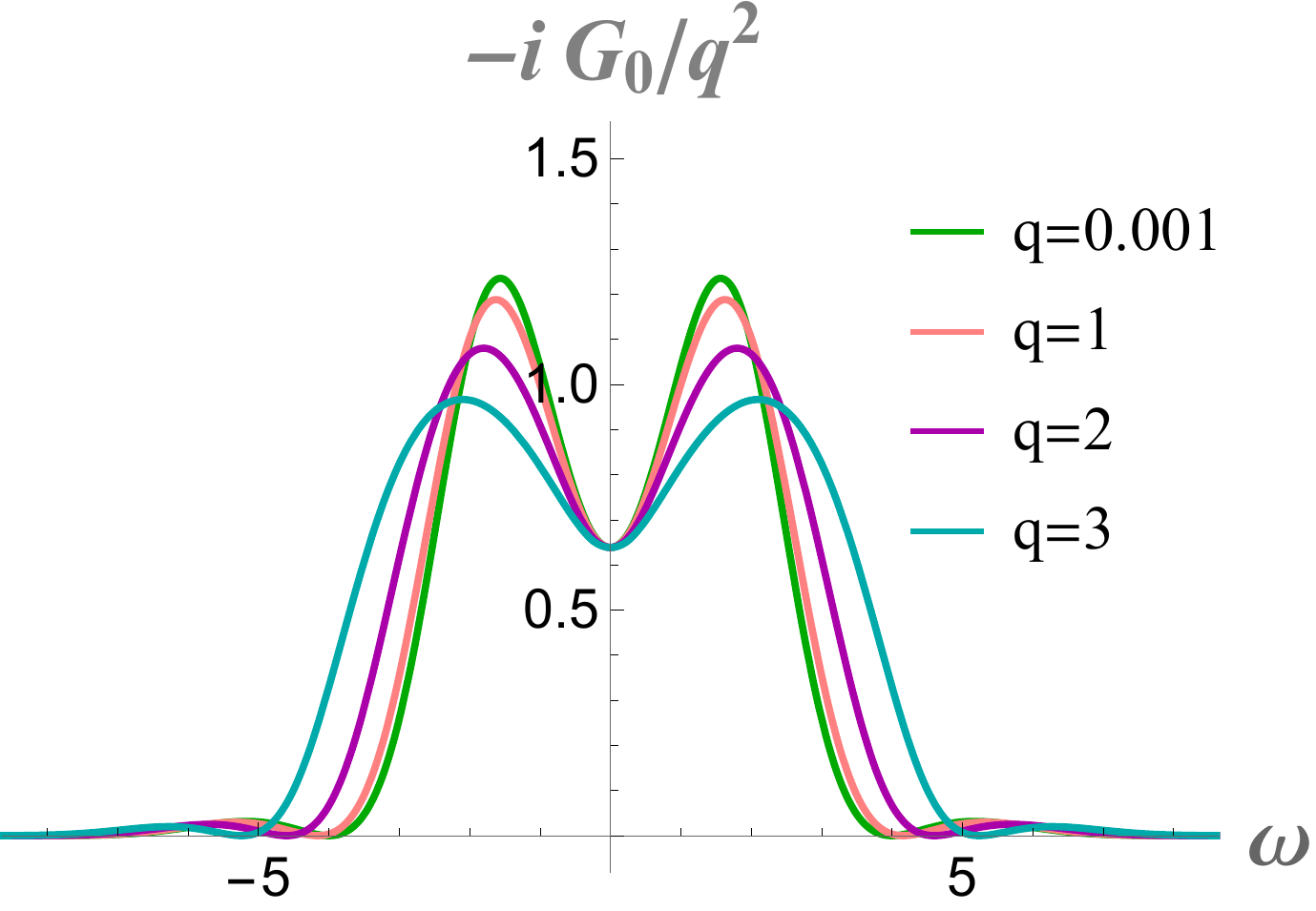}
        \caption{}
        \label{iG0va}

    \end{subfigure}
      \begin{subfigure}[h]{0.49\textwidth}
        \includegraphics[width=\textwidth]{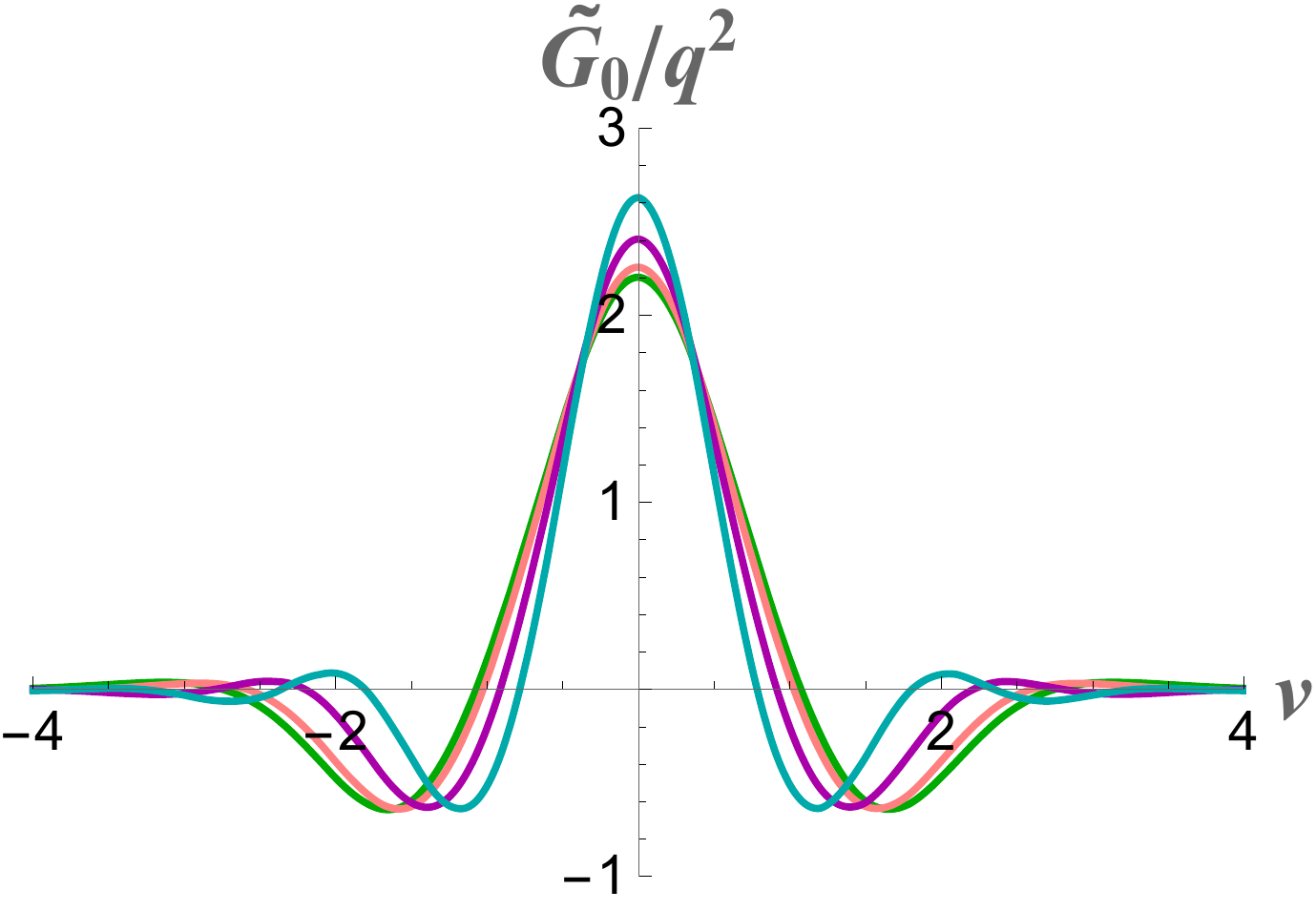}
        \caption{}
        \label{iG0vb}

    \end{subfigure}
    \caption{ (a) The $\omega$-dependence of $-iG_0(\omega)/q^2$ at fixed $q$-slices;
     (b) The inverse Fourier transform of $-iG_0(\omega)/q^2$ at fixed $q$-slices. }
    \label{IFTg0}
\end{figure}

Finally, the noise-noise correlator $-iG_0$ is displayed in Figure \ref{iG0wq} as a function of $\omega$ and $q^2$.
In Figure \ref{iG0va}, we plot the same function, but as a 2d slices at fixed representative values of $q^2$.
Since $-iG_0$ is proportional to $q^2$ up to sufficiently large momentum, we scale this dependence
out in Figures \ref{iG0wq} and \ref{iG0va}.  As is clear from the plots, $-iG_0$ initially
oscillates as function of $\omega$ but quickly vanishes at  large frequencies.

In order to better illustrate the coloured nature of the noise-noise correlator,  we perform an inverse
Fourier transform of $-iG_0$ with respect to the frequency, thus obtaining the time dependence of the correlator.
The analysis is performed for fixed values of  momentum $q$ and the results are displayed in Figure \ref{iG0vb}.
It might be interesting to additionally perform the inverse Fourier transform in the spatial momentum, so to obtain the full
space-time dependence of the correlator. Yet, this turns out to be numerically too expensive and we have
decided not to pursue this analysis.

%

\section{Summary and Outlook} \label{summary_outlook}

In this work we have further developed the off-shell SK holography.
While the core element of the formalism is the geometry proposed in   \cite{Glorioso:2018mmw}, our approach to solving the bulk EOMs is different from that of \cite{Glorioso:2018mmw}.  When discussing the hydrodynamic expansion, one has to be very careful with the non-commutativity of the hydrodynamic limit vs the near horizon limit.
Particularly, in order to reach a certain accuracy in the effective action, EOMs must be expanded to  higher orders in the momenta.
Our formalism completely avoids this subtlety since at no place it relies on the hydrodynamic expansion.

Starting from the off-shell SK holography, we have derived the effective action  \cite{Crossley:2015evo} for the charge diffusion and computed all order TCFs ($w_{1}...w_9$) parameterising it.
These TCFs display various types of behaviour as functions of momenta, without any clear pattern.
The evolution of $w_{1}...w_9$ from small momenta (the hydrodynamic limit) to very large momenta  can be thought as a flow of the effective action from IR to UV.
While we have not analysed it in any detail, the flow corresponds to {\it integrating in} of the heavy quasi-normal modes. It would be interesting to better understand this relation.
Putting the effective action on-shell,  we reproduced the prescription \cite{Son:2002sd} for the retarded two-point current-current correlators.

The constitutive relation \eqref{Jinoise} for the current follows from the off-shell effective action. It is parameterised by four TCFs: the diffusion, electric conductivity, magnetic  conductivity, and a thermal force. The latter is a new element emerging from the SK formalism.
The thermal force is responsible for fluctuations/noise in the current. Due to linearity of the Maxwell's theory in the bulk, the
noise is Gaussian, though coloured (non-local in space-time). We have demonstrated that by explicitly performing the inverse Fourier transform of the noise-noise correlator
to the real time.


Our new results for the diffusion TCF, electric and magnetic  conductivities are different from the ones reported earlier in \cite{Bu:2015ame}, even though they lead to identical two-point retarded correlators. The  disagreement originates from different holographic dictionaries used to define the off-shell currents. As discussed in Appendix \ref{discrepancy},  the off-shell formalism of \cite{Bu:2015ame} defines the current entirely in terms of the normalisable modes,  while  from the SK holography we now learn that there appear additional terms contributing to the current.

The Maxwell's theory in the bulk, even though much more complicated than a free scalar theory, provides a simple  Gaussian theory on the boundary. A much more challenging problem would be to consider stochastic neutral flows, which amounts to embedding the fluid/gravity correspondence \cite{Kovtun:2004de,Policastro:2001yc,Policastro:2002se,Policastro:2002tn,Son:2002sd,
Bhattacharyya:2008jc} (with dissipations only) into an EFT framework. This problem would require us to further develop the SK holography, particularly addressing
the questions of non-Gaussian noise, dynamical horizon, etc.

Another very interesting direction would be to learn about  hydrodynamic fluctuations associated with transport induced by chiral anomaly. While some relevant discussion could be found in \cite{Glorioso:2017lcn,Haehl:2013hoa,Iatrakis:2015fma,Lin:2018nxj,Liang:2020sgr}, the topic remains  largely unexplored.  From the  perspective of the SK holography, the problem could be addressed within the Maxwell-Chern-Simons theory in the bulk. Stochastic chiral hydrodynamics is expected to be
rich with many new phenomena.

\appendix

\section{The effective action from the basis decomposition} \label{general_structure}

In this appendix we demonstrate how the effective action \eqref{Leff3} is derived from \eqref{Leff1}.
Following the formalism of \cite{Bu:2015ame}, the bulk gauge fields $C_\mu$ can be linearly decomposed in terms of the basic tensor structures  built  from $B_{r\mu}$ and $B_{a\mu}$,
\begin{align}
&C_v= S_1 B_{rv}+ S_2 \partial_k B_{rk} + S_3 B_{av} + S_4 \partial_kB_{ak}, \nonumber\\
&C_i= V_1 B_{ri}+ V_2 \partial_i B_{rv} + V_3 \partial_i \partial_k B_{rk} + V_4 B_{ai} + V_5 \partial_i B_{av} + V_6 \partial_i \partial_k B_{ak}, \label{bdecomp}
\end{align}
where the decomposition coefficients (bulk-to-boundary propagators)
$S_i, V_i$ are $SO(3)$ scalar functionals of the spacetime derivative operators $\partial_\mu$, and functions of the radial coordinate $r$:
\begin{align}
S_i=S_i(r,\partial_v, \vec\partial),\qquad  V_i=V_i(r,\partial_v, \vec\partial).
\end{align}
The boundary conditions for $S_i,V_i$ are translated from those of $C_\mu$ \footnote{Our current prescription is
somewhat different from that of \cite{Bu:2015ame}. }:
\begin{align} \label{boundary_SV}
&r=\infty_1:\quad S_1=V_1=1, \quad S_3=V_4= \frac{1}{2}, \qquad {\rm others}=0, \nonumber \\
&r=\infty_2: \quad S_1=V_1=1, \quad S_3= V_4= -\frac{1}{2}, \quad \; {\rm others}=0.
\end{align}
In  Fourier space,  $(\partial_v,\vec\partial)\to (-i\omega, i\vec q)$,
these decomposition coefficients become
functions of the momenta,
\begin{align}
S_i(r,\partial_v,\vec\partial) \to S_i(r,\omega,q^2), \qquad V_i(r,\partial_v, \vec \partial) \to V_i(r,\omega,q^2).
\end{align}
With the help of \eqref{bdecomp} the original PDEs \eqref{dynamical_EOM} reduce to a system of linear ODEs.
Near $r=\infty_{1,2}$, the decomposition coefficients  $S_i, V_i$ can be expanded. Taking into account the boundary conditions \eqref{boundary_SV}, the expansion takes the form
\begin{align}
S_i(r\to \infty_1)= \cdots + \frac{s_{1i}}{r^2}+\cdots, \qquad  S_i(r\to \infty_2)= \cdots + \frac{s_{2i}}{r^2}+\cdots, \nonumber \\
V_i(r\to \infty_1)= \cdots + \frac{v_{1i}}{r^2}+\cdots, \qquad  V_i(r\to \infty_2)= \cdots + \frac{v_{2i}}{r^2}+\cdots.
\end{align}
where the coefficients $s_{1i}, v_{1i}$ and $s_{2i}, v_{2i}$ are respective normalizable modes.

Substituting the decomposition \eqref{bdecomp} into \eqref{Leff1}, the effective Lagrangian  in the $(r,a)$-basis reads
\begin{align} \label{Leff2}
\mathcal{L}_{\rm eff}=&-B_{rv} s_{a1} B_{rv} - B_{rv} s_{a2} \partial_k B_{rk} + B_{rk} v_{a1} B_{rk} + B_{rk} v_{a2} \partial_k B_{rv} + B_{rk} v_{a3} \partial_k \partial_l B_{rl} \nonumber \\
& +(- B_{rv} s_{a3} B_{av} - B_{av} s_{r1} B_{rv}) + (-B_{rv} s_{a4} \partial_k B_{ak} + B_{ak} v_{r2} \partial_k B_{rv}) \nonumber \\
& +(- B_{av} s_{r2} \partial_k B_{rk} + B_{rk} v_{a5} \partial_k B_{av}) - B_{av} s_{r3} B_{av} + (- B_{av} s_{r4} \partial_k B_{ak} + B_{ak} v_{r5} \partial_k B_{av}) \nonumber \\
&+(B_{rk} v_{a4} B_{ak} + B_{ak} v_{r1} B_{rk}) + (B_{rk} v_{a6} \partial_k \partial_l B_{al} + B_{ak} v_{r3} \partial_k \partial_l B_{rl}) + B_{ak} v_{r4} B_{ak} \nonumber \\
& + B_{ak} v_{r6} \partial_k \partial_l B_{al} + \frac{1}{2} \partial_k B_{ak} \partial_v B_{rv} - \frac{1}{2} B_{av} \partial_v \partial_k B_{rk}
+ \frac{1}{2}B_{av} {\vec\partial}^{\,2} B_{rv} - \frac{1}{2} B_{ak} \partial_v^2 B_{rk} \nonumber \\
& + \frac{1}{4} \mathcal{F}_{ajk} \mathcal{F}_{rjk} + B_{av}\partial_v^2 B_{rv}.
\end{align}
Here, the coefficients are represented in the $(r,a)$-basis:
\begin{align}
&s_{ai}= s_{1i}-s_{2i}, \qquad s_{ri}= \frac{1}{2}(s_{1i}+s_{2i}), \qquad i=1,2,3,4, \nonumber \\
&v_{ai}= v_{1i}- v_{2i}, \qquad v_{ri}= \frac{1}{2}(v_{1i}+v_{2i}), \qquad i=1,2,\cdots,6.
\end{align}
The first line of \eqref{Leff2} must vanish within the usual $(r,a)$ scheme.
Indeed within the holographic representation of the SK contour:
\begin{align} \label{check1}
s_{a1}=s_{a2}= v_{a1}= v_{a2}= v_{a3}=0.
\end{align}
To compare with  \cite{Crossley:2015evo},  \eqref{Leff2} has to be rewritten so that all the
$a$-fields are placed on the left, in all terms, say
\begin{align}
&\int d^4x B_{rv}(x) s_{a3}(\partial_t,\vec\partial)B_{av}(x)
=  \int d^4x B_{av}(x) s_{a3}(-\partial_v,-\vec\partial)B_{rv}(x).
\end{align}
Introducing $w_i= w_i(\partial_v, \vec\partial)$ as
\begin{align} \label{L_parameters}
&\frac{i}{2}w_1(\partial_v,\vec\partial)= -s_{r3}(\partial_v, \vec \partial), \nonumber \\ &\frac{i}{2}w_2(\partial_v, \vec\partial)= v_{r4}(\partial_v, \vec\partial), \nonumber \\
&\frac{i}{2} w_3(\partial_v, \vec\partial)= -v_{r6}(\partial_v, \vec\partial), \nonumber \\
& iw_4(\partial_v,\vec\partial)= -s_{r4}(\partial_v, \vec\partial)- v_{r5}(-\partial_v, -\vec\partial), \nonumber \\
& w_5(\partial_v, \vec\partial)= \partial_v^2 + \frac{1}{2} \vec\partial^2 - s_{r1}(\partial_v, \vec\partial) - s_{a3}(-\partial_v, -\vec\partial) \nonumber \\
& w_6(\partial_v, \vec\partial)\partial_v=- \frac{1}{2}-s_{r2}(\partial_v, \vec\partial) - v_{a5}(-\partial_v, -\vec\partial), \nonumber \\
& w_7(\partial_v,\vec\partial) = \frac{1}{2}\partial_v -s_{a4}(-\partial_v, -\vec\partial) -v_{r2}(\partial_v, \vec\partial) \nonumber \\
& w_8(\partial_v,\vec\partial) \partial_v = - \frac{1}{2}\partial_v^2 + v_{a4}(-\partial_v, -\vec\partial) + v_{r1}(\partial_v, \vec\partial) + \left[ v_{a6}(-\partial_v, -\vec\partial) + v_{r3}(\partial_v, \vec\partial) \right] {\vec\partial}^2, \nonumber \\
& w_9(\partial_v, \vec\partial)=\frac{1}{2}+ v_{a6}(-\partial_v, -\vec\partial) +v_{r3}(\partial_v, \vec\partial),
\end{align}
 the effective Lagrangian \eqref{Leff2} is cast into (\ref{Leff3}), consistently with \cite{Crossley:2015evo}.

\section{Validating   \eqref{glue_dCv}} \label{crosscheck_dCv_Cv}

In subsection \ref{match_cond_Cmu}, we derived the discontinuity relation \eqref{glue_dCv}.
Our goal here is to verify this condition by computing
 both left-hand side (LHS) and right-hand side (RHS) of \eqref{glue_dCv}, starting from the solutions constructed in the subsections \ref{sol_single_AdS} and \ref{sol_contour}.

The RHS of \eqref{glue_dCv} is proportional to $\nabla_M F^{Mr}$.  Recall that in general, $\nabla_M F^{Mr}$ has a very simple dependence on $r$, see
 \eqref{constraint_identity}. In \eqref{constraint_identity} the values of $\mathcal C^{\rm up}$ and $\mathcal C^{\rm dw}$ should be  determined from the solutions established in subsections \ref{sol_single_AdS} and \ref{sol_contour}. As has been emphasised towards the end of subsection \ref{sol_single_AdS},
among the four linearly independent solutions in the longitudinal sector, only the polynomial solution $\{C_v^{\rm pn}, C_x^{\rm pn}\}$ does not automatically satisfy the constraint equation \eqref{constraint_EF}.  Hence  $\mathcal C^{\rm up,\,dw}$ must be proportional to the coefficients $n_\parallel^{\rm up,\,dw}$ multiplying the polynomial solutions. More precisely,
by taking the near horizon limit of \eqref{constraint_identity}, we have
\begin{align}
\mathcal C^{\rm up}(k) = i\omega r_h^3 \tilde C_t^{{\rm pn}\, h} n_\parallel^{\rm up}, \qquad \qquad \mathcal C^{\rm dw}(k) = i\omega r_h^3 \tilde C_t^{{\rm pn}\, h} n_\parallel^{\rm dw}.
\end{align}
Thus the RHS of \eqref{glue_dCv} reads
\begin{align}
\lim_{\Delta \to 0}\int_{r_- - \Delta}^{r_+ + \Delta} dr \frac{\nabla_M F^{Mr}}{f(r)} & = \frac{\mathcal C^{\rm dw}(k)}{i\omega r_h^3} e^{i\omega \zeta_1(r_h- \epsilon)} - \frac{\mathcal C^{\rm up}(k)}{i\omega r_h^3} e^{i\omega \zeta_2(r_h- \epsilon)} \nonumber \\
& = \tilde C_t^{{\rm pn}\,h} \left[n_\parallel^{\rm dw} e^{i\omega \zeta_1(r_h -\epsilon)} - n_\parallel^{\rm up} e^{i\omega \zeta_2(r_h - \epsilon)} \right]. \label{drCv_RHS}
\end{align}

Based on the solutions found in \ref{sol_single_AdS} and \ref{sol_contour}, the discontinuity of $\partial_r C_v$ is
\begin{align}
\partial_r C_v (r_+) - \partial_r C_v(r_-) =& n_\parallel^{\rm up} \partial_r C_v^{\rm pn} \big|_{r=r_+} - n_\parallel^{\rm dw} \partial_r C_v^{\rm pn} \big|_{r=r_-} \nonumber \\
=& \frac{i\omega}{f(r)} n_\parallel^{\rm up} C_v^{\rm pn}(r) \bigg|_{r=r_+} - \frac{i\omega}{f(r)} n_\parallel^{\rm dw} C_v^{\rm pn}(r) \bigg|_{r=r_-} \nonumber \\
& + \tilde C_t^{{\rm pn}\, h} n_\parallel^{\rm up} e^{i\omega \zeta_2(r_h-\epsilon)} - \tilde C_t^{{\rm pn}\, h} n_\parallel^{\rm dw} e^{i\omega \zeta_1(r_h-\epsilon)}
\end{align}
While $C_v$ is continuous across the cutting slice, $C_r=-C_v/f(r)$ has a jump:
\begin{align}
\frac{C_v(r)}{f(r)}\bigg|_{r=r_+} - \frac{C_v(r)}{f(r)}\bigg|_{r=r_-} = \frac{n_\parallel^{\rm up}}{f(r)} C_v^{\rm pn}(r) \bigg|_{r=r_+} - \frac{n_\parallel^{\rm dw}}{f(r)} C_v^{\rm pn}(r) \bigg|_{r=r_-}.
\end{align}
So, the LHS of \eqref{glue_dCv} is computed as
\begin{align}
F^{rv}(r_+) - F^{rv} (r_-) & = \left[\frac{i\omega}{f(r)} - \partial_r \right] C_v \bigg|_{r=r_+} - \left[\frac{i\omega}{f(r)} - \partial_r \right] C_v \bigg|_{r=r_-} \nonumber \\
& = \tilde C_t^{{\rm pn}\, h} n_\parallel^{\rm dw} e^{i\omega \zeta_1(r_h-\epsilon)} -\tilde C_t^{{\rm pn}\, h} n_\parallel^{\rm up} e^{i\omega \zeta_2(r_h-\epsilon)}
\end{align}
which is exactly the same as  \eqref{drCv_RHS}.

\section{Son-Starinets prescription for retarded correlators revisited} \label{Son_Starinets_prescription}

At the early days of the fluid-gravity correspondence, Son and Starinets   \cite{Son:2002sd} proposed a prescription
for computing Minkowski-space retarded correlators. The prescription is formulated entirely within a single copy of the doubled BH-AdS,
and does not rely on SK holography. Yet, a proper derivation of the prescription from the SK holography
is missing, and  in this appendix we provide one. While there have been earlier works in this direction, particularly
\cite{Herzog:2002pc}, which considered SK matrix propagator for a scalar field starting from an eternal black hole in AdS space \cite{Maldacena:2001kr}, the SK geometry of \cite{Glorioso:2018mmw}  adopted here is different. Furthermore,
we are not aware of any derivation for the $U(1)$ field available in the literature.

The prescription of  \cite{Son:2002sd} relates the retarded correlators to the ingoing solution in a single BH AdS.
Starting from the SK holography, we can reproduce the result by
taking a few alternative paths.  First, the correlators could be
obtained from the off-shell effective Lagrangian \eqref{Leff3} by integrating out the dynamical  fields $\varphi_r$ and $\varphi_a$ \eqref{WAA}.
The result is the boundary
 generating functional $W[\mathcal A_{a\mu}, \mathcal A_{r\mu}]$ of the external fields only,
 form which the correlators could be read off straightforwardly.
 For the Lagrangian  quadratic in the dynamical fields, like \eqref{Leff3}, integrating out $\varphi_r$ and $\varphi_a$ could be done by imposing their classical EOMs.
On the bulk side, this corresponds to imposing the constraint equation. Putting the solutions \eqref{CvCx_up_dw1}
on-shell is equivalent to setting $n_\parallel^{\rm up}= n_\parallel^{\rm dw}=0$, which via \eqref{n-_sol} and \eqref{n+_sol} yields classical solutions for $\varphi_r$ and $\varphi_a$.

Alternatively, we could start with the constitutive relation \eqref{Jhydro}, and use the continuity equation, which leads
to  the retarded current-current correlators expressed in terms of the TCFs  \cite{Bu:2015ame}:
 \begin{align}
& G_R^{\perp \perp}= i\omega \sigma_e + q^2 \sigma_m, \qquad G_R^{vv}= \frac{q^2 \sigma_e}{-i\omega +q^2 \mathcal D}, \qquad G_R^{vx}= \frac{\omega q \sigma_e}{-i\omega +q^2 \mathcal D}, \nonumber \\
& G_R^{xx}= \frac{\omega^2 \sigma_e}{-i\omega +q^2 \mathcal D}. \label{G_R}
\end{align}
These expressions could be algebraically traced back to the ingoing solution in a single copy of BH-AdS.

Yet, we believe the most  illuminating derivation is to reconsider the problem from the very beginning,
starting within the on-shell SK holography, which offers a possibility to work directly with   gauge invariant fields
\begin{align}
E_\perp= \partial_\perp C_v- \partial_v C_\perp, \qquad \qquad E_x= \partial_x C_v - \partial_v C_x. \label{Ei_definition}
\end{align}
EOMs for the bulk electric fields $E_\perp$ and $E_x$ are
\begin{align}
&\partial_r\left[rf(r) \partial_r E_\perp\right] -2i\omega r \partial_r E_\perp -i\omega E_\perp -q^2 r^{-1} E_\perp =0, \nonumber \\
&\partial_r\left[ \frac{rf(r)}{\omega^2- r^{-2}f(r)q^2} \partial_r E_x \right] - \frac{2i\omega r}{\omega^2- r^{-2}f(r)q^2} \partial_r E_x + \partial_r \left[\frac{-i\omega r}{\omega^2- r^{-2}f(r)q^2} \right]E_x \nonumber \\
 &- \frac{r^{-1}q^2}{\omega^2- r^{-2}f(r) q^2} E_x=0. \label{Eperp_Ex_EOM_EF}
\end{align}
In the equation for $E_x$, there is a  singularity at $r=r_h(1-\omega^2/q^2)^{-1/4}$ for space-like momenta, which is however integrable \cite{CaronHuot:2006te}.

The on-shell bulk action \eqref{bulk_action_EF}  reads
\begin{align}
S_0=& -\frac{1}{2} \int \frac{d\omega dq}{(2\pi)^2} \left\{\frac{r}{i\omega} E_\perp(r,-k) E_\perp(r,k) + \frac{rf(r)}{\omega^2} E_\perp(r,-k) \partial_r E_\perp(r,k) \right. \nonumber \\
& \left. - \frac{i\omega r} {\omega^2-r^{-2}f(r)q^2} E_x(r,-k) E_x(r,k) + \frac{rf(r)}{\omega^2-r^{-2}f(r)q^2}E_x(r,-k) \partial_r E_x(r,k) \right\} \bigg|_{r=\infty_1} \nonumber \\
& +\frac{1}{2}\int \frac{d\omega dq}{(2\pi)^2} \left\{\frac{r}{i\omega} E_\perp(r,-k) E_\perp(r,k) + \frac{rf(r)}{\omega^2} E_\perp(r,-k) \partial_r E_\perp(r,k) \right. \nonumber \\
& \left. - \frac{i\omega r} {\omega^2-r^{-2}f(r)q^2} E_x(r,-k) E_x(r,k) + \frac{rf(r)}{\omega^2-r^{-2}f(r)q^2}E_x(r,-k) \partial_r E_x(r,k) \right\} \bigg|_{r=\infty_2}. \label{S0_Eperp+Ex}
\end{align}
Near each AdS boundary, $r\to \infty_s$ with $s=(1,2)$,
the bulk electric fields $E_\perp$ and $E_x$ behave as
\begin{align}
& E_{s,\perp}(r,k) \xrightarrow[]{r\to\infty_s} E_{s,\perp}^{(0)}(k) - \frac{i\omega} {r} E_{s,\perp}^{(0)}(k) + \frac{1}{2}(\omega^2-q^2) E_{s,\perp}^{(0)} (k) \frac{\log r}{r^2} + \frac{E_{s,\perp}^{(2)}(k)}{r^2}+ \cdots, \nonumber \\
& E_{s,x}(r,k) \xrightarrow[]{r\to\infty_s} E_{s,x}^{(0)}(k) - \frac{i\omega} {r} E_{s,x}^{(0)}(k) + \frac{1}{2}(\omega^2-q^2) E_{s,x}^{(0)} (k) \frac{\log r}{r^2} + \frac{E_{s,x}^{(2)}(k)}{r^2}+ \cdots.
\end{align}
Then, the generating functional of the boundary theory (i.e., the on-shell bulk action) becomes
\begin{align}
W[\mathcal A_{a\mu}, \mathcal A_{r\mu}]= S_0+ S_{\rm c.t.}= \int \frac{d\omega dq}{(2\pi)^2} \mathcal{L}_{\rm eff}^{\rm os}[\mathcal{F}_{alv}, \mathcal{F}_{rlv}],
\end{align}
where
\begin{align}
\mathcal{L}_{\rm eff}^{\rm os}= &\frac{1}{\omega^2} \left[ \mathcal{F}_{a\perp v}(-k) E_{r\perp}^{(2)}(k) + \mathcal{F}_{r\perp v}(-k) E_{a\perp}^{(2)}(k)  \right] \nonumber \\
& + \frac{1}{\omega^2-q^2} \left[ \mathcal{F}_{axv}(-k) E_{rx}^{(2)}(k) + \mathcal{F}_{rx v}(-k) E_{ax}^{(2)}(k)  \right] \nonumber \\
& + \frac{1}{2} \mathcal{F}_{a\perp v}(-k) \mathcal{F}_{r\perp v}(-k) + \frac{1}{2} \mathcal{F}_{ax\perp}(-k) \mathcal{F}_{rx\perp}(k)  + \frac{\omega^2+q^2}{2(\omega^2-q^2)} \mathcal{F}_{axv}(-k) \mathcal{F}_{rxv}(k). \label{W_Leff}
\end{align}
The EOMs \eqref{Eperp_Ex_EOM_EF} are solved similarly to the transverse sector $C_\perp$ in Section \ref{solve_bulk_dynamics}.
The piecewise solutions will be glued under matching conditions derived in subsection \ref{match_cond_Cmu} (imposing the constraint equation $\nabla_M F^{Mr}=0$)
\begin{align}
E_i(r_+)= E_i(r_-), \qquad  f(r_h-\epsilon)\left[\partial_r E_i(r_+)- \partial_r E_i(r_-)\right]=0, \qquad i= \perp, x. \label{matching_Ei}
\end{align}
Near the boundaries
\begin{align}
E_i(r\to \infty_1) = \mathcal{F}_{1i v}, \qquad E_i(r\to \infty_2) = \mathcal{F}_{2i v}, \qquad  i= \perp, x. \label{AdS_condition_Ei}
\end{align}
Over the entire contour of Figure \ref{holographic_SK_contour}, the solutions for $E_\perp$ and $E_x$ are
\begin{align}
& E_\perp^{\rm up}(r,\omega,q)= l_\perp E_\perp^{\rm ig}(r,\omega,q) + m_\perp E_\perp^{\rm ig}(r,-\omega,q) e^{2i\omega \zeta_2(r)}, \qquad \qquad r\in[r_h-\epsilon, \infty_2), \nonumber \\
& E_\perp^{\rm dw}(r,\omega,q)= l_\perp E_\perp^{\rm ig}(r,\omega,q) + m_\perp e^{-\beta\omega} E_\perp^{\rm ig}(r,-\omega,q) e^{2i\omega \zeta_1(r)}, \qquad r\in[r_h-\epsilon, \infty_1), \label{Eperp_up_dw1} \\
& E_x^{\rm up}(r,\omega,q)= l_x E_x^{\rm ig}(r,\omega,q) + m_x E_x^{\rm ig}(r,-\omega,q) e^{2i\omega \zeta_2(r)}, \qquad \qquad r\in[r_h-\epsilon, \infty_2), \nonumber \\
& E_x^{\rm dw}(r,\omega,q)= l_x E_x^{\rm ig}(r,\omega,q) + m_x e^{-\beta\omega} E_x^{\rm ig}(r,-\omega,q) e^{2i\omega \zeta_1(r)}, \qquad r\in[r_h-\epsilon, \infty_1), \label{Ex_up_dw1}
\end{align}
where the superposition coefficients $l_\perp$, $m_\perp$, $l_x$ and $m_x$ are
\begin{align}
& l_\perp = \frac{\mathcal{F}_{r\perp v}(k)}{E_\perp^{\rm ig(0)}(k)} + \frac{1}{2} \coth\frac{\beta \omega}{2} \frac{\mathcal{F}_{a\perp v} (k)} {E_\perp^{\rm ig(0)}(k)}, \qquad m_\perp = - \frac{\mathcal{F}_{a\perp v} (k)} {(1-e^{-\beta \omega}) E_\perp^{\rm ig(0)}(\bar k)}, \nonumber \\
& l_x = \frac{\mathcal{F}_{r xv}(k)}{E_x^{\rm ig(0)}(k)} + \frac{1}{2} \coth\frac{\beta \omega}{2} \frac{\mathcal{F}_{a xv} (k)} {E_x^{\rm ig(0)} (k)}, \qquad m_x = - \frac{\mathcal{F}_{a xv} (k)} {(1-e^{-\beta \omega}) E_x^{\rm ig(0)}(\bar k)}.
\end{align}
The near-boundary expansion of the ingoing solutions is
\begin{align}
E_\perp^{\rm ig}(r,k) \xrightarrow[]{r\to \infty} & E_\perp^{\rm ig(0)}(k) - \frac{i\omega E_\perp^{\rm ig(0)}(k)}{r} + \frac{1}{2}(\omega^2-q^2) E_\perp^{\rm ig(0)}(k) \frac{\log r} {r}   + \frac{E_\perp^{\rm ig(2)}(k)}{r^2}+ \cdots, \nonumber \\
E_x^{\rm ig}(r,k) \xrightarrow[]{r\to \infty} & E_x^{\rm ig(0)}(k) - \frac{i\omega E_x^{\rm ig(0)}(k)}{r} + \frac{1}{2}(\omega^2-q^2) E_x^{\rm ig(0)} (k) \frac{\log r} {r}  + \frac{E_x^{\rm ig(2)}(k)}{r^2}+ \cdots.
\end{align}
Substituting the superposition coefficients and representing the result in the $(r,a)$-basis,
the field's normalisable modes are
\begin{align}
E_{a\perp}^{(2)}(k)=& \frac{E_\perp^{\rm ig(2)}(\bar k)}{E_\perp^{\rm ig(0)}(\bar k)} \mathcal{F}_{a\perp v}(k),\nonumber \\
E_{r\perp}^{(2)}(k)=& \frac{1}{2} \coth\frac{\beta\omega}{2} \left[\frac{E_\perp^{\rm ig(2)}(k)}{E_\perp^{\rm ig(0)}(k)} - \frac{E_\perp^{\rm ig(2)}(\bar k)}{E_\perp^{\rm ig(0)}(\bar k)} \right] \mathcal{F}_{a\perp v}(k) + \frac{E_\perp^{\rm ig(2)}(k)}{E_\perp^{\rm ig(0)}(k)} \mathcal{F}_{r \perp v}(k), \nonumber \\
E_{ax}^{(2)}(k)= & \frac{E_x^{\rm ig(2)}(\bar k)}{E_x^{\rm ig(0)}(\bar k)} \mathcal{F}_{a xv}(k), \nonumber  \\
E_{rx}^{(2)}(k)= & \frac{1}{2} \coth\frac{\beta\omega}{2} \left[\frac{E_x^{\rm ig(2)}(k)}{E_x^{\rm ig(0)}(k)} - \frac{E_x^{\rm ig(2)}(\bar k)} {E_x^{\rm ig(0)}(\bar k)} \right] \mathcal{F}_{ax v}(k) + \frac{E_x^{\rm ig(2)}(k)}{E_x^{\rm ig(0)}(k)} \mathcal{F}_{r x v} (k).
\end{align}
Finally, \eqref{W_Leff} reads
\begin{align}
\mathcal{L}_{\rm eff}^{\rm os}=& \frac{1}{\omega^2} \mathcal{F}_{a\perp v}(-k) \frac{1}{2} \coth\frac{\beta \omega}{2} \left[ \frac{E_\perp^{\rm ig(2)}(k)} {E_\perp^{\rm ig(0)}(k)} - \frac{E_\perp^{\rm ig(2)}(\bar k)} {E_\perp^{\rm ig(0)}(\bar k)}  \right] \mathcal{F}_{a\perp v}(k) \nonumber \\
& + \frac{1}{\omega^2} \mathcal{F}_{a\perp v}(-k)\left[\frac{2E_\perp^{\rm ig(2)} (k)} {E_\perp^{\rm ig(0)}(k)} + \frac{1}{2}\omega^2 + \frac{1}{2}q^2 \right] \mathcal{F}_{r\perp v}(k) \nonumber \\
& + \frac{1}{\omega^2-q^2} \mathcal{F}_{axv}(-k) \frac{1}{2} \coth\frac{\beta \omega}{2} \left[ \frac{E_x^{\rm ig(2)}(k)} {E_x^{\rm ig(0)}(k)} - \frac{E_x^{\rm ig(2)}(\bar k)} {E_x^{\rm ig(0)}(\bar k)}  \right] \mathcal{F}_{axv}(k) \nonumber \\
& + \frac{1}{\omega^2-q^2} \mathcal{F}_{axv}(-k)\left[\frac{2E_x^{\rm ig(2)} (k)} {E_x^{\rm ig(0)}(k)} + \frac{1}{2}\omega^2+ \frac{1}{2}q^2 \right] \mathcal{F}_{rxv}(k). \label{W_Leff_final}
\end{align}
From the generating functional $W$, it is straightforward to read off all two-point correlation functions:
\begin{align}
& G_R^{\perp \perp}=\Pi^T(k), \qquad \qquad  G_S^{\perp \perp}= G^T(k), \nonumber \\
& G_R^{vv}= \frac{q^2}{\omega^2-q^2} \Pi^L(k), \quad G_R^{vx}=  \frac{\omega q} {\omega^2-q^2} \Pi^L(k), \quad G_R^{xx}= \frac{\omega^2}{\omega^2-q^2} \Pi^L(k), \nonumber \\
& G_S^{vv}= \frac{q^2}{\omega^2-q^2} G^L(k), \quad G_S^{vx}= \frac{\omega q} {\omega^2-q^2} G^L(k), \quad G_S^{xx}= \frac{\omega^2}{\omega^2-q^2} G^L(k),
\end{align}
where
\begin{align}
&\Pi^T(k)= \frac{2E_\perp^{\rm ig(2)} (k)} {E_\perp^{\rm ig(0)}(k)} + \frac{1}{2}\omega^2 + \frac{1}{2}q^2, \quad \Pi^L(\omega,q)= \frac{2E_x^{\rm ig(2)} (k)} {E_x^{\rm ig(0)}(k)} + \frac{1}{2}\omega^2+ \frac{1}{2}q^2, \nonumber \\
&G^T(k)= \frac{1}{2} \coth\frac{\beta \omega}{2} \left[ \frac{E_\perp^{\rm ig(2)} (k)} {E_\perp^{\rm ig(0)}(k)} - \frac{E_\perp^{\rm ig(2)} (\bar k)} {E_\perp^{\rm ig(0)}(\bar k)}  \right], \nonumber \\
&G^L(k)= \frac{1}{2} \coth\frac{\beta \omega}{2} \left[ \frac{E_x^{\rm ig(2)}(k)} {E_x^{\rm ig(0)}(k)} - \frac{E_x^{\rm ig(2)}(\bar k)} {E_x^{\rm ig(0)}(\bar k)}  \right]. \label{G_Pi}
\end{align}
From the EOMs \eqref{Eperp_Ex_EOM_EF}, it is clear that $E_\perp, E_x$ are  functions of $q^2$. So, \eqref{G_Pi} satisfy the FDRs:
\begin{align}
G^T(\omega,q)= \frac{1}{2}\coth\frac{\beta \omega}{2} {\rm Im}\left[\Pi^T(\omega,q) \right], \qquad G^L(\omega,q)= \frac{1}{2}\coth\frac{\beta \omega}{2} {\rm Im} \left[\Pi^L(\omega,q) \right].
\end{align}
We have reproduced the prescription of \cite{Son:2002sd} for the retarded correlators. A couple of comments are in order. The above derivations have not imposed any KMS-type conditions, rather they follow from the SK holography.
The original prescription of \cite{Son:2002sd} correctly but a-priori unjustifiably ignores the  horizon contribution
to the on-shell action.  From our derivation it is clear that only two AdS boundaries contribute to the boundary generating functional.

\section{Numerical results for  $w_5,w_7,w_8,w_9$} \label{numTCFs}

\begin{figure}
    \centering
    \begin{subfigure}[h]{0.49\textwidth}
        \includegraphics[width=\textwidth]{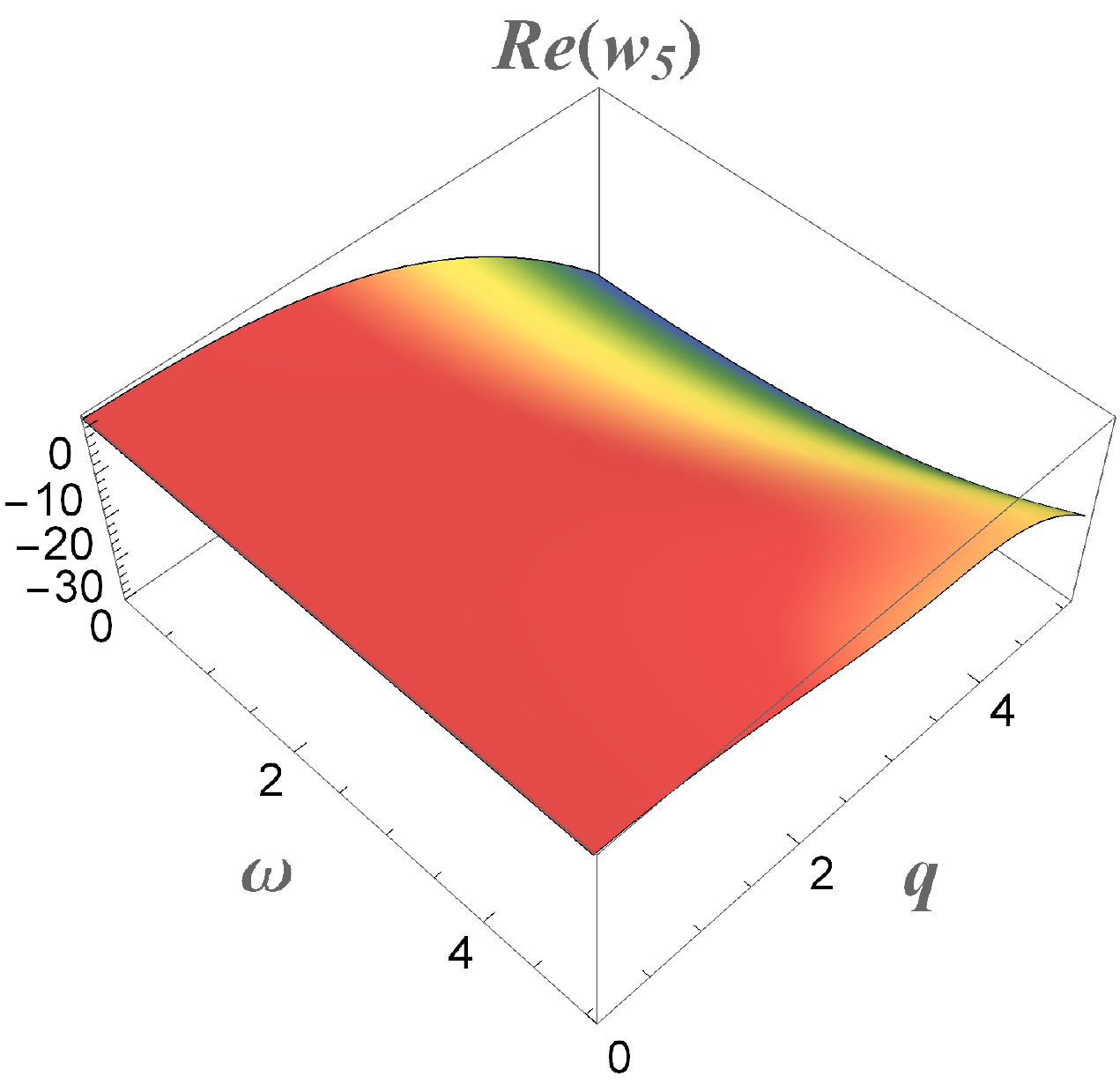}
        \caption{}
        \label{rw5p}

    \end{subfigure}
      \begin{subfigure}[h]{0.49\textwidth}
        \includegraphics[width=\textwidth]{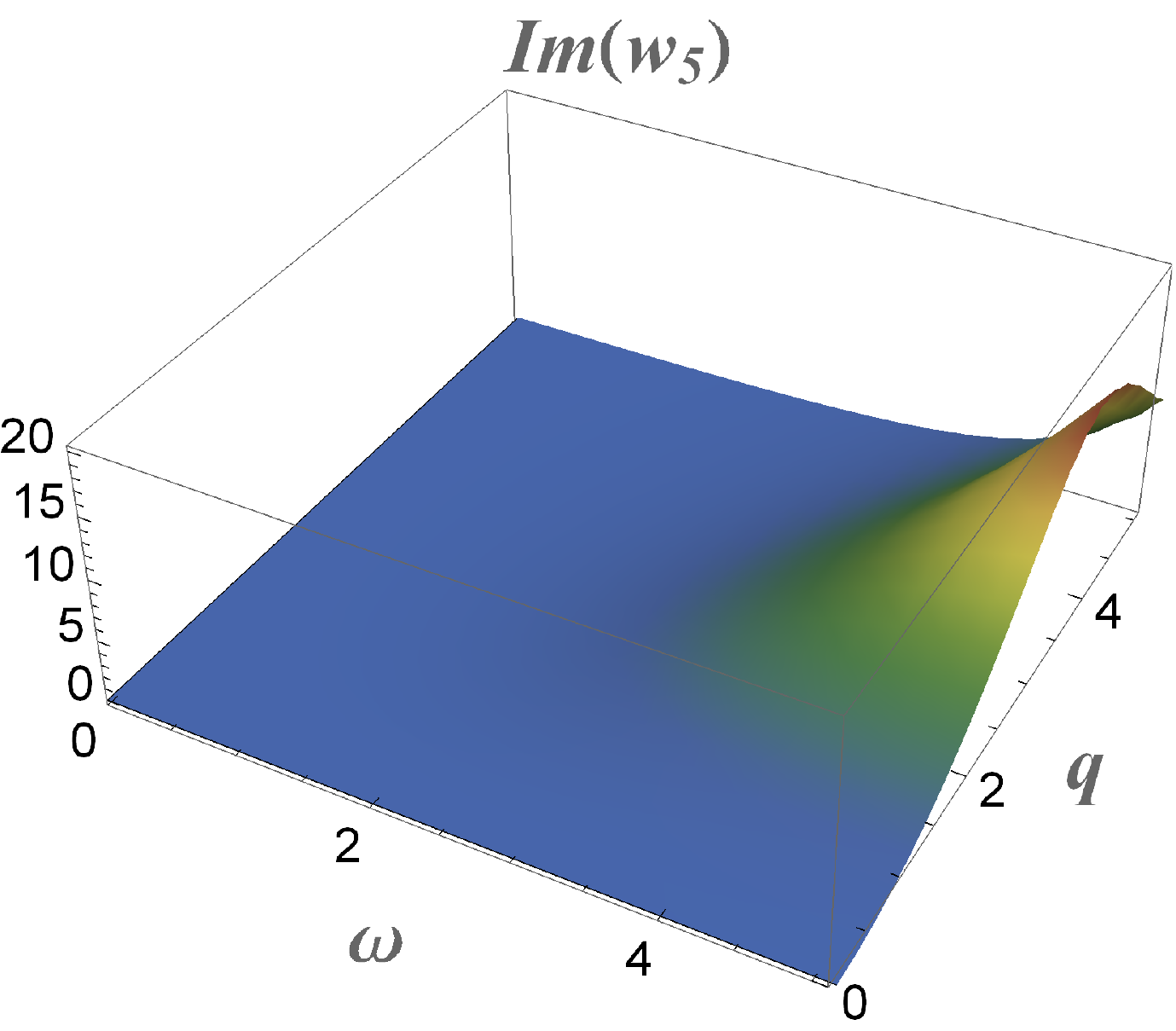}
        \caption{}

    \end{subfigure}
%
%
%
    \caption{Plots for $\omega$-, $q$-dependence of the coefficients (a) $Re(w_5(\omega,q))$, (b) $Im(w_5(\omega,q))$.}
    \label{w5p}
\end{figure}

\begin{figure}
    \centering
    \begin{subfigure}[h]{0.49\textwidth}
        \includegraphics[width=\textwidth]{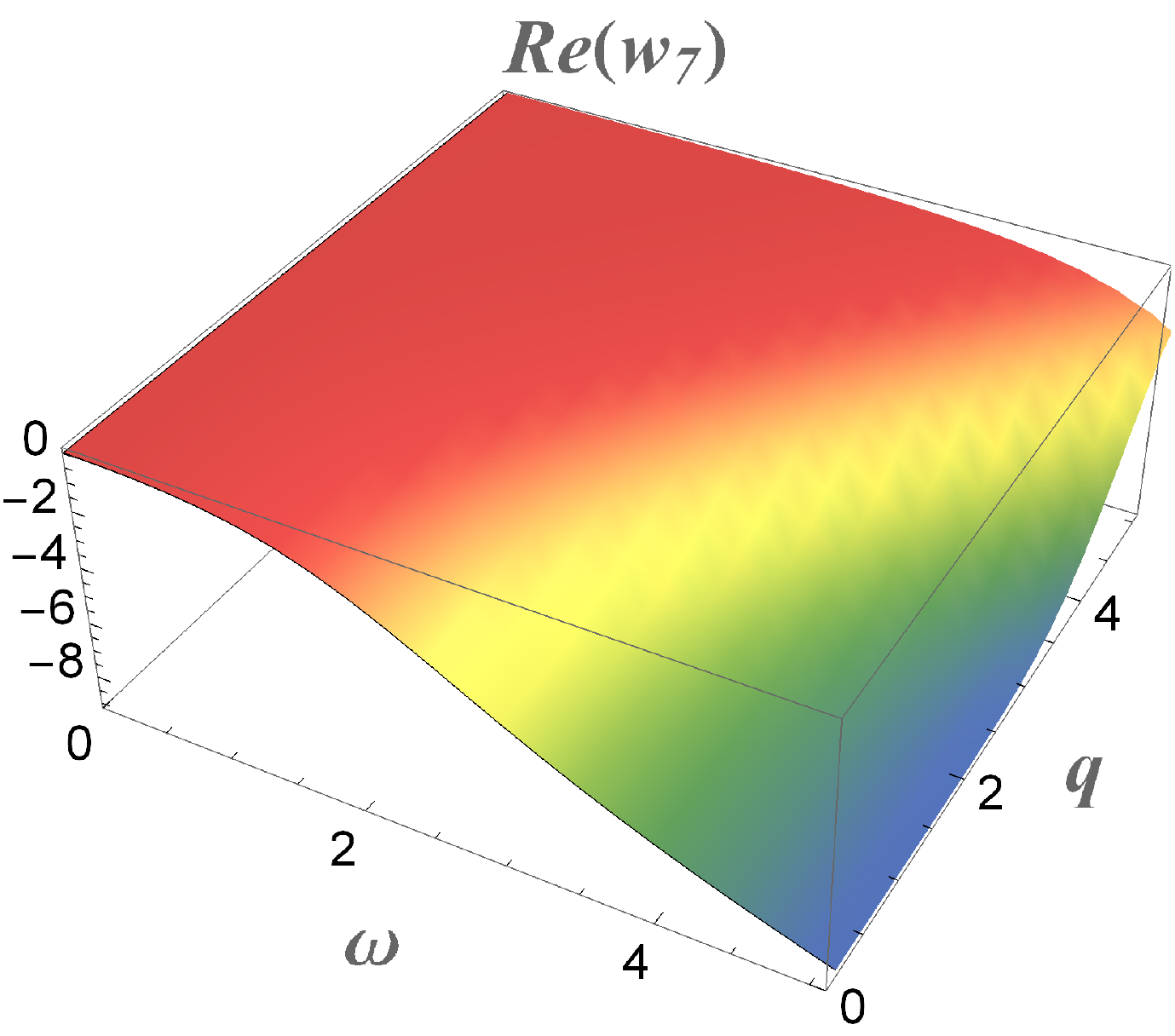}
        \caption{}
        \label{iw7p}
    \end{subfigure}
      \begin{subfigure}[h]{0.49\textwidth}
        \includegraphics[width=\textwidth]{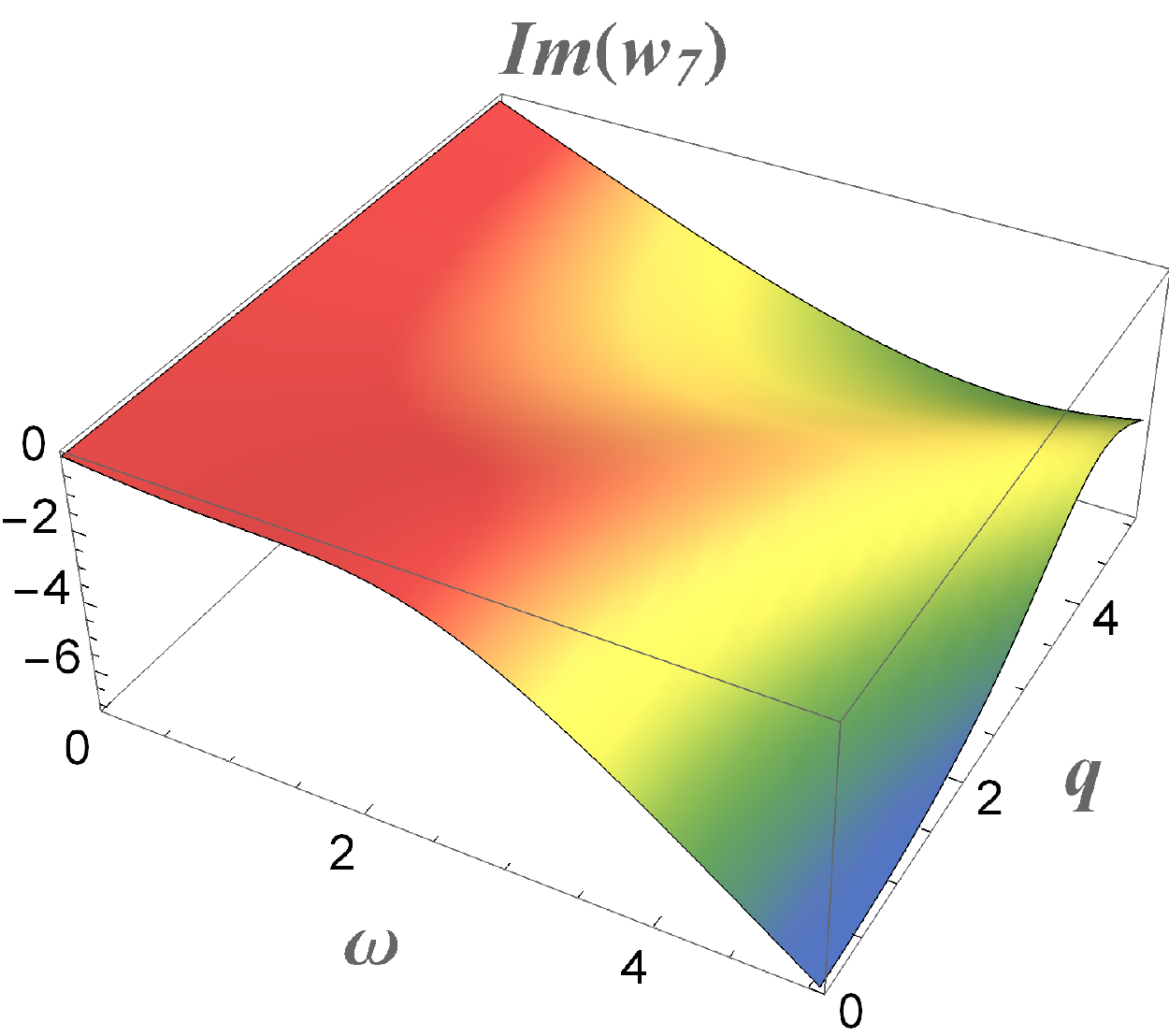}
        \caption{}
        \label{}
    \end{subfigure}

    \caption{Plots for $\omega$-, $q$-dependence of the coefficients (a) $Re(w_7(\omega,q))$, (b) $Im(w_7(\omega,q))$.}\label{w7p}
\end{figure}

\begin{figure}
    \centering
      \begin{subfigure}[h]{0.49\textwidth}
        \includegraphics[width=\textwidth]{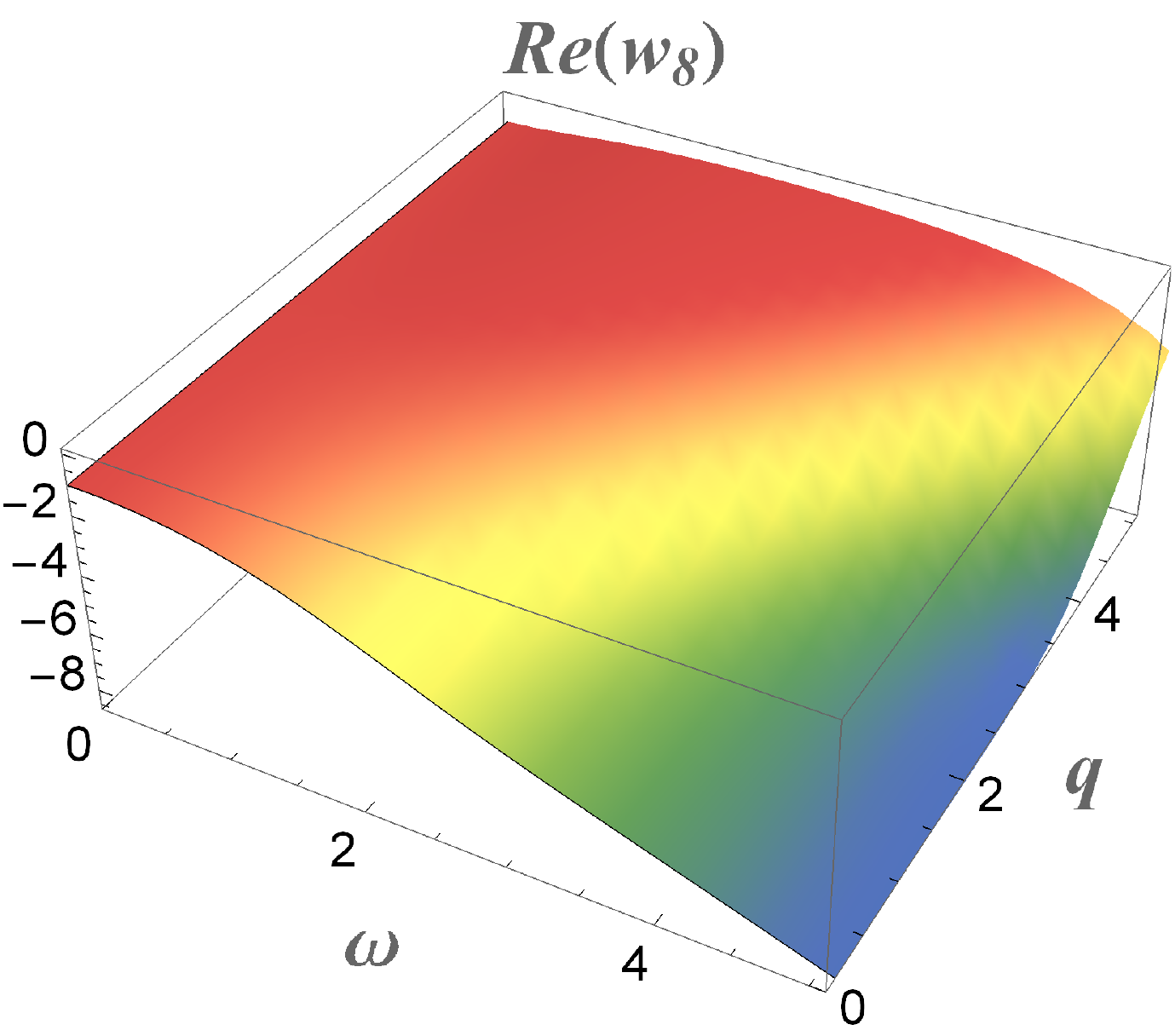}
        \caption{}
        \label{}
    \end{subfigure}
      \begin{subfigure}[h]{0.49\textwidth}
        \includegraphics[width=\textwidth]{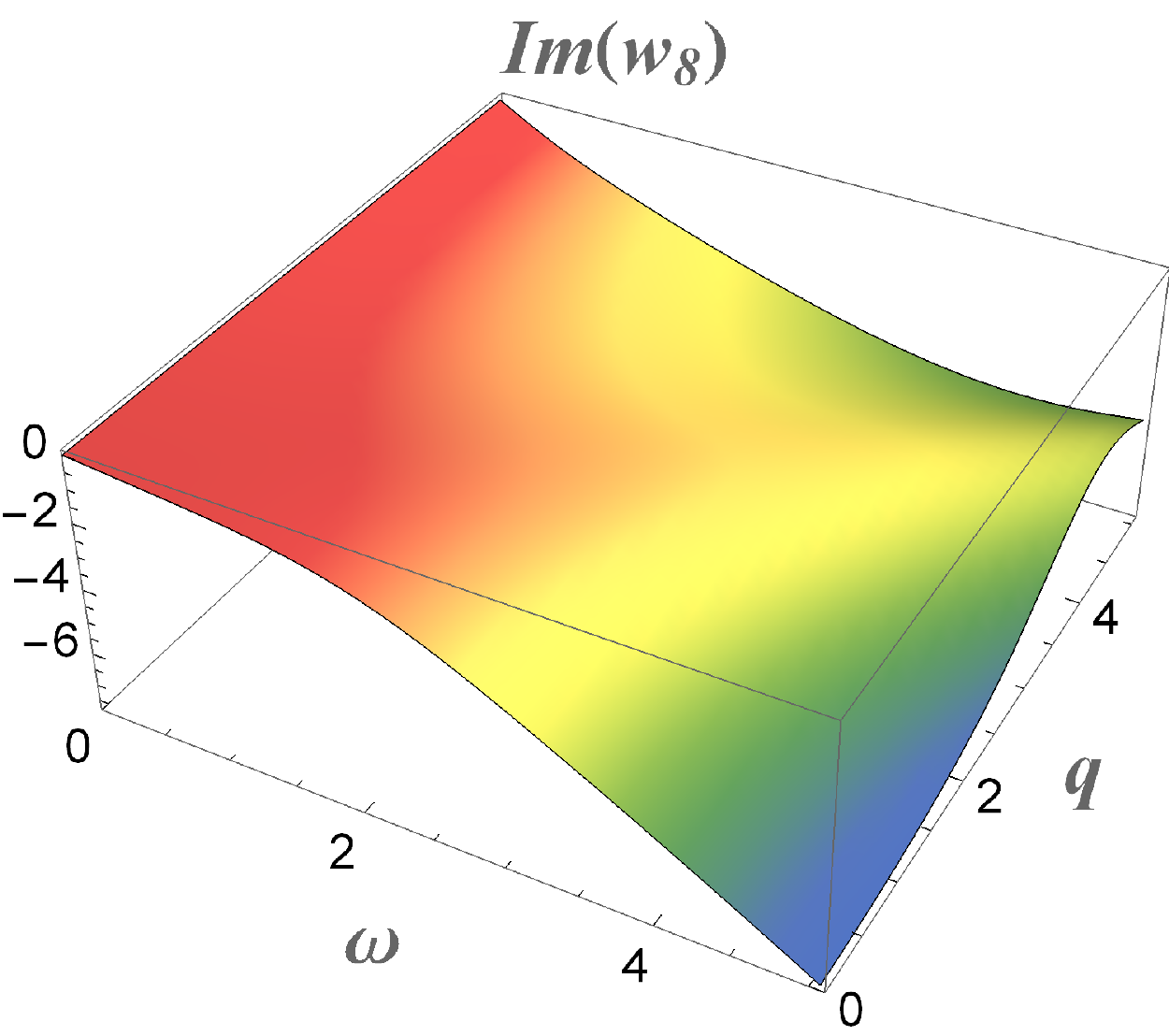}
        \caption{}
        \label{iw8p}
    \end{subfigure}
    \caption{Plots for $\omega$-, $q$-dependence of the coefficients (a) $Re(w_8(\omega,q))$, (b) $Im(w_8(\omega,q))$.}\label{w8p}
\end{figure}

\begin{figure}
    \centering
    \begin{subfigure}[h]{0.49\textwidth}
        \includegraphics[width=\textwidth]{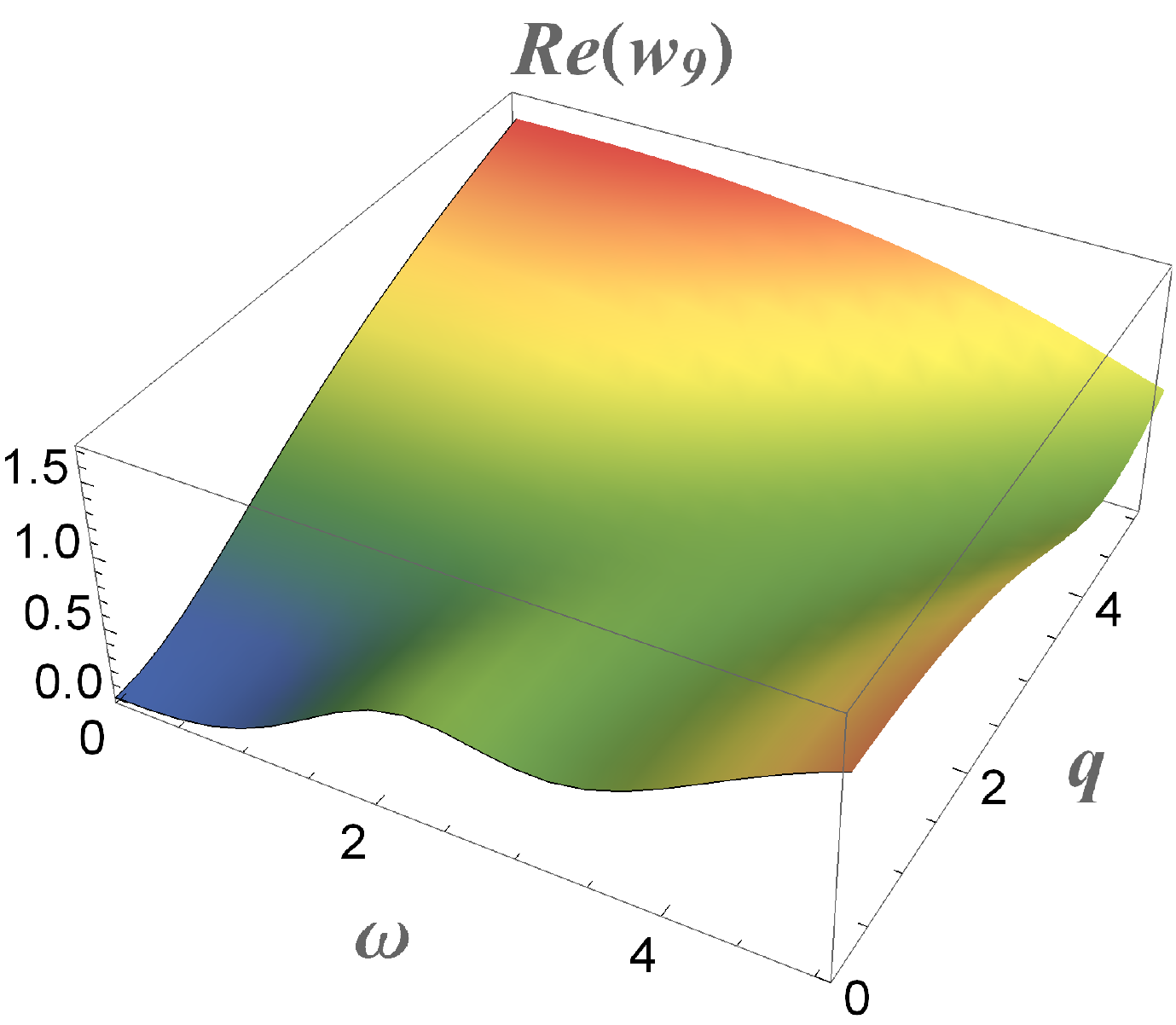}
        \caption{}
        \label{}
    \end{subfigure}
      \begin{subfigure}[h]{0.49\textwidth}
        \includegraphics[width=\textwidth]{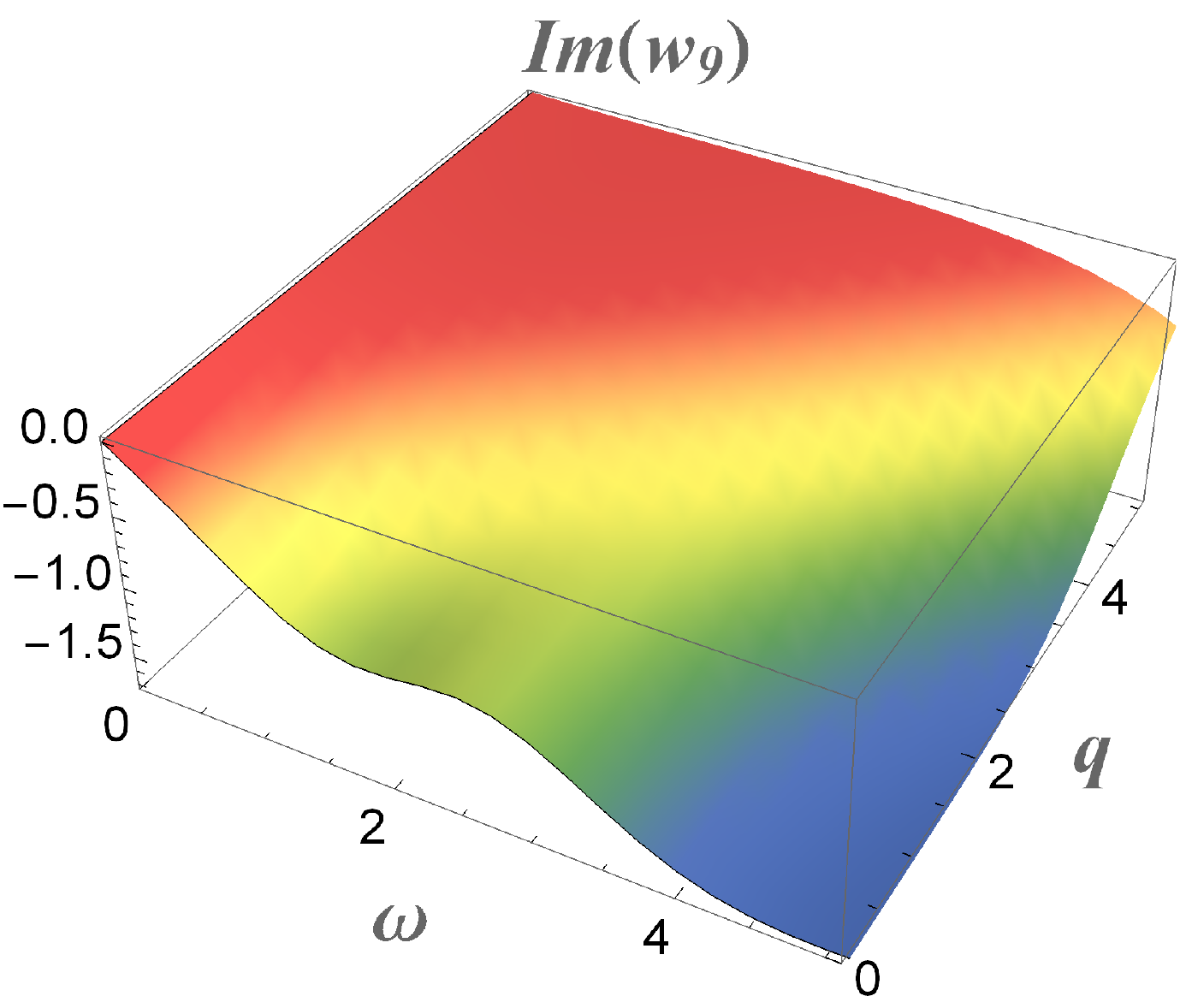}
        \caption{}
        \label{}
    \end{subfigure}
    \caption{Plots for $\omega$-, $q$-dependence of the coefficients (a) $Re(w_9(\omega,q))$, (b) $Im(w_9(\omega,q))$.}\label{w9p}
\end{figure}

The results for $\omega$-, $q$-dependence of the coefficients $w_5$, $w_7$, $w_8$ and $w_9$ are displayed as 3D plots in Figures \ref{w5p}, \ref{w7p}, \ref{w8p}, \ref{w9p} respectively.
There is no clear universal pattern in the functional dependencies of these TCFs.
They display different asymptotic behaviours at large momenta. For example,
imaginary part of $w_5$ develops a growing ridge-like structure in the $\omega\simeq q$ region;   imaginary parts of both $w_7$ and $w_8$ display a  decreasing ridge-like behaviour
also in the vicinity of $\omega\simeq q$ domain. For larger values of frequency (not shown in the plots), the amplitudes of all the TCFs ($w_5,w_7,w_8,w_9$) seem to keep on growing.
Each individual coefficient $w_i$ does not seem to have a clear physical interpretation
and this is the reason in the main part of the text we rather focus on the diffusion TCF $\mathcal D$ and
conductivities $\sigma_e$ and $\sigma_m$ only.


It is important to notice that  some of the TCFs $w_i$ vanish in the hydro limit.
Yet, all of the TCFs are non-zero at finite momenta and contribute non-trivially to the effective action. In a sense this constitutes an evolution
of the effective action from IR (hydro regime) to UV (all order/large momenta regime).

\section{Clarifying the origin of the discrepancy with  \cite{Bu:2015ame}} \label{discrepancy}

As discussed in the main text, our present results for the TCFs $\mathcal D$, $\sigma_e$ and $\sigma_m$ differ
from the ones obtained  previously in \cite{Bu:2015ame}.
In this Appendix, we identify the origin of this discrepancy.

The analysis of \cite{Bu:2015ame} is carried  in a single Schwarzschild-AdS$_5$. Instead of gluing the bulk
solutions  at the horizon implemented in the SK holography, regularity condition on the bulk fields was
imposed in \cite{Bu:2015ame}. This is equivalent to  setting all the $a$-type fields, $B_{a\mu}$, to zero from the
very start, at the level of equations of motion, thus making two segments of the SK contour identical.
This immediately implies that an analog of
the effective action along the SK contour vanishes, and, obviously, there is no possibility to vary it with
respect to $B_{a\mu}$. Hence, the off shell   formalism of \cite{Bu:2015ame} is not embeddable
into the  fundamental framework based on SK non-equilibrium field theory.

Next, we are going to explicitly demonstrate how setting $B_{a\mu}=0$ at the very beginning
leads to different  constitutive relations compared to the ones of the present paper, when
 $B_{a\mu}=0$ is imposed at the very end of the calculation.


The horizon regularity condition is equivalent to having no outgoing mode. That is $h_\perp=h_{||}=0$.  Hence, as mentioned, $B_{a\mu}=0$ and under this condition  the bulk solutions of the present work are essentially identical
\footnote{In \cite{Bu:2015ame}  a different choice of the residual gauge was implemented. It is, however, immaterial for the present discussion.} to those of \cite{Bu:2015ame}.
Following the prescription of \cite{Son:2002sd}, the  hydrodynamic current in  \cite{Bu:2015ame}
was identified with the normalisable modes of $C_\mu$. That is, using  present language,
\begin{align}
 J_v^{\rm BLS}  = 2 C_{rv}^{(2)}(k)|_{B_{a\mu}=0} + \cdots, \qquad J_i^{\rm BLS}= 2 C_{ri}^{(2)}(k)|_{B_{a\mu}=0}+ \cdots, \qquad i= \perp, x,
\end{align}
where  $\cdots$ stand for  contact terms,  the last six terms of \eqref{Leff1}.
This current is to be compared with the hydrodynamical current $J_{hydro}$ introduced in \eqref{Jrmu}.

In  the transverse sector the two currents are  equal,  $J^\perp_{hydro}=J_\perp^{\rm BLS} $.
Indeed,  the current $J^\perp_{hydro}$ derived  from the effective Lagrangian \eqref{Lperp1} is
\begin{align}
J^\perp_{hydro}= \frac{2 C_\perp^{\rm ig(2)}(k)}{C_\perp^{\rm ig(0)}(k)} B_{r\perp}(k) + \cdots, \label{Jperp_hydro}
\end{align}
which is exactly $2 C_{r\perp}^{(2)}(k)|_{B_{a\mu}=0}$, cf. \eqref{Cperp_norma_modes}.
The agreement within the transverse sector is related to the fact that the transverse current satisfies the
continuity equation  automatically, in this sense it is always on-shell.

The disagreement is entirely within the longitudinal sector and within the off-shell formalism only. On-shell,
the results agree
From the effective Lagrangian \eqref{Lparallel_10}, the hydrodynamic current  is
\begin{align}
 J^v_{hydro}  &=  C_{rv}^{(2)}|_{B_{a\mu}=0} + \left[B_{rv}\frac{\delta C_{av}^{(2)} }{ \delta B_{av}}-
 B_{rx}\frac{\delta C_{ax}^{(2)}}{ \delta B_{av}}\right]
 +\cdots, \nonumber \\
  J^x_{hydro} &=  C_{rx}^{(2)}|_{B_{a\mu}=0}+ \left[B_{rx}\frac{\delta C_{ax}^{(2)} }{ \delta B_{ax}}-
 B_{rv}\frac{\delta C_{av}^{(2)}}{ \delta B_{ax}}\right]
 + \cdots
\end{align}
Once the explicit expressions  \eqref{Crv_norma_mode} and \eqref{Crx_norma_mode} are used,
it is possible to demonstrate that $J^{||}_{hydro} \neq J_{||}^{\rm BLS}$ with the difference being proportional to $n=n^{\rm{up}}_{||}=n^{\rm{dw}}_{||}$m ($n^{\rm{up}}_{||}=n^{\rm{dw}}_{||}$ when $B_{a\mu}=0$ ).
On-shell, $n=0$,  and the results agree
and lead to the very same current-current correlators discussed in the previous Appendix.

\section*{Acknowledgements}

We would like to thank Xin Gao, Song He, Shu Lin, Gao-Liang Zhou and Tianchun Zhou for useful discussions. YB was supported by the Natural Science Foundation of China (NSFC) under the grant No.11705037. TD and ML were supported by the Israeli Science Foundation (ISF) grant \#1635/16 and the BSF grants \#2012124 and \#2014707. TD was supported in part by the JRG Program at the APCTP through the Science and Technology Promotion Fund and Lottery Fund of the Korean
Government and also by the Korean Local Governments --- Gyeongsangbuk-do Province and Pohang City.

\providecommand{\href}[2]{#2}\begingroup\raggedright\endgroup

\end{document}